\documentclass[preprint,12pt]{elsarticle}

\journal{Journal of Computational Physics}

\usepackage[utf8x]{inputenc}         
\usepackage{ucs}                     

\usepackage[ngerman,english]{babel}  
\usepackage{colortbl}                
\usepackage[table]{xcolor}
\usepackage{booktabs}
\usepackage{tabularx}                
  \newcolumntype{L}[1]{>{\raggedright\arraybackslash}p{#1}}
  \newcolumntype{C}[1]{>{\centering\arraybackslash}p{#1}}
  \newcolumntype{R}[1]{>{\raggedleft\arraybackslash}p{#1}}
  
\usepackage{xcolor}

\definecolor{hzdr-blue}    {rgb}{0,0.345098039216,0.611764705882}
\definecolor{hzdr-orange}  {rgb}{0.811764705882,0.407843137255,0}
\definecolor{hzdr-darkblue}{rgb}{0,0.243137254902,0.43137254902}
\definecolor{hzdr-gray1}   {gray}{0.317647058824}
\definecolor{hzdr-gray2}   {gray}{0.611764705882}
\definecolor{hzdr-gray3}   {gray}{0.725490196078}
\definecolor{hzdr-gray4}   {gray}{0.8}
\definecolor{hzdr-gray5}   {gray}{0.85}
\definecolor{hzdr-health}  {rgb}{0.83137254902,0.176470588235,0.0705882352941}
\definecolor{hzdr-struct}  {rgb}{0.811764705882,0.407843137255,0}
\definecolor{hzdr-energy}  {rgb}{0.901960784314,0.686274509804,0.0666666666667}
\definecolor{hzdr-earth}   {rgb}{0.078431372549,0.301960784314,0.156862745098}
\definecolor{hzdr-keytec}  {rgb}{0.662745098039,0.709803921569,0.0352941176471}
\definecolor{hzdr-aero}    {rgb}{0,0.635294117647,0.878431372549}
\definecolor{hzdr-lila}    {rgb}{0.39215686,0.039215686,0.47058824}

\usepackage{amsmath,amssymb,amsthm,amsfonts,mathrsfs}
\usepackage{units}

    \newcommand{\abs}[1]{\left| #1 \right|}
    \newcommand{\norm}[1]{\lVert #1 \rVert}

    \renewcommand{\Re}[1]{\operatorname{Re}\left( #1 \right)}
    \renewcommand{\Im}[1]{\operatorname{Im}\left( #1 \right)}
    \newcommand{\re}[1]{{#1}_{\scriptstyle\operatorname{Re}}}
    \newcommand{\im}[1]{{#1}_{\scriptstyle\operatorname{Im}}}
    \newcommand{\rei}[2]{{#1}_{#2;\,{\scriptstyle\operatorname{Re}}}}
    \newcommand{\imi}[2]{{#1}_{#2;\,{\scriptstyle\operatorname{Im}}}}
    \newcommand{\reim}[1]{{#1}_{{\scriptstyle\operatorname{Re}},\,{\scriptstyle\operatorname{Im}}}}

    \newcommand{\C}{\mathbb{C}}
    \newcommand{\phasor}[1]{#1_\C}
    
    \newcommand{\phasoramp}[1]{\hat{#1}_\C}

    \renewcommand{\vec}[1]{\boldsymbol{#1}}
    
    \newcommand{\grad}[1]{\nabla #1}
    
    \newcommand{\laplace}[1]{\nabla^2 #1}
    \renewcommand{\div}[1]{\nabla \cdot #1}
    
    \newcommand{\rot}[1]{\nabla \times #1}
    
    \newcommand{\shortpd}[2]{\partial_{#2} #1}
    
    \newcommand{\shortpdd}[2]{\partial_{#2 #2} #1}

    \newcommand{\intvar}[1]{\; \mathrm{d}#1}
    \newcommand{\vintvar}[1]{\cdot \mathrm{d}\vec{#1}}
    \newcommand{\integral}[4]{\int_{#1}^{#2} #3 \intvar{#4}}
    \newcommand{\vintegral}[4]{\int_{#1}^{#2} #3 \vintvar{#4}}

    \newcommand{\Sum}[3]{\sum_{#1}^{#2} #3}

    \newcommand{\avg}[2]{\left<#1\right>_{#2}}

    \newcommand{\bc}[2]{\left. #1 \right|_{#2}}
    \newcommand{\jump}[1]{\left[ #1 \right]}
    \newcommand{\jumpP}[1]{{#1}^{+}}
    \newcommand{\jumpN}[1]{{#1}^{-}}

    \newcommand{\iter}[2]{#1^{\, \left\{ #2 \right\}}}
    
    \newcommand{\overbar}[1]{\mkern 1.5mu\overline{\mkern-1.5mu#1\mkern-1.5mu}\mkern 1.5mu}
    \newcommand{\degree}{\ensuremath{^\circ}}

\usepackage{tikz}
\usepackage{tikzscale}
\usepackage{tikz-3dplot}
\usepackage{environ}

\usetikzlibrary{calc}
\usetikzlibrary{arrows}
\usetikzlibrary{babel,quotes}
\usetikzlibrary{angles}
\usetikzlibrary{positioning}
\usetikzlibrary{intersections}
\usetikzlibrary{patterns}
\usetikzlibrary{decorations.markings}
\usetikzlibrary{decorations.pathmorphing}
\usetikzlibrary{decorations.pathreplacing}
\usetikzlibrary{shapes.arrows}

\usetikzlibrary{external}
\newcommand{\forceInlcudeExternalPdf}{}

\newcommand{\includetikz}[2][0.99]{%
  \ifx\forceInlcudeExternalPdf\undefined
    \tikzsetnextfilename{#2}%
    \newcommand{\tikzresizefactor}{#1}%
    \resizebox{#1\linewidth}{!}{\includegraphics{#2.tikz}}
  \else
    \resizebox{#1\linewidth}{!}{\includegraphics{#2.pdf}}%
  \fi
}

\NewEnviron{includeannotatetikz}[2][0.99]{%
  \newcommand{\tikzresizefactor}{#1}%

  \tikzexternaldisable
  \tikzsetnextfilename{#2-annotated}
  \resizebox{\tikzresizefactor\linewidth}{!}{%
    \begin{tikzpicture}[scale=1.0/\tikzresizefactor]
      \node[%
        anchor=south west,
        inner sep=0pt,%
        text width=\linewidth
      ] (image) at (0,0) {%
        \includegraphics[width=\linewidth]{#2}
      };
      \begin{scope}[x={(image.south east)},y={(image.north west)}]
        {\BODY};
      \end{scope}
    \end{tikzpicture}
  }
}

\makeatletter
\newcommand{\pgfcalcxy}[4]{%
  \path (#2); \pgfgetlastxy{\gettikzxyxpt}{\gettikzxyypt};%
  \pgfmathsetmacro{#3}{\gettikzxyxpt/\pgf@xx*#1};%
  \pgfmathsetmacro{#4}{\gettikzxyypt/\pgf@yy*#1};%
}
\newcommand{\pgfcalclengthx}[3]{%
  \pgfmathsetmacro{#1}{#3\pgf@xx/#2};%
}
\newcommand{\pgfcalclengthy}[3]{%
  \pgfmathsetmacro{#1}{#3\pgf@yy/#2};%
}
\makeatother

\makeatletter
\pgfarrowsdeclare{centero}{centero}
{
  \pgfarrowsleftextend{+-.5\pgflinewidth}
  \pgfutil@tempdima=0.4pt%
  \advance\pgfutil@tempdima by.2\pgflinewidth%
  \pgfarrowsrightextend{4.5\pgfutil@tempdima}
}
{
  \pgfutil@tempdima=0.4pt%
  \advance\pgfutil@tempdima by.2\pgflinewidth%
  \pgfsetdash{}{+0pt}
  \pgfpathcircle{\pgfqpoint{4.5\pgfutil@tempdima}{0bp}}{4.5\pgfutil@tempdima}
  \pgfusepathqstroke
}
\makeatother

\makeatletter
\pgfarrowsdeclare{center*}{center*}
{
  \pgfarrowsleftextend{+-.5\pgflinewidth}
  \pgfutil@tempdima=0.4pt%
  \advance\pgfutil@tempdima by.2\pgflinewidth%
  \pgfarrowsrightextend{4.5\pgfutil@tempdima}
}
{
  \pgfutil@tempdima=0.4pt%
  \advance\pgfutil@tempdima by.2\pgflinewidth%
  \pgfsetdash{}{+0pt}
  \pgfpathcircle{\pgfqpoint{4.5\pgfutil@tempdima}{0bp}}{4.5\pgfutil@tempdima}
  \pgfusepathqfillstroke
}
\makeatother

\newlength{\hatchspread}
\newlength{\hatchthickness}
\newlength{\hatchshift}
\newcommand{\hatchcolor}{}
\tikzset{hatchspread/.code={\setlength{\hatchspread}{#1}},
         hatchthickness/.code={\setlength{\hatchthickness}{#1}},
         hatchshift/.code={\setlength{\hatchshift}{#1}},
         hatchcolor/.code={\renewcommand{\hatchcolor}{#1}}}
\tikzset{hatchspread=3pt,
         hatchthickness=0.4pt,
         hatchshift=0pt,
         hatchcolor=black}
\pgfdeclarepatternformonly[\hatchspread,\hatchthickness,\hatchshift,\hatchcolor]
  {custom horizontal lines}
  {\pgfqpoint{0pt}{\dimexpr\hatchshift}}
  {\pgfqpoint{100pt}{\dimexpr\hatchshift+2\hatchthickness+0.2pt}}
  {\pgfqpoint{100pt}{\dimexpr\hatchspread}}
  {
    \pgfsetlinewidth{\hatchthickness}
    \pgfpathmoveto{\pgfqpoint{0pt}{\dimexpr\hatchshift+\hatchthickness+0.1pt}}
    \pgfpathlineto{\pgfqpoint{100pt}{\dimexpr\hatchshift+\hatchthickness+0.1pt}}
    \pgfsetstrokecolor{\hatchcolor}
    \pgfusepath{stroke}
  }
\pgfdeclarepatternformonly[\hatchspread,\hatchthickness,\hatchshift,\hatchcolor]
  {custom vertical lines}
  {\pgfqpoint{\dimexpr\hatchshift}{0pt}}
  {\pgfqpoint{\dimexpr\hatchshift+2\hatchthickness+0.2pt}{100pt}}
  {\pgfqpoint{\dimexpr\hatchspread}{100pt}}
  {
    \pgfsetlinewidth{\hatchthickness}
    \pgfpathmoveto{\pgfqpoint{\dimexpr\hatchshift+\hatchthickness+0.1pt}{0pt}}
    \pgfpathlineto{\pgfqpoint{\dimexpr\hatchshift+\hatchthickness+0.1pt}{100pt}}
    \pgfsetstrokecolor{\hatchcolor}
    \pgfusepath{stroke}
  }
\pgfdeclarepatternformonly[\hatchspread,\hatchthickness,\hatchshift,\hatchcolor]
  {custom north west lines}
  {\pgfqpoint{\dimexpr-2\hatchthickness}{\dimexpr-2\hatchthickness}}
  {\pgfqpoint{\dimexpr\hatchspread+2\hatchthickness}{\dimexpr\hatchspread+2\hatchthickness}}
  {\pgfqpoint{\dimexpr\hatchspread}{\dimexpr\hatchspread}}
  {
    \pgfsetlinewidth{\hatchthickness}
    \pgfpathmoveto{\pgfqpoint{0pt}{\dimexpr\hatchspread+\hatchshift}}
    \pgfpathlineto{\pgfqpoint{\dimexpr\hatchspread+0.15pt+\hatchshift}{-0.15pt}}
    \ifdim \hatchshift > 0pt
      \pgfpathmoveto{\pgfqpoint{0pt}{\hatchshift}}
      \pgfpathlineto{\pgfqpoint{\dimexpr0.15pt+\hatchshift}{-0.15pt}}
    \fi
    \pgfsetstrokecolor{\hatchcolor}
    \pgfusepath{stroke}
  }
\pgfdeclarepatternformonly[\hatchspread,\hatchthickness,\hatchshift,\hatchcolor]
  {custom north east lines}
  {\pgfqpoint{\dimexpr-2\hatchthickness}{\dimexpr-2\hatchthickness}}
  {\pgfqpoint{\dimexpr\hatchspread+2\hatchthickness}{\dimexpr\hatchspread+2\hatchthickness}}
  {\pgfqpoint{\dimexpr\hatchspread}{\dimexpr\hatchspread}}
  {
    \pgfsetlinewidth{\hatchthickness}
    \pgfpathmoveto{\pgfqpoint{\dimexpr\hatchshift-0.15pt}{-0.15pt}}
    \pgfpathlineto{\pgfqpoint{\dimexpr\hatchspread+0.15pt}{\dimexpr\hatchspread-\hatchshift+0.15pt}}
    \ifdim \hatchshift > 0pt
      \pgfpathmoveto{\pgfqpoint{-0.15pt}{\dimexpr\hatchspread-\hatchshift-0.15pt}}
      \pgfpathlineto{\pgfqpoint{\dimexpr\hatchshift+0.15pt}{\dimexpr\hatchspread+0.15pt}}
    \fi
    \pgfsetstrokecolor{\hatchcolor}
    \pgfusepath{stroke}
   }

\usepackage{framed}

  \newtheoremstyle{myboxstyle}
    {3pt}         
    {3pt}         
    {\itshape}        
    {}          
    {\normalfont\bfseries}      
    {:}         
    { }         
    {}          

  \theoremstyle{myboxstyle}
  \newtheorem*{remark}{Remark}
  \newtheorem*{question}{Question}
  \newtheorem*{todo}{Task}

\usepackage[bookmarksopen,bookmarksdepth=2,colorlinks]{hyperref}
\usepackage[capitalise,english]{cleveref} 

\usepackage[font=normalsize]{caption}
\usepackage[font=footnotesize,subrefformat=brace]{subcaption}
\DeclareCaptionLabelFormat{andtable}{#1~#2  \&  \tablename~\thetable}

\begin{document}


\begin{frontmatter}

  \title{Efficient solution of 3D electromagnetic eddy-current problems within the finite volume framework of \textit{OpenFOAM}}

  \author[HZDR]{Pascal Beckstein}
  \author[HZDR]{Vladimir Galindo}
  \author[FAMENA]{Vuko Vukčević}

  \address[HZDR]{Helmholtz-Zentrum Dresden-Rossendorf, Institute of Fluid Dynamics, Department of Magnetohydrodynamics, Bautzner Landstr. 400, Dresden, Germany}
  \address[FAMENA]{University of Zagreb, Faculty of Mechanical Engineering and Naval Architecture, Ivana Lučića 5, Zagreb, Croatia}

  \begin{abstract}

    Eddy-current problems occur in a wide range of industrial and metallurgical applications where conducting material is processed inductively. Motivated by realising coupled multi-physics simulations, we present a new method for the solution of such problems in the finite volume framework of \textit{foam-extend}, an extended version of the very popular \textit{OpenFOAM} software. The numerical procedure involves a semi-coupled multi-mesh approach to solve Maxwell's equations for non-magnetic materials by means of the Coulomb gauged magnetic vector potential $\vec{A}$ and the electric scalar potential $\phi$. The concept is further extended on the basis of the impressed and reduced magnetic vector potential and its usage in accordance with Biot-Savart's law to achieve a very efficient overall modelling even for complex three-dimensional geometries. Moreover, we present a special discretisation scheme to account for possible discontinuities in the electrical conductivity. To complement our numerical method, an extensive validation is completing the paper, which provides insight into the behaviour and the potential of our approach.

  \end{abstract}


\end{frontmatter}


\section{Introduction}

  Eddy-current problems can be found in a wide range of industrial and metallurgical applications. The basic idea of such processes is to use alternating electromagnetic fields, originating from an inductor like a powered coil, to excite eddy-currents in electrically conducting material. Such induced currents produce secondary effects like electromagnetic forces and heat. Depending on the arrangement and geometry of one or more excitation coils in the proximity of the conductor, various different force fields may be tailored for special tasks. The spectrum of possibilities is further extended by the time-dependent behaviour of the driving source current density. Electromagnetic forces are mainly used for the processing of liquid materials e.g. for stirring, mixing, levitation or retention. The scope of possible applications for electromagnetic heat sources is versatile, too. This comprises processes like welding, hardening, melting or casting.

  The design of induction processing applications and its electromagnetic effects is often very difficult based only on experimental investigation or measurement. Insights from numerically modelling of eddy-current problems are thus very desirable. Especially in the field of liquid metal and semiconductor processing, most industrial applications rely on a complex interaction of hydrodynamic and electromagnetic effects. Covering multiple physical effects and their interaction is however challenging for numerical models and computer simulations, particularly in three-dimensional space.

  Electromagnetic phenomena are most commonly formulated and solved using the finite element method (FEM). Numerous different formulations with its own characteristics exist based on primary and secondary variables \cite{ARTICLE_Carpenter_1977, ARTICLE_Biro_1999, ARTICLE_Biro_Preis_2000, ARTICLE_Xu_Simkin_2004, ARTICLE_Biro_Valli_2007}. The physics is governed by the time-dependent Maxwell's equations \cite{BOOK_Stratton_Electromagnetic_Theory_1941}, which are defined on an unbounded domain. Depending on the application, more or less additional simplifications may be recognised. We will concentrate only on highly conducting, non-magnetic materials, such as non-ferrous metals or semi-conductors at high temperatures. The unboundedness of Maxwell's equations is approximately captured by means of a sufficiently large computational domain.

  For many industrial induction processes involving liquids, alternating magnetic fields are used with oscillation frequencies of $1\,\unit{kHz}$ and above. In those cases, solving a time-dependent electromagnetic problem may not be convenient as very small time-scales have to be resolved, while the time-scales of the coupled phenomena like fluid dynamics or thermodynamics may be independent. If the difference in the order of magnitude of the time-scales is sufficient, a quasi-steady description of Maxwell's equations is advantageous.

  In computational fluid dynamics (CFD), the finite volume method (FVM) is the favourable solution in contrast to FEM due to its conservative property \cite{BOOK_FerzigerPeric_Computational_Methods_2002}. Even though a combination of FEM and FVM for coupled magnetohydrodynamic (MHD) applications is possible \cite{ARTICLE_Spitans_Baake_Nacke_Jakovics_2014}, its realisation may suffer from reduced efficiency due to additional overhead. This is especially true for simulations with involved, time-dependent geometric changes, where recurring interpolation and grid generation may become a limiting factor. Staying in one single framework, either FEM or FVM, avoids such overhead.

  Our claim is to propose a method to solve three-dimensional eddy-current problems based on unstructured, polyhedral FVM, which can be implemented and combined readily within existing CFD-software. This development is motivated by our recent investigation of free-surface flows under the influence of electromagnetic forces in the context of the Ribbon Growth on Substrate (RGS) process \cite{ARTICLE_Beckstein_Galindo_Gerbeth_2015, INPROCEEDINGS_Beckstein_Gerbeth_Galindo_2016}.

  The main difficulty thereby is that for unstructured FVM, the size of the computational stencil is limited. Compact numerical stencils however conflict with a proper implicit discretisation of differential operators like $\grad{(\div{(\,)})}$ or $\rot{(\,)}$. Hence, a suitable formulation of the electromagnetic problem should preferably rely on differential operators which are typical for CFD and known from the Navier-Stokes-Equations \cite{BOOK_FerzigerPeric_Computational_Methods_2002}. We therefor use a description of Maxwell's equations based on the Coulomb gauged magnetic vector potential $\vec{A}$ and the electric scalar potential $\phi$ \cite{ARTICLE_Biro_Valli_2007}. An alternative formulation can be found in Djambazov et al. \cite{ARTICLE_Djambazov_Bojarevics_Pericleous_Croft_2015}. A closer look however reveals that their implementation is not feasible for very large problems.

  In the complex, quasi-steady formulation of the electromagnetic problem, the resulting governing equation system poses an additional challenge for a solution within a finite volume framework. A fully coupled approach of the whole system for a similar, geophysical problem was presented in \cite{ARTICLE_Haber_Ascher_Aruliah_Oldenburg_2000, ARTICLE_Aruliah_Ascher_Haber_Oldenburg_2001} for simple, structured meshes. Even though it is possible to solve the whole system fully coupled, it is not only difficult to implement for unstructured meshes, but it will also require tremendous amounts of memory.

  The CFD code \textit{foam-extend} \cite{MISC_foam-extend}, an extended version of the very popular \textit{OpenFOAM} \cite{PHDTHESIS_Jasak_1996, ARTICLE_Weller_Tabor_Jasak_Fureby_1998, MISC_OpenFOAM}, provides a special framework for the solution of coupled problems. The implementation for coupled equations is based on block-matrices, which use tensor-valued matrix coefficients. The development in this field is still very active (cf. \cite{ARTICLE_Darwish_Sraj_Moukalled_2009, INPROCEEDINGS_Jareteg_Vukcevic_Jasak_2014}) and is motivated mainly to address the pressure-velocity coupling originating from the Navier-Stokes equations.

  To avoid the problems of a fully coupled discretisation, we propose a semi-coupled approach in \textit{foam-extend} which treats the weak coupling between $\vec{A}$ and $\phi$ explicitly in a segregated manner (similar to the SIMPLE algorithm \cite{ARTICLE_Patankar_Spalding_1972, BOOK_FerzigerPeric_Computational_Methods_2002}), while the strong coupling between the complex components of the phasor amplitude is addressed implicitly. The partially block-coupled solution renders techniques like source term linearisation (cf. \cite{ARTICLE_Djambazov_Bojarevics_Pericleous_Croft_2015}) obsolete and is much more robust at higher frequencies. Our semi-coupled proposal relies on the discretisation of a non-conducting region around the conducting region of interest. To maintain an effective method, a special multi-mesh implementation has been developed, where two overlapping finite volume meshes are being used.

  We will extend our proposal on the basis of the impressed and reduced magnetic vector potential and its usage in accordance with Biot-Savart's law to achieve a very efficient overall modelling even for complex three-dimensional geometries. Although the whole idea of this step is not new and has been used extensively in literature \cite{ARTICLE_Biro_Preis_2000, ARTICLE_Xu_Simkin_2004, BOOK_BinnsLawrensonTrowbridge_Electric_Magnetic_Fields_1992, MISC_Opera3D} within different finite element frameworks, it is particularly helpful within the finite volume framework. The reason for this is simply that high quality, unstructured finite volume meshes of complex geometries are much more difficult to create than, for example, finite element meshes. Mesh skewness and non-orthogonality may degrade results as only compact numerical stencils are used.

  The application of inductors modelled on the basis of Biot-Savart's law offers a great potential to reduce the size and geometrical complexity of the non-conducting region to a minimum, but it comes in general at high computational costs. In our solution concept, this application is very limited, not intended to be used on bulk volume data, and it is also not involved in any iterative steps. Even though we have to deal with a simplified non-conducting region, we expect our proposal to be much faster compared to e.g. \cite{ARTICLE_Djambazov_Bojarevics_Pericleous_Croft_2015}, where Biot-Savart is iteratively used in a fully segregated approach.

  With the formulation of the electromagnetic problem based on the magnetic vector potential $\vec{A}$, charge conservation of induced currents is not intrinsically satisfied. In fact, the electric scalar potential $\phi$, more specifically its gradient, is adjusted to achieve a solenoidal current density field based on a derived Poisson-type differential equation. The corresponding equation is strictly valid only for a continuous electrical conductivity. For cases with large jumps in the electrical conductivity, e.g. at boundaries of different materials, the conducting region has to be either split in several sub-regions or special care has to be taken during discretisation. In order to avoid additional regions with possibly complex geometries and to keep the setup of simulation cases simple, it is much more convenient to account for discontinuities in the process of discretisation. For instance in \cite{ARTICLE_Haber_Ascher_Aruliah_Oldenburg_2000} a generalised current density has been used to address this problem for structured meshes.

  In order to address the discontinuity of the electrical conductivity, we follow the Ghost-Fluid-Method (GFM) procedure as presented in \cite{PHDTHESIS_Vukcevic_2016,ARTICLE_Vukcevic_Jasak_Gatin_2017} for free-surface flow simulations in unstructured FVM. The GFM defines one-sided extrapolations in the discretisation process, thus embedding the discontinuity into discretisation schemes \cite{ARTICLE_Fedkiw_Aslam_Xu_1999, ARTICLE_Huang_Carrica_Stern_2007}. The method has been successfully used by Desjardins et. al. \cite{ARTICLE_Desjardins_Moureau_Pitsch_2008} for incompressible two-phase flows in a finite difference framework, while just recently, Lalanne et. al. \cite{ARTICLE_Lalanne_Villegas_Tanguy_Risso_2015} have provided a detailed overview of the treatment of tangential stresses in such flows. An approach very similar to the GFM is called embedded free-surface method \cite{ARTICLE_Johansen_Colella_1998, ARTICLE_Crockett_Colella_Graves_2010}, where the discontinuities are taken into account using possibly higher order discretisation. In fact, Wang et. al. \cite{ARTICLE_Wang_Glimm_Samulyak_Jiao_Diao_2013} note that the difference between the GFM and embedded free-surface method is simply in the treatment (discretisation) of discontinuities. Most of the research activity regarding the GFM and embedded free-surface method concerns two-phase flows, and to our knowledge, no one has applied these methods for electromagnetic eddy-current problems with a discontinuous electrical conductivity, yet. Note that in this work, we have chosen to refer to this newly developed approach as embedded discretisation because we feel that the name Ghost-Fluid-Method would be misleading due to the absence of a fluid.

  The paper is organised in the following way: In \cref{sec:eddy,sec:harmonic}, governing equations, coupling mechanisms and time-harmonic excitation will be discussed. Afterwards, in \cref{sec:efficiency}, we will present our semi-coupled multi-mesh approach. In \cref{sec:embedded}, the embedded discretisation scheme for jump discontinuities in the electrical conductivity is explained. To validate our whole numerical method, we will then investigate simulation results for two test cases in \cref{sec:validation}, and finally give a short summary. Our main goal is thereby devoted to providing a comprehensive overview including all important aspects to solve eddy-current problems with the help of \textit{foam-extend}.


\section{Induced eddy-currents}
\label{sec:eddy}

\begin{figure}
    \centering
    \resizebox{\linewidth}{!}{\includetikz{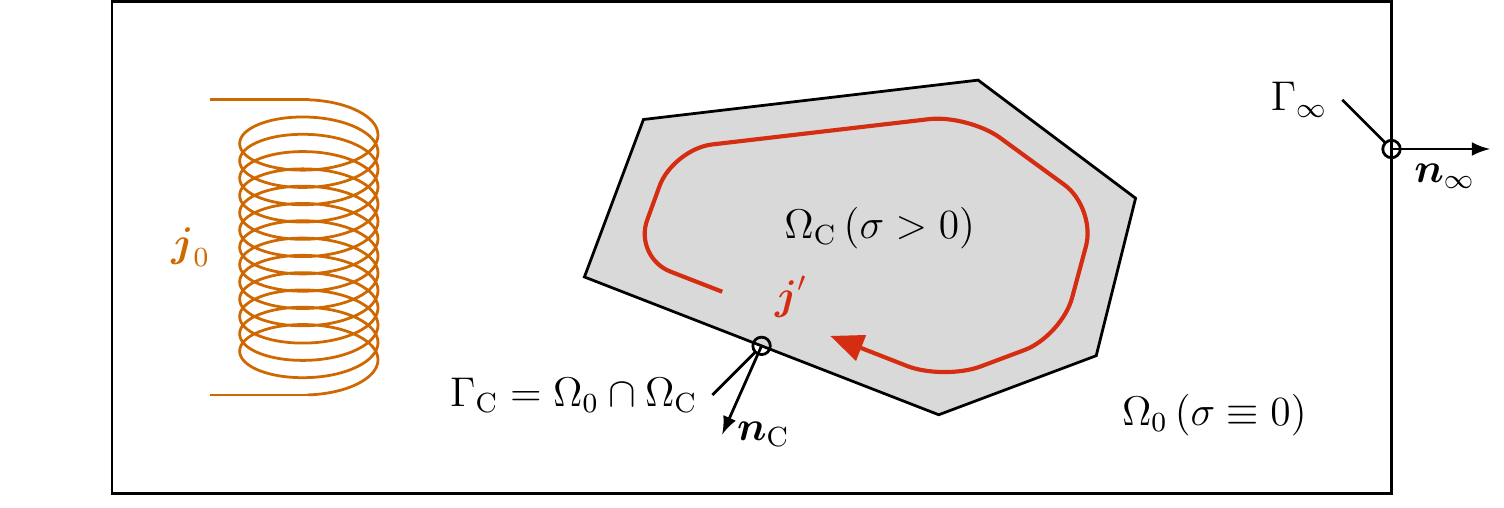}}
    \centering
    \caption{Typical setup for electromagnetic processing of conducting materials. An alternating source current in the induction coil induces eddy-currents in the conductor. The non-conducting region reaches up to a sufficiently large distance away from the coil and the conducting region.}
    \label{fig:domain:setup}
  \end{figure}

  \Cref{fig:domain:setup} shows the scheme of a typical setup, which is widely used for electromagnetic processing of materials: An excitation coil, situated in a non-conducting region $\Omega_\mathrm{0} \left(\sigma \equiv 0\right)$ and represented by an alternating current density $\vec{j}_\mathrm{0}$, is used to induce eddy-currents in a conducting region $\Omega_\mathrm{C} \left(\sigma > 0\right)$ close to or usually inside the inductor. The conducting domain may contain multiple solid or liquid sub-regions with each locally varying electrical conductivity $\sigma$. Both regions share the conductor interface $\Gamma_\mathrm{C} = \Omega_\mathrm{0} \cap \Omega_\mathrm{C}$ and the non-conducting domain is bounded by $\Gamma_\mathrm{\infty}$.

\subsection{Governing equations}
\label{subsec:eddy:equations}

  For highly conducting, non-magnetic materials, the local rate of change of electric charges and displacement currents may be neglected. This quasi-static approximation is common for various magnetohydrodynamic applications \cite{BOOK_Moreau_Magnetohydrodynamics_1990, BOOK_Davidson_An_Introduction_To_Magnetohydrodynamics} and can be formulated by means of a simplified differential form of Maxwell's equations:
  \begin{subequations}
    \label{seqn:maxwell}
    \begin{align}
      \rot{\vec{B}} &= \mu_\mathrm{0} \left(\vec{j}_\mathrm{0} + \vec{j}'\right) \label{seqn:maxwell:ampere} \\
      \div{\vec{B}} &= 0                                                         \label{seqn:maxwell:gaussmag} \\
      \rot{\vec{E}} &= -\shortpd{\vec{B}}{t}                                     \label{seqn:maxwell:faraday} \\
      \div{\vec{E}} &= \rho_\mathrm{E}/\epsilon_\mathrm{0}                       \label{seqn:maxwell:gauss} \text{,}
    \end{align}
  \end{subequations}
  where $\vec{B}$ and $\vec{E}$ denote the magnetic flux density and the electric field intensity, respectively, for the physical time $t$. The distribution of electrical charges is represented by $\rho_\mathrm{E}$ and the total current density $\vec{j} = \vec{j}_\mathrm{0} + \vec{j}'$ in \cref{seqn:maxwell:ampere} comprises a purely induced part $\vec{j}'$ and the part which represents the externally applied current sources $\vec{j}_\mathrm{0}$ (cf. \cref{fig:domain:setup}). For media with an isotropic electrical conductivity, the induced current density $\vec{j}'$ is proportionally related to the electric field intensity via Ohm's law
  \begin{equation}
    \text{$\Omega_\mathrm{C}$:} \quad \vec{j}' = \sigma \vec{E} \label{eqn:ohm} \text{.}
  \end{equation}

  Even in the presence of conducting liquids convected with a flow velocity $U$ on a length scale $L$, equation \cref{eqn:ohm} is still an adequate approximation in the range of small Magnetic Reynolds numbers
  \begin{equation}
    \mathrm{Re}_\mathrm{m} = \mu_\mathrm{0}\sigma U L \ll 1 \label{eqn:magneticreynolds} \text{,}
  \end{equation}
  for which flow-induced currents may be neglected \cite{BOOK_Moreau_Magnetohydrodynamics_1990,BOOK_Davidson_An_Introduction_To_Magnetohydrodynamics}. Due to the solenoidal character of the magnetic flux density and taking into account Faraday's of induction (\ref{seqn:maxwell:faraday}), the Coulomb-gauged \cite{BOOK_BinnsLawrensonTrowbridge_Electric_Magnetic_Fields_1992} magnetic vector potential $\vec{A}$ and the electric scalar potential $\phi$ can be introduced as follows:
  \begin{subequations}
    \label{seqn:potentials}
    \begin{align}
      \vec{B} &= \rot{\vec{A}} \text{,}\quad \div{\vec{A}} = 0     \label{seqn:potentials:magpot} \\
      \vec{E} &= -\left(\shortpd{\vec{A}}{t} + \grad{\phi} \right) \label{seqn:potentials:elepot} \text{.}
    \end{align}
  \end{subequations}

  Consequently, Ampere's circuital law (\ref{seqn:maxwell:ampere}) and \cref{seqn:potentials:magpot} may be combined to one new governing equation for $\vec{A}$:
  \begin{equation}
    \text{$\Omega$:} \quad \rot{\rot{\vec{A}}} = \mu_\mathrm{0} \left(\vec{j}' + \vec{j}_\mathrm{0}\right) \label{eqn:magpot} \text{.}
  \end{equation}
  Gauss's law (\ref{seqn:maxwell:gauss}) is only necessary if the distribution of electrical charges $\rho_\mathrm{E}$ is of interest.

  Exploiting the vector Laplacian identity $\rot{\rot{\vec{A}}} = \grad{\left(\div{\vec{A}}\right)} - \laplace{\vec{A}}$, the gauge condition from (\ref{seqn:potentials:magpot}), the definition of the electric scalar potential (\ref{seqn:potentials:elepot}) and Ohm's law (\ref{eqn:ohm}), \cref{eqn:magpot} can be further elaborated to:
  \begin{equation}
    \text{$\Omega$:} \quad \laplace{\vec{A}} = \mu_\mathrm{0} \sigma \left(\shortpd{\vec{A}}{t} + \grad{\phi} \right) - \mu_\mathrm{0}\vec{j}_\mathrm{0} \label{eqn:magpot:omega} \text{.}
  \end{equation}

  To obey flux continuity of the magnetic field, the Coulomb-gauged $\vec{A}$ is itself continuous across the conductor boundary $\Gamma_\mathrm{C}$. Hence, \cref{eqn:magpot:omega} is valid everywhere in $\Omega = \Omega_\mathrm{0} \cup \Omega_\mathrm{C}$. Depending on $\sigma$, it may nevertheless appear in two different shapes
  \begin{subequations}
    \label{seqn:magpot}
    \begin{alignat}{3}
      \text{$\Omega_\mathrm{0}$:} \quad && \laplace{\vec{A}} &= -\mu_\mathrm{0}\vec{j}_\mathrm{0}                                       \label{seqn:magpot:omega0} \\
      \text{$\Omega_\mathrm{C}$:} \quad && \laplace{\vec{A}} &=  \mu_\mathrm{0} \sigma \left(\shortpd{\vec{A}}{t} + \grad{\phi} \right) \label{seqn:magpot:omegaC} \text{,}
    \end{alignat}
  \end{subequations}
  since the implications
  \begin{subequations}
    \label{seqn:domainsetup}
    \begin{alignat}{3}
      \text{$\Omega_\mathrm{0}$:} \quad && \vec{j}_\mathrm{0} \neq \vec{0} \quad & \Rightarrow\quad \sigma = 0                   \label{seqn:domainsetup:omega0} \\
      \text{$\Omega_\mathrm{C}$:} \quad &&                      \sigma > 0 \quad & \Rightarrow\quad \vec{j}_\mathrm{0} = \vec{0} \label{seqn:domainsetup:omegaC}
    \end{alignat}
  \end{subequations}
  hold for a setup similar to \cref{fig:domain:setup}.

  Using the vector identity $\div{\left(\sigma \shortpd{\vec{A}}{t}\right)} = \grad{\sigma} \cdot \shortpd{\vec{A}}{t} + \sigma\shortpd{\left(\div{\vec{A}}\right)}{t}$ and the Coulomb-gauge condition, an additional equation for $\phi$ can be derived from \cref{seqn:potentials:elepot} to ensure conservation of charges ($\div{\vec{j}'} = 0$) in the conducting region:
  \begin{equation}
    \text{$\Omega_\mathrm{C}$:}\quad \div{\left(\sigma\grad{\phi}\right)} = -\grad{\sigma} \cdot \shortpd{\vec{A}}{t} \label{eqn:elepot} \text{.}
  \end{equation}
  \Cref{eqn:elepot} is strictly valid only if $\sigma$ is continuous everywhere in $\Omega_\mathrm{C}$. \Cref{sec:embedded} is solely dedicated to our corresponding embedded implementation for $\sigma$ for polyhedral (unstructured) meshes in \textit{foam-extend}.

  If the electric field $\vec{E}$ in $\Omega_\mathrm{0}$ is of interest, it can be calculated from the electric scalar potential $\phi$ and $\laplace{\phi} = 0$, which follows from \cref{seqn:maxwell:gauss} and \cref{seqn:potentials:elepot} if we assume $\rho_\mathrm{E} \equiv 0$.

\subsection{Boundary conditions}
\label{subsec:eddy:boundaries}

  For the sake of simplicity, boundary conditions for \cref{seqn:magpot,eqn:elepot} will only be detailed for the case illustrated in \cref{fig:domain:setup}, where the conducting region $\Omega_\mathrm{C}$ is entirely surrounded by the non-conducting region $\Omega_\mathrm{0}$. For different boundaries, like e.g. existing symmetry planes, further information can be found in specialised books such as \cite{BOOK_BinnsLawrensonTrowbridge_Electric_Magnetic_Fields_1992}.

  Maxwell's equations (\ref{seqn:maxwell:ampere}) are actually defined for an unbounded domain. From the elliptic \cref{eqn:magpot:omega} or \cref{seqn:magpot:omega0} it is however apparent that the amplitude of $\vec{A}$ will rapidly decay with increasing distance from current sources. That is, extending the non-conducting region only up to a sufficiently large, finite distance from the location of $\Gamma_\mathrm{C}$ and $\vec{j}_\mathrm{0}$, will still result in a good approximation for a numerical model. Recalling the solenoidal nature of $\vec{B}$, it is obvious to assume closing magnetic field lines within our region of interest. This can be imposed assuming tangentially magnetic boundary conditions on the far-field boundary $\Gamma_\mathrm{\infty}$: $\vec{n}_\mathrm{\infty} \cdot \vec{B} = 0$, where $\vec{n}_\mathrm{\infty}$ denotes the corresponding normal vector. Using the Coulomb-gauge condition, an adequate constraint for $\vec{A}$ reads:
  \begin{equation}
    \text{$\Gamma_\mathrm{\infty}$:}\quad \vec{n}_\mathrm{\infty} \times \vec{A} = \vec{0} \text{,} \quad \vec{n}_\mathrm{\infty} \cdot \grad{\vec{A}} = \vec{0} \label{eqn:magpot:bc:gammainfty} \text{.}
  \end{equation}
  Expressed in words, this means the tangential components of $\vec{A}$ and its normal-derivative need to vanish on $\Gamma_\mathrm{\infty}$ to achieve a consistent behaviour. A simpler (but stiffer) constraint for $\vec{A}$ at $\Gamma_\mathrm{\infty}$ is to use homogeneous Dirichlet-conditions $\bc{\vec{A}}{\Gamma_\mathrm{\infty}} = \vec{0}$, assuming a fully decayed magnetic field as approximation.

  On the conductor surface $\Gamma_\mathrm{C}$, as already mentioned above, the Coulomb-gauged vector potential $\vec{A}$ remains continuous:
  \begin{equation}
    \text{$\Gamma_\mathrm{C}$:}\quad \bc{\vec{A}}{\Gamma_\mathrm{C}} = \bc{\vec{A}}{\Gamma_\mathrm{C} \in \Omega_\mathrm{0}} = \bc{\vec{A}}{\Gamma_\mathrm{C} \in \Omega_\mathrm{C}} \label{eqn:magpot:bc:gammaC} \text{.}
  \end{equation}

  For \cref{eqn:elepot}, suitable boundary conditions on $\Gamma_\mathrm{C}$ with its normal vector $\vec{n}_\mathrm{C}$ can be identified based on the requirement of a vanishing normal component of the induced current density: $\vec{n}_\mathrm{C} \cdot \vec{j}' = \vec{0}$. The condition can be rewritten by means of an inhomogeneous Neumann-condition for $\phi$
  \begin{equation}
    \text{$\Gamma_\mathrm{C}$:}\quad \vec{n}_\mathrm{C} \cdot \grad{\phi} = -\vec{n}_\mathrm{C} \cdot \shortpd{\vec{A}}{t} \label{eqn:elepot:bc:gammaC} \text{,}
  \end{equation}
  which relates the normal gradient of the electric scalar potential to the temporal derivative of the magnetic vector potential in normal direction.

\subsection{Lorentz-force and Joule-heat}
\label{subsec:eddy:forceheat}

  The induced eddy-current density in the conducting region produces secondary effects which are fundamental for induction processing of materials.

  On the one hand, the existence of a current in the direction perpendicular to the magnetic field results in the Lorentz-force:
  \begin{equation}
    \vec{F} = \vec{j}' \times \vec{B} = \vec{j}' \times \left( \rot{\vec{A}} \right) \label{eqn:lorentz} \text{.}
  \end{equation}

  On the other hand, the current distribution in $\Omega_\mathrm{C}$ may introduce large amounts of thermal energy which is being transmitted contactlessly. This power source is known as Joule heating:
  \begin{equation}
    Q = \vec{j}' \cdot \vec{E} = \frac{|\vec{j}^{'}|^2}{\sigma} \label{eqn:joule} \text{.}
  \end{equation}

\subsection{Two-dimensional cases}
\label{subsec:eddy:2D}

  For a two-dimensional case (e.g. plane/wedge), the magnetic vector potential has only one component, e.g. in $z$-direction $\vec{A} = A \,\vec{e}_z$, and the spatial derivatives for all fields in that direction are equal to zero ($\shortpd{.}{z}=0$) by definition. From \cref{eqn:magpot} for $\Omega$ we subsequently find that the induced current may only contain components in the direction of $\vec{A}$, too:
  \begin{equation}
    \vec{j}' = j' \,\vec{e}_z = -j_\mathrm{0} \,\vec{e}_z - 1/\mu_\mathrm{0} \left( \shortpdd{A}{x} + \shortpdd{A}{y} \right) \label{eqn:2D:current} \vec{e}_z \text{.}
  \end{equation}

  In order to keep \cref{eqn:2D:current} self-consistent, the source current density $\vec{j}_\mathrm{0}$ needs to be restricted accordingly. Furthermore, the originally inhomogeneous Neumann-condition (\ref{eqn:elepot:bc:gammaC}) becomes homogeneous, since also the time-derivative of $\vec{A}$ is parallel to the $z$-direction ($\shortpd{\vec{A}}{t} \parallel \vec{e}_z$) while the latter is always perpendicular to the normal vector of the conductor boundary ($\vec{e}_z \perp \vec{n}_\mathrm{C}$):
  \begin{equation}
    \vec{n}_\mathrm{C} \cdot \grad{\phi} = 0 \label{eqn:2D:bc:gammaC} \text{.}
  \end{equation}

  Similarly, the right hand side (RHS) of \cref{eqn:elepot} vanishes completely due to $\shortpd{\sigma}{z}=0$ and $\vec{A} \parallel \vec{e}_z$. With \cref{eqn:2D:bc:gammaC}, we can finally identify $\phi \equiv 0$ as solution. This means that in the two-dimensional case only the magnetic vector potential is relevant and \cref{eqn:magpot:omega} in $\Omega$ simply reduces to
  \begin{equation}
    \laplace{A} = \mu_\mathrm{0} \sigma\shortpd{A}{t} - \mu_\mathrm{0}j_\mathrm{0} \label{eqn:2D:magpot} \text{.}
  \end{equation}

\subsection{Supplementary notes}
\label{subsec:eddy:notes}

  Together with the boundary conditions from \crefrange{eqn:magpot:bc:gammainfty}{eqn:elepot:bc:gammaC}, the governing \cref{seqn:magpot,eqn:elepot} constitute a closed differential system for both potentials $\vec{A}$ and $\phi$ in the three-dimensional space. As the electric scalar potential is only necessary in the conducting region, this form is often related to as $\vec{A},\phi$\,-$\vec{A}$-formulation in literature (e.g. \cite{ARTICLE_Biro_1999, ARTICLE_Biro_Valli_2007}). Especially due to the absence of a curl-operator, the presented description is perfectly suitable for a discretisation based on an unstructured finite volume method with compact stencils. Special numerical schemes are only necessary for cases of discontinuous electrical conductivity. All involved terms are well known from the Navier-Stokes equations \cite{BOOK_FerzigerPeric_Computational_Methods_2002}. That is, numerical codes like \textit{OpenFOAM}, which are actually tailored for Computational Fluid Dynamics (CFD) applications, directly qualify as tools to solve such time-dependent, electromagnetic problems. So far, the main difference compared to classical CFD cases lies in the existence of the non-conducting region $\Omega_\mathrm{0}$. Two-dimensional cases make less computational demand due to the trivial solution of the electric scalar potential $\phi$. No additional numerical treatment or special implementations are necessary in this case.


\section{Time-harmonic excitation}
\label{sec:harmonic}

  In most induction applications, the excitation coil in \cref{fig:domain:setup} is driven by a time-harmonic source current density
  \begin{equation}
    \vec{j}_\mathrm{0} = \hat{\vec{j}}_\mathrm{0} \cos{\left(\omega_\mathrm{0}t - \alpha_\mathrm{0}\right)} \label{eqn:current0}
  \end{equation}
  with a case-dependent angular frequency $\omega_\mathrm{0}$ and phase shift $\alpha_\mathrm{0}$. Assuming that a harmonic solution of \cref{seqn:magpot,eqn:elepot} exists (for both potentials represented by $\psi$), we may introduce the complex phasor function
  \begin{equation}
    \phasor{\psi}(\vec{x},t) = \phasoramp{\psi}(\vec{x}) \, e^{i\,\omega_\mathrm{0} t} \quad\text{with}\quad \phasoramp{\psi}(\vec{x}) = \hat{\psi}(\vec{x}) \, e^{-i\,\alpha_\psi(\vec{x})} \label{eqn:phasor} \text{,}
  \end{equation}
  such that the harmonic solution is obtained from its real part
  \begin{equation}
    \psi(\vec{x},t) = \Re{\phasor{\psi} (\vec{x},t)} \label{eqn:phasorreal} \text{.}
  \end{equation}
  The phasor amplitude $\phasoramp{\psi}$ contains the physical amplitude $\hat{\psi}$ and the local phase shift $\alpha_\psi$. Using \cref{eqn:phasor} we may transform the electromagnetic system into the frequency domain. The result is a quasi-steady description in the complex plane since time-derivatives can be evaluated analytically:
  \begin{equation}
    \shortpd{\textcolor{black}{\psi(\vec{x},t)}}{t} = \Re{ i\,\omega_\mathrm{0} \, \phasor{\psi}(\vec{x},t)} \label{eqn:phasordt} \text{.}
  \end{equation}

  In order to maintain readability, we may additionally introduce the notations
  \begin{align}
    \re{\psi} = \Re{\phasoramp{\psi}} \quad\text{and}\quad \im{\psi} = \Im{\phasoramp{\psi}} \label{eqn:reim}
  \end{align}
  for the complex parts of the phasor amplitude $\phasoramp{\psi}$. If we hereinafter refer to the potentials $\vec{A}$, $\phi$ in the context of the quasi-steady formulation, we will always imply that actually their complex phasor amplitude is addressed.

\subsection{Quasi-steady formulation}
\label{subsec:harmonic:steady}

  Based on the ansatz function from \cref{eqn:phasor} for $\vec{A}$, $\phi$ and $\vec{j}_\mathrm{0}$, and the definitions from \cref{eqn:reim}, the resulting set of governing equations and boundary conditions can be evaluated for real and imaginary parts, separately:
  \begingroup
    \allowdisplaybreaks
    \begin{subequations}
      \label{seqn:harmonic}
      \begin{alignat}{3}
        \text{$\Omega_\mathrm{0}$:} \quad && 1/\mu_\mathrm{0} \laplace{\re{\vec{A}}} &= - \rei{\vec{j}}{\mathrm{0}}                                      \nonumber \\
                                    \quad && 1/\mu_\mathrm{0} \laplace{\im{\vec{A}}} &= - \imi{\vec{j}}{\mathrm{0}}                                      \label{seqn:harmonic:magpot:omega0} \\
        \text{$\Omega_\mathrm{C}$:} \quad && 1/\mu_\mathrm{0} \laplace{\re{\vec{A}}} &= - \sigma \omega_\mathrm{0}\im{\vec{A}} + \sigma \re{\grad{\phi}} \nonumber \\
                                    \quad && 1/\mu_\mathrm{0} \laplace{\im{\vec{A}}} &= + \sigma \omega_\mathrm{0}\re{\vec{A}} + \sigma \im{\grad{\phi}} \label{seqn:harmonic:magpot:omegaC} \\
                                          &&                                         &                                                                   \nonumber \\
        \text{$\Gamma_\mathrm{\infty}$:} \quad && \vec{n}_\mathrm{\infty} \times \re{\vec{A}} = 0 \text{,} &\quad \vec{n}_\mathrm{\infty} \cdot \grad{\re{\vec{A}}} = 0 \nonumber \\
                                         \quad && \vec{n}_\mathrm{\infty} \times \im{\vec{A}} = 0 \text{,} &\quad \vec{n}_\mathrm{\infty} \cdot \grad{\im{\vec{A}}} = 0 \label{seqn:harmonic:magpot:bc:gammainfty} \\
                                          &&                                         &                                                                   \nonumber \\
        \text{$\Gamma_\mathrm{C}$:} \quad && \bc{\re{\vec{A}}}{\Gamma_\mathrm{C}} &= \bc{\re{\vec{A}}}{\Gamma_\mathrm{C} \in \Omega_\mathrm{0}} = \bc{\re{\vec{A}}}{\Gamma_\mathrm{C} \in \Omega_\mathrm{C}} \nonumber \\
                                    \quad && \bc{\im{\vec{A}}}{\Gamma_\mathrm{C}} &= \bc{\im{\vec{A}}}{\Gamma_\mathrm{C} \in \Omega_\mathrm{0}} = \bc{\im{\vec{A}}}{\Gamma_\mathrm{C} \in \Omega_\mathrm{C}} \label{seqn:harmonic:magpot:bc:gammaC} \\
                                          &&                                         &                                                                   \nonumber \\
        \text{$\Omega_\mathrm{C}$:} \quad && \div{\left(\sigma\grad{\re{\phi}}\right)} &= + \grad{\sigma} \cdot \omega_\mathrm{0}\im{\vec{A}}            \nonumber \\
                                    \quad && \div{\left(\sigma\grad{\im{\phi}}\right)} &= - \grad{\sigma} \cdot \omega_\mathrm{0}\re{\vec{A}}            \label{seqn:harmonic:elepot} \\
                                          &&                                         &                                                                   \nonumber \\
        \text{$\Gamma_\mathrm{C}$:} \quad && \vec{n}_\mathrm{C} \cdot \grad{\re{\phi}} &= + \vec{n}_\mathrm{C} \cdot \omega_\mathrm{0}\im{\vec{A}}       \nonumber \\
                                    \quad && \vec{n}_\mathrm{C} \cdot \grad{\im{\phi}} &= - \vec{n}_\mathrm{C} \cdot \omega_\mathrm{0}\re{\vec{A}}       \label{seqn:harmonic:elepot:bc:gammaC} \text{.}
      \end{alignat}
    \end{subequations}
  \endgroup
  Due to both complex parts, the whole system is now twice as large compared to the transient version. In total, there are eight scalar-valued equations and eight unknown, scalar-valued variables. Six occur from the phasor of the magnetic vector potential and two from the phasor of the electric scalar potential.

\subsection{Time-averaged Lorentz-force and Joule-heat}
\label{subsec:harmonic:forceheat}

  At high frequencies, the time scale of the electromagnetic problem may be orders of magnitude smaller than characteristic time-scales of induced effects. In such cases, especially if we think of coupled multi-physics simulations, it might be favourable to use time-averaged quantities for the secondary effects according to:
  \begin{equation}
    \avg{\psi}{t} = \frac{1}{T_\mathrm{0}} \integral{0}{T_\mathrm{0}}{\psi}{t} \quad\text{with}\quad T_\mathrm{0} = \frac{2\pi}{\omega_\mathrm{0}} \label{eqn:harmonic:average} \text{.}
  \end{equation}

  The time-average of the Lorentz-force (\ref{eqn:lorentz}) can be expressed in terms of the complex parts of the phasor amplitude (\ref{eqn:phasor}) as:
  \begin{equation}
    \avg{\vec{F}}{t} = \frac{1}{2} \left( \re{\vec{j}'} \times \left( \rot{\re{\vec{A}}} \right) + \im{\vec{j}'} \times \left( \rot{\im{\vec{A}}} \right) \right) \label{eqn:harmonic:lorentz} \text{.}
  \end{equation}
  Similarly, a time-averaged formulation for the calculation of the Joule heat (\ref{eqn:joule}) reads:
  \begin{equation}
    \avg{Q}{t} = \frac{1}{2\sigma} \left( \re{\vec{j}'}^2 + \im{\vec{j}'}^2 \right) \label{eqn:harmonic:joule} \text{.}
  \end{equation}

\subsection{Coupling mechanisms}
\label{subsec:harmonic:coupling}

  Examining the system of \cref{seqn:harmonic} more closely reveals three distinct coupling mechanisms. The first mechanism concerns the connection of the non-conducting $\Omega_\mathrm{0}$ and conducting region $\Omega_\mathrm{C}$ and was basically mentioned as \cref{seqn:magpot:omega0,seqn:magpot:omegaC} were split from \cref{eqn:magpot:omega}. It is hence a matter of the domain discretisation whether this type of coupling even appears. That is, if we describe and numerically solve both \cref{seqn:harmonic:magpot:omega0,seqn:harmonic:magpot:omegaC} in one single region $\Omega$ analogously to \cref{eqn:magpot:omega}:
  \begin{align}
    \text{$\Omega$:}&\quad & 1/\mu_\mathrm{0} \laplace{\re{\vec{A}}} &= - \sigma \omega_\mathrm{0}\im{\vec{A}} + \sigma \re{\grad{\phi}} - \rei{\vec{j}}{\mathrm{0}} \nonumber \\
                    &\quad & 1/\mu_\mathrm{0} \laplace{\im{\vec{A}}} &= + \sigma \omega_\mathrm{0}\re{\vec{A}} + \sigma \im{\grad{\phi}} - \imi{\vec{j}}{\mathrm{0}} \label{eqn:harmonic:magpot:omega} \text{,}
  \end{align}
  this coupling can be avoided and the transition conditions from \cref{seqn:harmonic:magpot:bc:gammaC} may be dropped. As long as $\Gamma_\mathrm{C}$ is resolved by the numerical grid, FVM-CFD codes like \textit{OpenFOAM}, which rely on data stored at the cell-centres of the mesh, can be used straightforwardly to solve \cref{eqn:magpot:omega} in $\Omega$. There is no differential operator acting on $\sigma$, which is always discontinuous at $\Gamma_\mathrm{C}$, and the terms on the RHS will vanish except for their associated sub-region according to \cref{seqn:domainsetup}. In contrast to \cref{eqn:magpot:omega}, solving \cref{eqn:harmonic:magpot:omega} with a CFD code is however much more challenging due to another coupling effect.

  This second coupling is also visible in \cref{seqn:harmonic:magpot:omegaC} and arose from the transformation into the complex plane, where \cref{eqn:magpot:omega} was intermediately expressed as complex vector-valued Helmholtz-equation by subsequent evaluation of the time-derivatives. This introduced a cross-wise linkage between real and imaginary part for each component of the phasor amplitude of $\vec{A}$. If the rightmost terms of \cref{seqn:harmonic:magpot:omegaC} are notionally ignored, we can identify pairs of Helmholtz- or respectively Poisson-like equations for each component of the magnetic vector potential, whose inhomogeneous parts are sources of the respective other complex part. The strength of the coupling thereby depends on the source-coefficient $\omega_\mathrm{0}\sigma$. Hence, the coupling-effect becomes more and more dominant for an increasing frequency in combination with a specific material.

  The third coupling mechanism is more branched and lies of course in the dependency of $\vec{A}$ on $\phi$ and vice versa. Firstly, \cref{seqn:harmonic:magpot:omegaC} connects each component of $\vec{A}$ with the electric scalar potential due to the gradient-operator on the RHS. Secondly, there is a relation between $\phi$ and the components of the magnetic vector potential from the inner product on the RHS of \cref{seqn:harmonic:elepot,seqn:harmonic:elepot:bc:gammaC}. As explained in \cite{ARTICLE_Haber_Ascher_Aruliah_Oldenburg_2000, ARTICLE_Aruliah_Ascher_Haber_Oldenburg_2001}, and in contrast to the strong coupling of the complex parts of the phasor amplitude of $\vec{A}$, \cref{seqn:harmonic:magpot:omegaC,seqn:harmonic:elepot} are only weakly coupled.

  The two last coupling mechanisms can be illustrated with the example of the conducting domain, if \cref{seqn:harmonic:magpot:omegaC,seqn:harmonic:elepot} are written in pseudo-matrix form
  \begin{equation}
    \def\arraystretch{1.4}
    \def\arraycolsepOld{\arraycolsep}
    \arraycolsep=3pt
    \newcommand\scalemath[2]{\scalebox{#1}{\mbox{\ensuremath{\displaystyle #2}}}}
    \text{$\Omega_\mathrm{C}$:}\,
    \scalemath{0.69}{
      \newcommand{\la}{1/\mu_\mathrm{0} \laplace{.}}
      \newcommand{\dg}{\div{(\sigma\grad{.}})}
      \newcommand{\so}{\omega_\mathrm{0}\sigma}
      \newcommand{\gr}[1]{\sigma\shortpd{.}{#1}}
      \newcommand{\os}[1]{\omega_\mathrm{0}\shortpd{\sigma}{#1}}
      \begin{bmatrix}
            \dg &       0 &       0 &       0 & -\os{x} & -\os{y} & -\os{z} &       0 \\
        -\gr{x} &     \la &       0 &       0 &    +\so &       0 &       0 &       0 \\
        -\gr{y} &       0 &     \la &       0 &       0 &    +\so &       0 &       0 \\
        -\gr{z} &       0 &       0 &     \la &       0 &       0 &    +\so &       0 \\
              0 &    -\so &       0 &       0 &     \la &       0 &       0 & -\gr{x} \\
              0 &       0 &    -\so &       0 &       0 &     \la &       0 & -\gr{y} \\
              0 &       0 &       0 &    -\so &       0 &       0 &     \la & -\gr{z} \\
              0 & +\os{x} & +\os{y} & +\os{z} &       0 &       0 &       0 &     \dg
      \end{bmatrix}
      \cdot
      \begin{bmatrix}
      \re{\phi} \\
      \rei{\vec{A}}{x} \\
      \rei{\vec{A}}{y} \\
      \rei{\vec{A}}{z} \\
      \imi{\vec{A}}{x} \\
      \imi{\vec{A}}{y} \\
      \imi{\vec{A}}{z} \\
      \im{\phi}
      \end{bmatrix}
      = \vec{0} \text{,}
    }
    \arraycolsep=\arraycolsepOld
    \label{eqn:largeblockmatrix}
  \end{equation}
  where all entries with lower dots ``$.$'' are meant to be applied to the corresponding solution field as differential operators, and $\vec{A}_{x,y,z}$ each denote one Cartesian vector-component of $\vec{A}$.


\section{Efficient finite volume solution}
\label{sec:efficiency}

  In this section we will present an efficient solution for the quasi-steady problem (\ref{seqn:harmonic}). From several numerical tests with the CFD code \textit{foam-extend} for various different setups we found that the coupling in the system is dominated by the strong coupling between the complex components of $\vec{A}$. Especially for higher frequencies, this has a great impact on convergence behaviour and general solubility of the discretised linear equation system.

  Although it is possible to solve the whole system fully coupled (cf. \cite{ARTICLE_Haber_Ascher_Aruliah_Oldenburg_2000, ARTICLE_Aruliah_Ascher_Haber_Oldenburg_2001}), it requires tremendous amounts of memory. For a numerical mesh with a size of $N$, the size of the matrix to resolve the fully coupled system is of $8N \times 8N$. Even though this matrix is sparse, preserving the sparseness pattern defined by mesh faces, this is a factor of $64$ compared to the discretisation of a scalar valued problem. Furthermore, a fully coupled approach as presented in \cite{ARTICLE_Haber_Ascher_Aruliah_Oldenburg_2000, ARTICLE_Aruliah_Ascher_Haber_Oldenburg_2001} would require to solve $\phi$ in $\Omega_\mathrm{0}$ even if it is not explicitly required. Finally also the transition condition from \cref{eqn:elepot:bc:gammaC} and the vanishing electrical conductivity in $\Omega_\mathrm{0}$ has to be dealt with.

  A fully coupled CFD simulation (cf. \cite{INPROCEEDINGS_Jareteg_Vukcevic_Jasak_2014}) in \textit{foam-extend} with three velocity components and the pressure results in a matrix of size $4N \times 4N$, which is still four times smaller than the discretisation of a system like \cref{seqn:harmonic} based on the same mesh size $N$. Due to the presence of the non-conducting region, a typical electromagnetic problem is likely to be much larger than a comparable CFD case as the region of interest is only $\Omega_\mathrm{C}$ in the former.

\subsection{Semi-coupled multi-mesh approach}
\label{subsec:efficiency:solution}

  A segregated solution of $\vec{A}$ and $\phi$ greatly reduces the peak memory consumption and gives us the opportunity to solve \cref{seqn:harmonic:magpot:omega0,seqn:harmonic:magpot:omegaC} combined in one single region $\Omega$ according to \cref{eqn:harmonic:magpot:omega}, while \cref{seqn:harmonic:elepot} is solved only in $\Omega_\mathrm{C}$. Concurrently, a simultaneous region- and equation-coupling in \textit{foam-extend} is avoided.

  We use a multi-mesh approach based on two overlapping finite volume meshes. The base-mesh represents the domain $\Omega$ and a sub-mesh models $\Omega_\mathrm{C}$. The special feature of both these meshes is that the mesh geometry of $\Omega_\mathrm{C}$ in the base-mesh exactly coincides with the mesh geometry of the sub-mesh. Both meshes share the same origin of co-ordinates, too. This enables us to use direct, bi-directional mapping for bulk numerical data. Interpolation is only necessary for a mapping from the base-mesh to $\Gamma_\mathrm{C}$ in the sub-mesh.

  The solution of $\vec{A}$ and $\phi$ is realised by means of repeating iterative steps. Let us assume a trivial solution of $\grad{\phi}$ in $\Omega_\mathrm{C}$ before the first iteration. The reverse mapping of the gradient of the electric scalar potential from the sub-mesh representing  $\Omega_\mathrm{C}$ back to the base-mesh representing $\Omega$ can be formulated as:
  \begin{subequations}
    \label{seqn:reversemapping}
    \begin{alignat}{3}
      \text{$\Omega_\mathrm{C}$:} \quad && &\qquad\,\,\,\iter{\reim{\grad{\phi}}}{0}(\vec{x}_\mathrm{C}) \equiv \vec{0} \\[2ex]
      \text{$\Omega$:}            \quad && \reim{\grad{\phi}}^*(\vec{x}) &=
      \left\{
        \begin{array}{ll}
          \iter{\reim{\grad{\phi}}}{{m-1}}(\vec{x}_\mathrm{C} = \vec{x}), & \vec{x} \in \Omega_\mathrm{C} \ \\
          \vec{0},                                                        & \vec{x} \in \Omega_\mathrm{0}
        \end{array}
      \right.\text{,}
    \end{alignat}
  \end{subequations}
  where $m$ is the integer iteration counter and the superscripted curly braces $\iter{(.)}{m}$ shall denote the approximate solution at the $m$-th iteration step. The field $\grad{\phi}^*$ is the representation of $\grad{\phi}$ in $\Omega$. The part of $\grad{\phi}^*$ which lies in $\Omega_\mathrm{0}$ is not coupled to the solution of $\vec{A}$. Although it physically represents a part of the electric field in the non-conducting domain, we are hence free to ignore it.

  \begin{figure}
    \centering
    \begin{subfigure}[c]{\textwidth}
      \resizebox{\linewidth}{!}{\includetikz{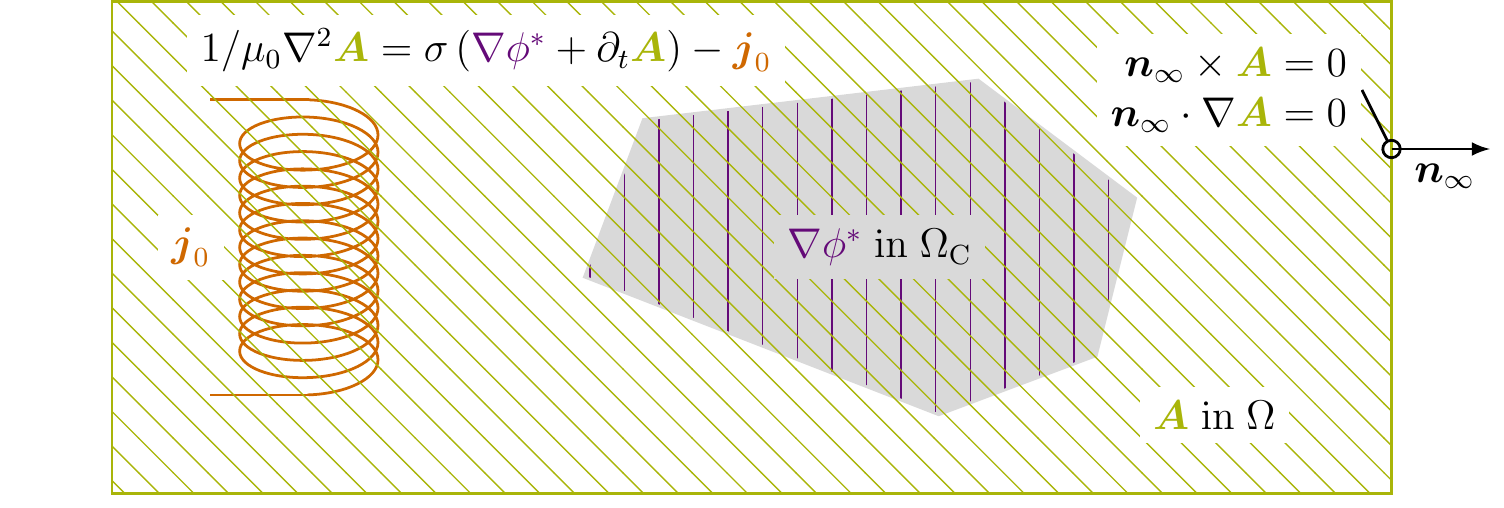}}
      \caption{Solution of the magnetic vector potential in $\Omega$. The gradient of the electric scalar potential is reverse mapped and treated explicitly. Note that each component of $\vec{A}$ can be solved separately.}
      \label{sfig:domain:magpot:magpot}
    \end{subfigure} \\\vspace*{0.5em}
    \begin{subfigure}[c]{\textwidth}
      \resizebox{\linewidth}{!}{\includetikz{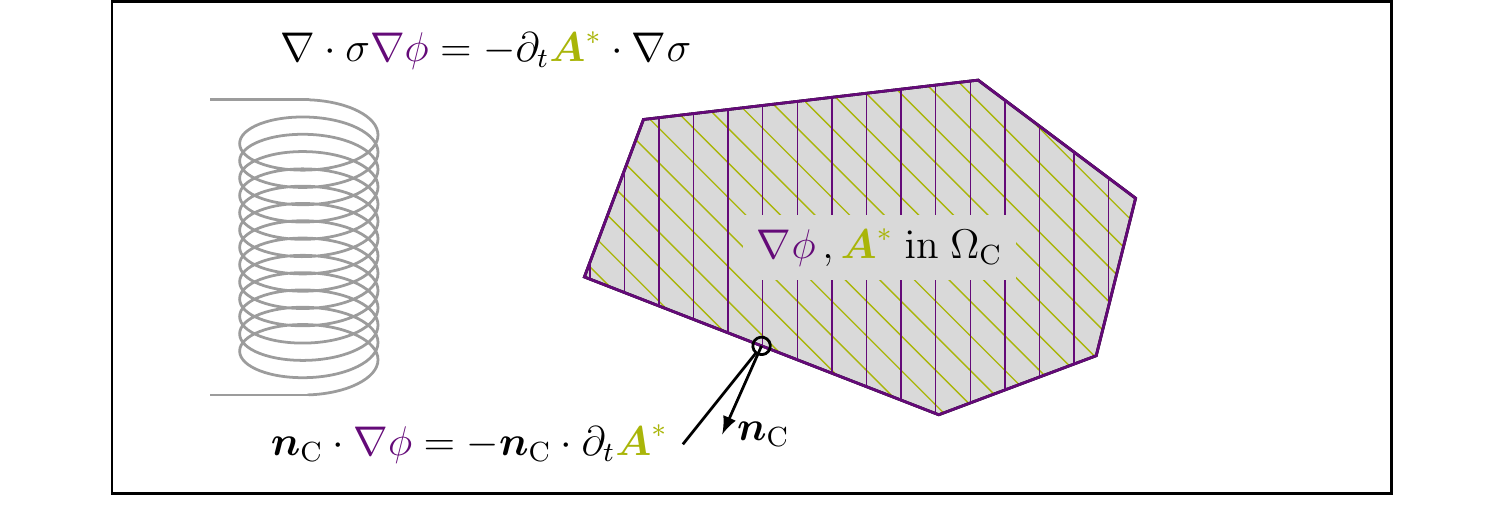}}
      \caption{Correction of the gradient of the electric scalar potential in $\Omega_\mathrm{C}$. The magnetic vector potential is forward mapped and explicitly included.}
      \label{sfig:domain:magpot:gradelepot}
    \end{subfigure}
    \caption{Semi-coupled (segregated) solution of the electromagnetic problem based on the total magnetic vector potential.}
    \label{sfig:domain:magpot}
  \end{figure}

  Provisioning $\grad{\phi}^*$ from a previous iteration ($m-1$) in $\Omega$, we may use it to gain a better approximation of $\vec{A}$ in the current iteration step ($m$). This procedure is illustrated in \cref{sfig:domain:magpot:magpot}. As explained above, this is done for each complex vector-component $\vec{A}_k$ separately, as they uncouple as soon as $\grad{\phi}$ is treated explicitly. The strong coupling of the complex parts of each vector-component $\vec{A}_k$ is resolved implicitly on the basis of block-matrices within the framework of \textit{foam-extend}. Using the matrix-notation as of \cref{eqn:largeblockmatrix}, a formulation of \cref{eqn:harmonic:magpot:omega} for each complex vector-component $\vec{A}_k$ reads:
  \begin{align}
    \def\arraystretch{1.4}
    \def\arraycolsepOld{\arraycolsep}
    \arraycolsep=3pt
    \text{$\Omega$:}\quad
      \newcommand{\la}{1/\mu_\mathrm{0} \laplace{.}}
      \newcommand{\dg}{\div{(\sigma\grad{.}})}
      \newcommand{\so}{\omega_\mathrm{0}\sigma}
      \begin{bmatrix}
         \la &    +\so \\
        -\so &     \la
      \end{bmatrix}
      \cdot
      \begin{bmatrix}
      \iter{\rei{\vec{A}}{k}}{m} \\
      \iter{\imi{\vec{A}}{k}}{m}
      \end{bmatrix}
      =
      \sigma
      \begin{bmatrix}
      \re{\grad{\phi}}^* \cdot \vec{e}_k \\
      \im{\grad{\phi}}^* \cdot \vec{e}_k
      \end{bmatrix}
      -
      \begin{bmatrix}
      \rei{\vec{j}}{\mathrm{0}} \cdot \vec{e}_k \\
      \imi{\vec{j}}{\mathrm{0}} \cdot \vec{e}_k
      \end{bmatrix}
    \arraycolsep=\arraycolsepOld \nonumber \\[2ex]
    \text{for}\quad k = x,y,z  \label{eqn:smallblockmatrix} \text{.}
  \end{align}
  While the left hand side (LHS) reflects the implicit coupling, the RHS only contains explicit sources. These known quantities in the $m$-th iteration comprise the $k$-th component of the mapped $\grad{\phi}^*$ and $\vec{j}_\mathrm{0}$. If we assume that fields in $\Omega$ are numerically represented in $N$ discrete data points, the size of the solution vector of \cref{eqn:smallblockmatrix} is $2N$. The size of a matrix, which is necessary to discretise the LHS of \cref{eqn:smallblockmatrix} is consequently $2N \times 2N$ and thus $16$ times smaller than a corresponding matrix for a fully coupled system like \cref{eqn:largeblockmatrix} would be.

  Given the solution of $\vec{A}$ in the $m$-th iteration the gradient of the electric scalar potential can be corrected in $\Omega_\mathrm{C}$. Therefor the magnetic vector potential in the conducting region is needed for \cref{seqn:harmonic:elepot}. The forward mapping from the base-mesh representing $\Omega$ to the sub-mesh representing $\Omega_\mathrm{C}$ is given by:
  \begin{alignat}{3}
    \text{$\Omega_\mathrm{C}$:} \quad && \reim{\vec{A}}^*(\vec{x}_\mathrm{C}) &=
    \left\{
      \begin{array}{ll}
        \iter{\reim{\vec{A}}}{{m}}(\vec{x} = \vec{x}_\mathrm{C}),                         & \vec{x}_\mathrm{C} \in \Omega_\mathrm{C} \wedge \vec{x}_\mathrm{C} \not\in \Gamma_\mathrm{C} \ \\[2ex]
        \bc{\iter{\reim{\vec{A}}}{{m}}}{\Gamma_\mathrm{C}}(\vec{x} = \vec{x}_\mathrm{C}), & \vec{x}_\mathrm{C} \in \Gamma_\mathrm{C}
      \end{array}
    \right. \label{eqn:forwardmapping} \text{,}
  \end{alignat}
  where $\vec{A}^*$ is the representation of $\vec{A}$ in $\Omega_\mathrm{C}$ in the $m$-th iteration. The boundary values $\bc{\vec{A}}{\Gamma_\mathrm{C}}$ on the conductor boundary cannot be mapped directly. This results from the fact that in \textit{foam-extend} discretised data is stored at cell centres, while $\Gamma_\mathrm{C}$ is represented by faces. Hence, $\bc{\vec{A}}{\Gamma_\mathrm{C}}$ is obtained either from interpolation in $\Omega$ or from extrapolation in $\Omega_\mathrm{C}$.

  With $\vec{A}$ explicitly available in $\Omega_\mathrm{C}$ in terms of $\vec{A}^*$ we can correct $\grad{\phi}$ according to \cref{seqn:harmonic:elepot}:
  \begin{alignat}{3}
    \text{$\Omega_\mathrm{C}$:} \quad && \div{\left(\sigma\iter{\grad{\re{\phi}}}{m}\right)} &= + \grad{\sigma} \cdot \omega_\mathrm{0}\im{\vec{A}}^* \nonumber \\
                                \quad && \div{\left(\sigma\iter{\grad{\im{\phi}}}{m}\right)} &= - \grad{\sigma} \cdot \omega_\mathrm{0}\re{\vec{A}}^* \label{eqn:blockpartPhi} \text{.}
  \end{alignat}
  This can be done for both complex parts sequentially. An illustration of the correction can be found in \cref{sfig:domain:magpot:gradelepot}.

  The whole procedure from \cref{seqn:reversemapping} to (\ref{eqn:forwardmapping}) can be repeated until convergence for both potentials $\vec{A}$ and $\phi$ is achieved.

\subsection{Impressed \& Reduced magnetic vector potential}
\label{subsec:efficiency:magpotsplit}

  In the last section we have presented a semi-coupled multi-mesh approach for the solution of $\vec{A}$ and $\phi$, where \cref{eqn:harmonic:magpot:omega} is solved in $\Omega$ and \cref{seqn:harmonic:elepot} in $\Omega_\mathrm{C}$. Following our proposal, only the complex components of the phasor amplitude of the magnetic vector potential are solved in a coupled manner. In this regard, \cref{eqn:harmonic:magpot:omega} requires special attention in contrast to \cref{eqn:magpot:omega}. Further explanations will however be given only referring to the general time-dependent formulation of \cref{seqn:magpot,eqn:elepot}, as all of the following methods can be applied to the harmonic, quasi-steady formulation from \cref{seqn:harmonic} in the same way.

  We want to stress again that the solution of the \cref{eqn:magpot:omega} in $\Omega$ is dependent on the discretisation of the non-conducting region $\Omega_\mathrm{0}$ which contains the excitation coils carrying the source current density $\vec{j}_\mathrm{0}$. Especially for complex inductor geometries this may lead to a large number of necessary cells. Furthermore, high field gradients may occur close to inductors and they would need to be represented by means of a reasonable local mesh resolution. To avoid meshing of the inductors within $\Omega_\mathrm{0}$, and thus reducing the number of cells to a minimum, it is possible to firstly split the magnetic vector potential $\vec{A}$ into an impressed $\vec{A}_\mathrm{0}$ and a reduced part $\vec{A}'$ \cite{ARTICLE_Xu_Simkin_2004}:
  \begin{equation}
    \vec{A} = \vec{A}_\mathrm{0} + \vec{A}' \label{eqn:magpotsplit} \text{.}
  \end{equation}
  The former corresponds to the part which is defined by the source current density $\vec{j}_\mathrm{0}$:
  \begin{equation}
    \text{$\Omega$:} \quad 1/\mu_\mathrm{0} \laplace{\vec{A}_\mathrm{0}} = - \vec{j}_\mathrm{0} \label{eqn:impressedmagpot:omega} \text{,}
  \end{equation}
  while the latter is caused solely by the induced current density $\vec{j}'$:
  \begin{equation}
    \text{$\Omega$:} \quad 1/\mu_\mathrm{0} \laplace{\vec{A}'} = \sigma \left(\shortpd{\vec{A}_\mathrm{0}}{t} + \shortpd{\vec{A}'}{t} + \grad{\phi} \right) \label{eqn:reducedmagpot:omega} \text{.}
  \end{equation}
  \Cref{eqn:reducedmagpot:omega} originates from the electric field $\vec{E} = -\left(\shortpd{\vec{A}_\mathrm{0}}{t} + \shortpd{\vec{A}'}{t} + \grad{\phi} \right)$. In a second step, we can exploit Biot-Savart's law \cite{BOOK_Stratton_Electromagnetic_Theory_1941}
  \begin{equation}
    \vec{A}_\mathrm{0}\!\left(\vec{x}\right) = \frac{\mu_\mathrm{0}}{4\pi} \integral{\Omega}{}{\frac{\vec{j}_\mathrm{0}\!\left(\vec{r}\right)}{\abs{\vec{x}-\vec{r}}}}{V\!\!\left(\vec{r}\right)} \label{eqn:biot-savart}
  \end{equation}
  to algebraically determine the impressed magnetic vector potential instead of solving \cref{eqn:impressedmagpot:omega}. This can be done either by discretising $\vec{j}_\mathrm{0}$ separately or by an exact integration of a parametrised geometry.

  The downside of this approach is however that the calculation of the integral from \cref{eqn:biot-savart} may get extremely expensive in terms of computational demand, especially if $\vec{j}_\mathrm{0}$ is distributed over a large volume. Let us just assume that $\vec{j}_\mathrm{0}$ would exist in a number of $N_\mathrm{0}$ discrete locations ($\vec{r}$) and we were interested in $\vec{A}_\mathrm{0}$ at a number of $N$ locations in $\Omega$. Then the necessary number of operations to evaluate \cref{eqn:biot-savart} would consequently be in the order of $N_\mathrm{0} \times N$. The problem becomes even more pronounced if one needs to parallelise this procedure as lots of communication needs to be performed.

  For some cases (e.g. \cite{ARTICLE_Weber_Galindo_Stefani_Weier_Wondrak_2013, ARTICLE_Djambazov_Bojarevics_Pericleous_Croft_2015}), the existence of the surrounding region might be a greater obstacle than a significantly higher computational effort. In such circumstances, one can go one step ahead and utilise Biot-Savart's law to calculate the reduced potential $\vec{A}'$ from $\vec{j}' = \sigma \vec{E}$. While this indeed renders the non-conducting region superfluous, a closer inspection shows that this idea is inconvenient for the quasi-steady formulation from \cref{seqn:harmonic} of harmonically alternating fields. The cause for this stems from the inherent coupling-mechanisms (cf. \cref{subsec:harmonic:coupling,subsec:efficiency:solution}), which then had to be resolved iteratively, if at all possible for higher frequencies. And even in the case of slowly varying fields ($\shortpd{\vec{A}}{t} \approx 0$) there is still the coupling between $\vec{A}$ and $\phi$ which had to be captured sequentially.

  Taking into account both advantages and disadvantages of Biot-Savart's law, we will now explain what we found to be the most promising procedure for a realisation in \textit{foam-extend}. We would like to draw the readers attention to an interesting effect in \cref{eqn:reducedmagpot:omega}. A structural comparison with \cref{eqn:magpot:omega} reveals that the source term ``$+\sigma\shortpd{\vec{A}_\mathrm{0}}{t}$'' in the latter replaces the source term ``$-\vec{j}_\mathrm{0}$'' in the former. That is, the source for the total magnetic vector potential partly lies in the non-conducting region, while the source of the reduced magnetic vector potential is only located in the conducting region.  This comes to our advantage if we are only interested in the distribution of induced eddy-currents or secondary effects like Lorentz-force (\ref{eqn:lorentz}) and Joule-heat (\ref{eqn:joule}). In such a case we can simply refrain from solving $\vec{A}_\mathrm{0}$ in $\Omega_\mathrm{0}$ as it does not contribute to the induced fields.

  \begin{figure}
    \centering
    \begin{subfigure}[c]{\textwidth}
      \resizebox{\linewidth}{!}{\includetikz{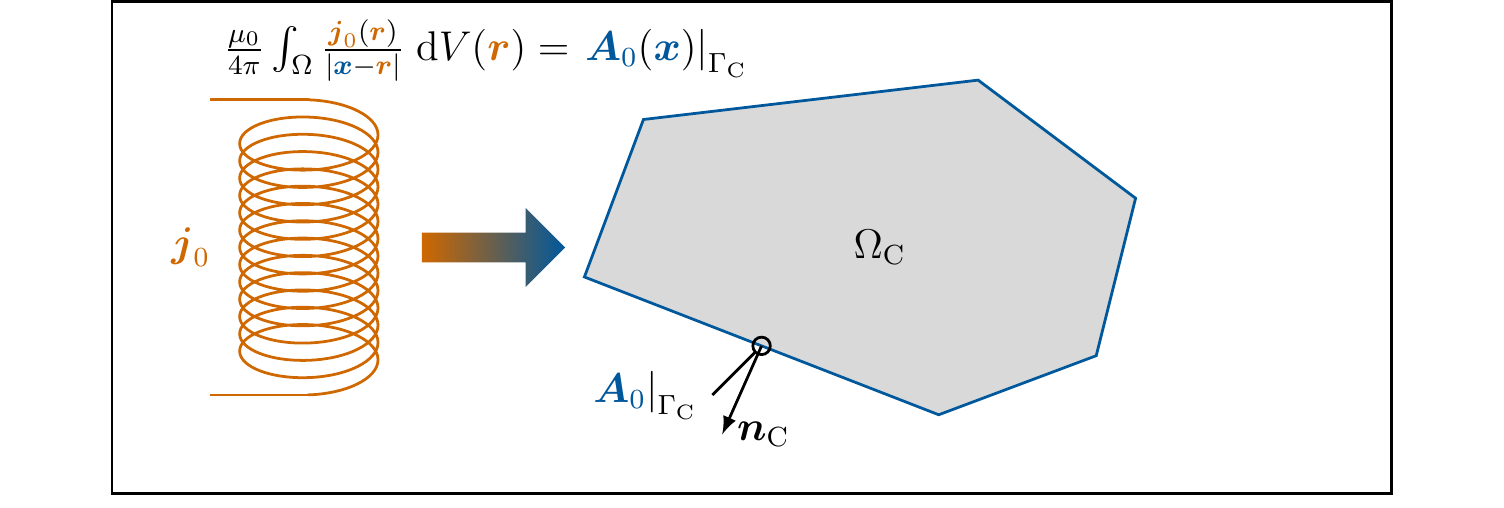}}
      \caption{Calculation of the impressed magnetic vector potential on $\Gamma_\mathrm{C}$. Only boundary values are calculated to keep the computational costs of Biot-Savart's law as low as possible.}
      \label{sfig:domain:impressedmagpot:impressedmagpot:gammaC}
    \end{subfigure} \\\vspace*{0.5em}
    \begin{subfigure}[c]{\textwidth}
      \resizebox{\linewidth}{!}{\includetikz{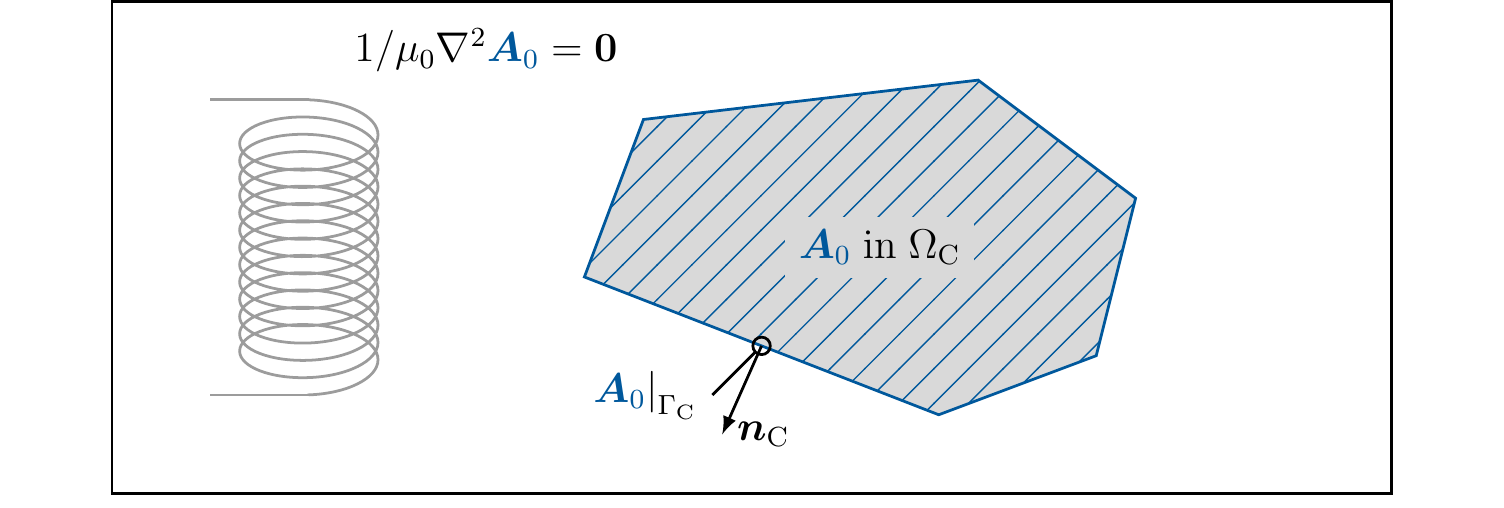}}
      \caption{Solution of the impressed magnetic vector potential in $\Omega_\mathrm{C}$. For complex shaped coils, the solution based on its defining differential equation is computationally less demanding. The part in $\Omega_\mathrm{0}$ is only necessary if the electric field needs to be known there.}
      \label{sfig:domain:impressedmagpot:omegaC}
    \end{subfigure}
    \caption{Calculation of the impressed magnetic vector potential with Biot-Savart's law.}
    \label{sfig:domain:impressedmagpot}
  \end{figure}

  Moreover, primarily to keep the computational costs of Biot-Savart's law as low as possible, we are free to exploit the usage of \cref{eqn:biot-savart} only to determine values of the impressed magnetic vector potential $\bc{\vec{A}_\mathrm{0}}{\Gamma_\mathrm{C}}$ on the conductor boundary. Applying these in terms of a Dirichlet-boundary condition for \cref{eqn:impressedmagpot:omega} readily allows us to solve $\vec{A}_\mathrm{0}$ in $\Omega_\mathrm{C}$ as the source term of $\vec{j}_\mathrm{0}$ on the RHS vanishes in the conducting region. Both necessary steps are illustrated and commented in \cref{sfig:domain:impressedmagpot}. Furthermore, \cref{eqn:impressedmagpot:omega} only needs to be solved once as long as no geometric changes occur and no iterations are required.

  \begin{figure}
    \centering
    \begin{subfigure}[c]{\textwidth}
      \resizebox{\linewidth}{!}{\includetikz{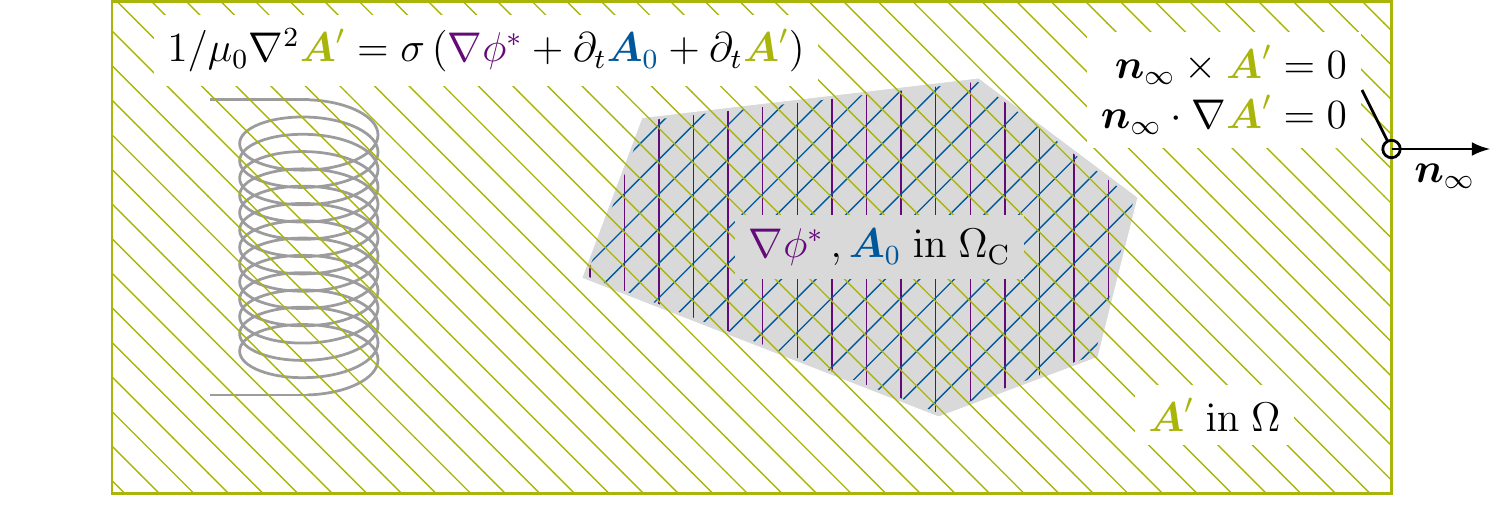}}
      \caption{Solution of the reduced magnetic vector potential in $\Omega$. The impressed magnetic vector potential prior to the iteration and the gradient of the electric scalar potential is reverse mapped and treated explicitly. Each component of $\vec{A}$ can be solved separately.}
      \label{sfig:domain:reducedmagpot:reducedmagpot}
    \end{subfigure} \\\vspace*{0.5em}
    \begin{subfigure}[c]{\textwidth}
      \resizebox{\linewidth}{!}{\includetikz{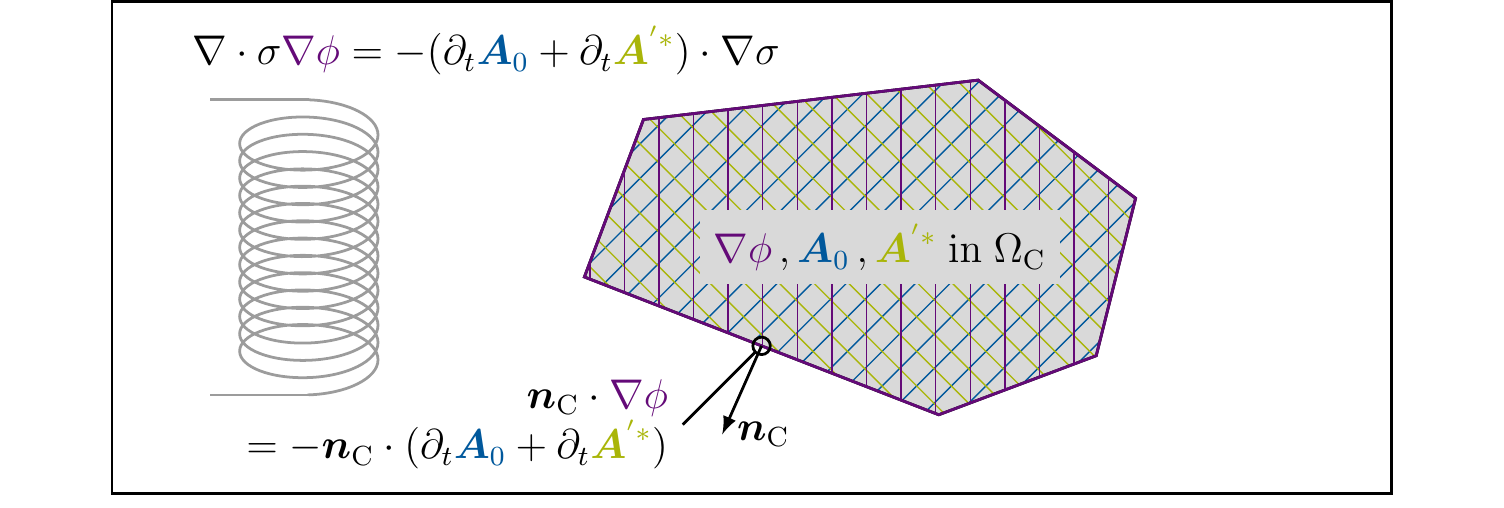}}
      \caption{Correction of the gradient of the electric scalar potential in $\Omega_\mathrm{C}$. The reduced and the impressed magnetic vector potential are forward mapped and explicitly included.}
      \label{sfig:domain:reducedmagpot:gradelepot}
    \end{subfigure}
    \caption{Semi-coupled (segregated) solution of the electromagnetic problem based on the reduced magnetic vector potential, analogously to \cref{sfig:domain:magpot}.}
    \label{sfig:domain:reducedmagpot}
  \end{figure}

  With the solution of the impressed magnetic vector potential in $\Omega_\mathrm{C}$ we also found the solution of the corresponding impressed part of the electric field from its time-derivative. The total magnetic vector potential may be calculated from \cref{eqn:magpotsplit} and does not need to be stored. For the segregated solution of $\vec{A}'$ and $\phi$ we can proceed in analogy to the solution of $\vec{A}$ and $\phi$ as proposed in \cref{subsec:efficiency:solution} (cf. \cref{sfig:domain:magpot}). Starting with an initial guess for the electric scalar potential $\phi^*$ we can solve \cref{eqn:reducedmagpot:omega} and correct $\phi$ based on \cref{eqn:elepot}. The principle based on $\vec{A}_\mathrm{0}$ is pictured in \cref{sfig:domain:reducedmagpot}. From a direct comparison with \cref{sfig:domain:magpot} one can comprehend that we just need to superpose the impressed magnetic vector potential wherever the total magnetic vector potential occurs.

\subsection{Inductors based on Biot-Savart's law}
\label{subsec:biot-svart}

\begin{figure}
    \centering
    \resizebox{0.7\textwidth}{!}{\includetikz{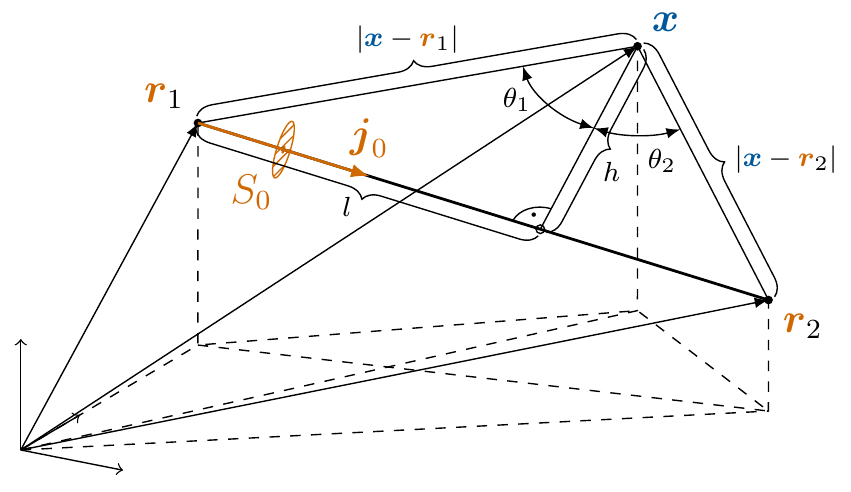}}
    \caption{Geometrical relations at a finite wire between two points $\vec{r}_1$ and $\vec{r}_2$ and a point $\vec{x}$. The current density $\vec{j}_\mathrm{0}$ carried by the wire with a small cross section $S_\mathrm{0}$ causes a magnetic field density $\vec{B}_\mathrm{0}$ in $\vec{x}$ which may be expressed in terms of the magnetic vector potential $\vec{A}_\mathrm{0}(\vec{x})$.}
    \label{fig:biot-savart:finite-wire}
  \end{figure}

  For an excitation coil with a sufficiently small cross section $S_\mathrm{0}$ compared to its characteristic length, the integral of \cref{eqn:biot-savart} may be limited to an integration over a path in three-dimensional space. If this type of modelling is taken in consideration, it must be further ensured that also the closest distance between the location of the source current density $\vec{j}_\mathrm{0}$ and the conducting region $\Omega_\mathrm{C}$ is sufficient. That guaranteed, we are moreover free to approximate complex inductor paths by means of a finite number of straight lines. These line segments represent pieces of straight, current carrying wire as shown in \cref{fig:biot-savart:finite-wire}.

  Coils with larger cross sections, up to certain extent, can be modelled in a similar way by using bundles of paths to reproduce the envelopes. We therefor need to assume that the current density is concentrated on the inductor surface. The quality of this modelling assumption however depends on the skin-depth $\delta_\mathrm{Coil} = \sqrt{2/\left(\sigma_\mathrm{Coil}\,\omega_\mathrm{0}\,\mu_\mathrm{0}\right)}$ and holds especially for high frequencies in combination with a highly conducting coil-material like copper. Coils which are used for induction processing are very often internally cooled, which even supports the usage of path bundles.

  Integrating $\vec{j}_\mathrm{0}$ according to \cref{eqn:biot-savart} over a finite wire from the origin $\vec{r}_1$ to the destination $\vec{r}_2$ for a point of interest $\vec{x}$ yields:
  \begin{equation}
    \vec{A}_\mathrm{0} = \frac{\mu_\mathrm{0}}{4\pi} \left[ \ln\left(\frac{\cos{\theta_2} + \sin{\theta_2} + 1}{\cos{\theta_2} - \sin{\theta_2} + 1}\right)
                                                          - \ln\left(\frac{\cos{\theta_1} - \sin{\theta_1} + 1}{\cos{\theta_1} + \sin{\theta_1} + 1}\right)
                                                     \right] S_\mathrm{0} \vec{j}_\mathrm{0} \label{eqn:biot-svart:edge} \text{.}
  \end{equation}
  Both angles $\theta_{1,2}$ can be obtained from the geometric relations
  \begin{subequations}
    \label{seqn:biot-svart:edge:sincos}
    \begin{alignat}{3}
      \sin{\theta_1} &= \frac{l}{\abs{\vec{x} - \vec{r}_1}} \quad && \quad \sin{\theta_2} &&= \frac{\abs{\vec{r}_2 - \vec{r}_1} - l}{\abs{\vec{x} - \vec{r}_2}} \\[2ex]
      \cos{\theta_1} &= \frac{h}{\abs{\vec{x} - \vec{r}_1}} \quad && \quad \cos{\theta_2} &&= \frac{h}{\abs{\vec{x} - \vec{r}_2}} \text{,}
    \end{alignat}
  \end{subequations}
  while the length $l$ and the perpendicular $h$ are given by:
  \begin{subequations}
    \label{seqn:biot-svart:edge:lh}
    \begin{alignat}{2}
      l &= \frac{\left( \vec{x} - \vec{r}_1 \right) \cdot \left( \vec{r}_2 - \vec{r}_1 \right)}{\abs{\vec{r}_2 - \vec{r}_1}} \\[2ex]
      h &= \sqrt{\abs{\vec{x} - \vec{r}_1}^2 - l^2} \text{.}
    \end{alignat}
  \end{subequations}
  For most practical cases, a current $I_\mathrm{0}$ or harmonic current amplitude $\hat{I}_\mathrm{0}$ is directly known instead of $\vec{j}_\mathrm{0}$. In such cases, the product $S_\mathrm{0} \vec{j}_\mathrm{0}$ in \cref{eqn:biot-svart:edge} can be substituted:
  \begin{equation}
    S_\mathrm{0} \vec{j}_\mathrm{0} = \frac{\vec{r}_2 - \vec{r}_1}{\abs{\vec{r}_2 - \vec{r}_1}} I_\mathrm{0} = \frac{\vec{r}_2 - \vec{r}_1}{\abs{\vec{r}_2 - \vec{r}_1}} \hat{I}_\mathrm{0} \cos{\left(\omega_\mathrm{0} t - \alpha_\mathrm{0}\right)}\text{.}
  \end{equation}
  Let us finally recall from \cref{eqn:current0} that the real and imaginary parts of the current amplitude can be obtained according to:
  \begin{subequations}
    \label{seqn:biot-svart:phase}
    \begin{alignat}{2}
      \rei{I}{\mathrm{0}} &= \cos{\alpha_\mathrm{0}} \, \hat{I}_\mathrm{0} \\
      \imi{I}{\mathrm{0}} &= -\sin{\alpha_\mathrm{0}} \, \hat{I}_\mathrm{0} \text{,}
    \end{alignat}
  \end{subequations}
  if one uses the quasi-steady formulation from \cref{seqn:harmonic}. Biot-Savart's law (\ref{eqn:biot-savart}) can be applied separately for both complex parts.


\section{Embedded discretisation of charge conservation}
\label{sec:embedded}

  A consistent finite volume discretisation of \cref{eqn:elepot} is not straight-forward if we allow discontinuities in $\sigma$ at material boundaries $\Gamma_\mathrm{M}$ within the conducting domain $\Omega_\mathrm{C}$. Charge conservation ($\div{\vec{j}'}=0$) implies that the current density flux $\vec{n}_\mathrm{M} \cdot \vec{j}'$ remains continuous at these boundaries, where $\vec{n}_\mathrm{M}$ shall denote the corresponding normal vector. Ohm's law (\ref{eqn:ohm}) directly implies that a discontinuity in $\sigma \vec{n}_\mathrm{M}$ is compensated by an inverse proportional discontinuity in the electric field in normal direction $\vec{n}_\mathrm{M} \cdot \vec{E}$.

  More specifically, a sudden rise in electrical conductivity $\sigma$ across $\Gamma_\mathrm{M}$ is compensated by a corresponding local charge accumulation $\rho_\mathrm{E}$. This leads to a jump of the electric field $\vec{n}_\mathrm{M} \cdot \vec{E}$ according to Gauss's law (\ref{seqn:maxwell:gauss}). Due to the continuous magnetic vector potential $\vec{n}_\mathrm{M} \cdot \vec{A} \cdot $, this discontinuity is however only inherited by $\vec{n}_\mathrm{M} \cdot \grad{\phi}$.

\subsection{Jump conditions}
\label{subsec:embedded:jumpcond}

  In order to express these discontinuities across $\Gamma_\mathrm{M}$ mathematically, we will use a linear jump operator
    \begin{equation}
      \jump{\psi} = \jumpP{\psi} - \jumpN{\psi} \label{eqn:jump}
    \end{equation}
  in the same way as it is used in \cite{ARTICLE_Vukcevic_Jasak_Gatin_2017}. The superscripts ``$+$'' and ``$-$'' each denote values of a general function $\psi$, infinitesimally close to one side of the material boundary. The sign of the jump is not relevant for our purposes and jump locations always coincide with the location of faces of the finite volume mesh.

  Recalling the continuous current density flux due to charge conservation
  \begin{equation}
    \jump{\vec{n}_\mathrm{M} \cdot \vec{j}'} = \jump{\vec{n}_\mathrm{M} \cdot \sigma\vec{E}} = 0 \label{eqn:jump:current}
  \end{equation}
  and the absence of discontinuities in the magnetic vector potential
  \begin{equation}
    \jump{\vec{A}} = 0 \label{eqn:jump:magpot} \text{,}
  \end{equation}
  we obtain a set of jump conditions from \cref{seqn:potentials:elepot}:
  \begin{subequations}
    \label{seqn:jumpcond}
    \begin{align}
      \jump{\sigma} &= \jumpP{\sigma} - \jumpN{\sigma} \label{seqn:jumpcond:sigma} \\
      \jump{\phi} &= 0 \label{seqn:jumpcond:elepot} \\
      \jump{\vec{n}_\mathrm{M} \cdot \sigma\grad{\phi}} &= - \jump{\sigma}\vec{n}_\mathrm{M} \cdot \shortpd{\vec{A}}{t} \label{seqn:jumpcond:elepotgrad} \text{,}
    \end{align}
  \end{subequations}
  if we additionally assume that the electric scalar potential is continuous. The jump in electrical conductivity $\jump{\sigma}$ depends only on the material combination at $\Gamma_\mathrm{M}$ and is thus known.

\subsection{Discretisation of Gradient \& Laplacian}
\label{subsec:embedded:gradlaplacian}

\begin{figure}
    \centering
    \resizebox{0.6\textwidth}{!}{\includetikz{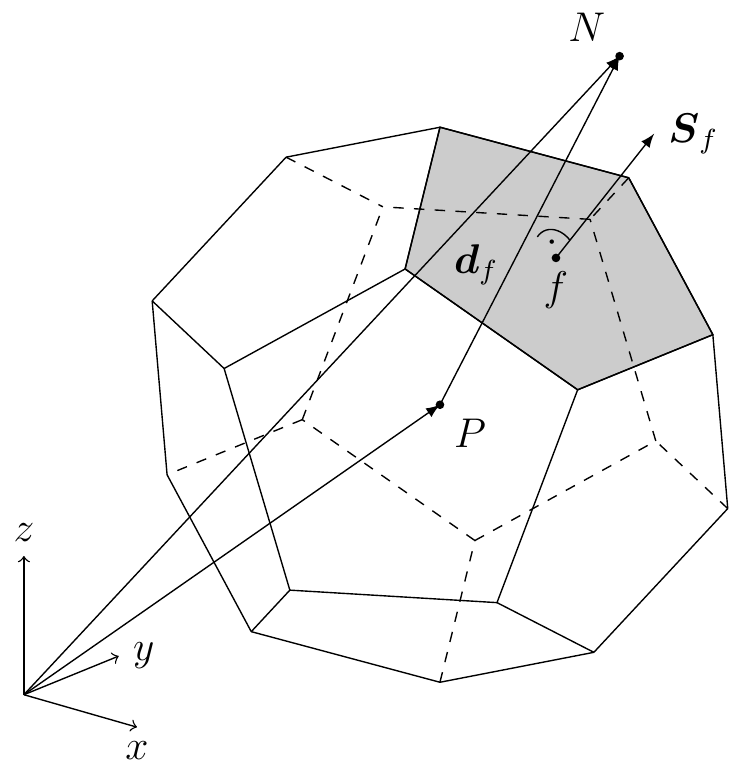}}
    \caption{Polyhedral control volume (CV) with owner centroid ($P$), neighbour cell centre ($N$), face centre ($f$), surface normal vector $\vec{S}_f$ and cell-centre distance $\vec{d}_f$. Each CV shares exactly one flat face with its immediate neighbouring cell (in the style of \cite[Fig. 1]{ARTICLE_Tukovic_Jasak_2012}).}
    \label{fig:embedded:polyhedra}
  \end{figure}

  The primary step to achieve spatial discretisation in accordance with the finite volume method (cf. \cite{BOOK_FerzigerPeric_Computational_Methods_2002, BOOK_VersteegMalalasekera_An_Introduction_To_Computational_Fluid_Dynamics_2007}) based on the notations of \textit{OpenFOAM} \cite{PHDTHESIS_Jasak_1996, ARTICLE_Weller_Tabor_Jasak_Fureby_1998}, is to divide $\Omega_\mathrm{C}$ into non-overlapping, polyhedral control volumes (CV), which are each bounded by flat faces (cf. \cref{fig:embedded:polyhedra}). Every internal face (indicated with subscript $f$) of the resulting finite volume mesh is only in contact with one owner CV (denoted with subscript $P$) and one neighbour CV (denoted with subscript $N$). The control volumes may be also referred to as cells. Faces and cells have centroids named according to the corresponding subscript ($f$, $P$ or $N$). The field data is numerically stored at the cell centroids.

  The finite volume method is based on the assumption of linear variation of fields being discretised. A discrete data value which is stored at the centroid of a CV corresponds to the cell-average. In this framework, both the gradient-term and the Laplacian-term may be evaluated using Gauss' theorem by integrating over the control volume. More specifically, an approximation with second-order accuracy can be achieved for the gradient of $\phi$ \cite{PHDTHESIS_Jasak_1996} in \cref{eqn:magpot:omega} as
  \begin{equation}
    \integral{V_\mathrm{CV}}{}{\grad{\phi}}{V} = \vintegral{\partial V_\mathrm{CV}}{}{\phi}{S} \approx \Sum{f}{}{\vec{S}_f \phi_f} \label{eqn:fv:gradientphi} \text{,}
  \end{equation}
  and for the Laplacian of \cref{eqn:elepot} according to
  \begin{equation}
    \integral{V_\mathrm{CV}}{}{\div{\sigma\grad{\phi}}}{V} = \vintegral{\partial V_\mathrm{CV}}{}{\sigma\grad{\phi}}{S} \approx \Sum{f}{}{\vec{S}_f \cdot \left(\sigma\grad{\phi}\right)_f} \label{eqn:fv:laplaciansigmaphi} \text{,}
  \end{equation}
  where the sum is evaluated for all faces of the CV and the subscript $f$ denotes values at the corresponding centroids of the face. The surface normal vector $\vec{S}_f = \vec{n}_f S_f$ for each flat face is the product of its normal and the corresponding face area $S_f$ and points always towards the outside of the CV. We will now discuss the calculation of the missing face values at a material boundary from the cell values on both sides.

\subsection{Material boundary faces}
\label{subsec:embedded:faces}

  Some of the internal faces of the domain $\Omega_\mathrm{C}$, but no boundary faces, may be part of the material boundary $\Gamma_\mathrm{M}$. Let us now examine the situation across one single face at $\Gamma_\mathrm{M}$ as it has been done in \cite{ARTICLE_Vukcevic_Jasak_Gatin_2017} for a free-surface boundary. It is important to note that the essential idea of the following procedure is to revise the hypothesis of linear variation across such faces, from which we know that it is locally inaccurate, to find a suitable replacement from the jump conditions (\ref{seqn:jumpcond}) instead. These relations contain one condition for $\phi$ and a second condition for $\grad{\phi}$, which may serve us as additional information to derive one-sided extrapolations for both. Compared to \cite{ARTICLE_Vukcevic_Jasak_Gatin_2017}, we only have to deal with a jump in $\grad{\phi}$, while $\phi$ itself remains continuous.

  Rewriting \cref{seqn:jumpcond:elepot,seqn:jumpcond:elepotgrad} using the definitions from \cref{eqn:jump}, reads:
  \begin{subequations}
    \label{seqn:facejumpcond}
    \begin{align}
      \jumpP{\phi_f} - \jumpN{\phi_f} &= 0 \\
      \jumpP{\sigma_f} \jumpP{\left(\vec{n} \cdot \grad{\phi}\right)}_f - \jumpN{\sigma_f} \jumpN{\left(\vec{n} \cdot \grad{\phi}\right)}_f &= \left(\jumpN{\sigma}_f - \jumpP{\sigma}_f\right) \shortpd{\vec{n}_f \cdot \vec{A}_f}{t} \text{.}
    \end{align}
  \end{subequations}
  Let us further restrict ourselves to the case where ``$+$'' corresponds to the neighbour side ($N$) and ``$-$'' corresponds to the owner side ($P$) of the face. The gradients normal to the face can then be calculated using the spatial change of $\phi$ between cell centre and face centre for both sides individually, which yields:
  \begin{subequations}
    \label{seqn:facegradients}
    \begin{align}
      \jumpN{\left(\vec{n} \cdot \grad{\phi}\right)}_f &\approx \frac{\left( \phi_f - \phi_P \right)}{\overline{Pf}} \label{seqn:facegradients:minus} \\
      \jumpP{\left(\vec{n} \cdot \grad{\phi}\right)}_f &\approx \frac{\left( \phi_N - \phi_f \right)}{\overline{fN}} \label{seqn:facegradients:plus} \text{.}
    \end{align}
  \end{subequations}
  Although \cref{seqn:facegradients:plus,seqn:facegradients:minus} are strictly valid only for orthogonal meshes ($\vec{n}_f \parallel \overline{fN} \parallel \overline{Pf}$), they still represent a good approximation for small deviations. For severely non-orthogonal meshes, a correction may be introduced in the same way as the discretisation of the Laplacian is corrected \cite{PHDTHESIS_Jasak_1996}. We will hereafter only consider the orthogonal case. With the assumption of constant $\sigma$ along $\overline{fN}$ and $\overline{Pf}$ according to
  \begin{subequations}
    \label{seqn:cellsigma}
    \begin{align}
      \jumpN{\sigma_f} &= \sigma_P \label{seqn:cellsigma:minus} \\
      \jumpP{\sigma_f} &= \sigma_N \label{seqn:cellsigma:plus} \text{,}
    \end{align}
  \end{subequations}
  it is possible to derive an expression for the face value $\phi_f = \jumpN{\phi_f} = \jumpP{\phi_f}$ of the electric scalar potential from \cref{seqn:facejumpcond,seqn:facegradients}:
  \begin{equation}
    \phi_f = \frac{\sigma_P}{\bar{\sigma}_f} f_x \, \phi_P + \frac{\sigma_N}{\bar{\sigma}_f} (1-f_x) \, \phi_N + \frac{\sigma_N - \sigma_P}{\bar{\sigma}_f} f_x \, (1-f_x) \frac{\overline{Pf}+\overline{fN}}{\abs{\vec{S}_f}} F^{\shortpd{\vec{A}}{t}}_f \label{eqn:facephi} \text{,}
  \end{equation}
  with $F^{\shortpd{\vec{A}}{t}}_f = \vec{S}_f \cdot \shortpd{\vec{A}_f}{t}$ being the face flux of the rate of change of the magnetic vector potential. In agreement with \cite{PHDTHESIS_Jasak_1996}, we have introduced the central-differencing weight $f_x = \overline{fN}/(\overline{Pf}+\overline{fN})$ (cf. \cref{fig:embedded:polyhedra}) and the linear interpolated face value of the electrical conductivity
  \begin{equation}
    \bar{\sigma}_f = f_x \, \sigma_P + (1-f_x) \, \sigma_N \label{eqn:linearfacesigma} \text{.}
  \end{equation}
  Evaluating $\phi_f$ from \cref{eqn:facephi} now allows us to calculate the face contribution for the gradient in \cref{eqn:fv:gradientphi} based on the jump conditions from \cref{seqn:jumpcond}. It is important to note that even for electrostatic problems ($\shortpd{\vec{A}}{t} = \vec{0}$), where the rightmost term in \cref{eqn:facephi} vanishes, a spacial interpolation is required to obey current conservation at a material boundary. Additionally, we would like to emphasise that the normal-gradient of $\sigma$, which is a part of $\grad{\sigma}$ on the RHS of \cref{eqn:elepot}, is always homogeneous on both sides of the material boundary. The reason for this lies in the assumption of \cref{seqn:cellsigma}. For the discretisation of $\grad{\sigma}$ similar to the discretisation of $\grad{\phi}$ from \cref{eqn:fv:gradientphi}, the face values $\sigma_f$ are different on both sides and need be taken from \cref{seqn:cellsigma}. That is, a special gradient scheme is necessary for the gradient of $\sigma$.

\subsection{One-sided extrapolation}
\label{subsec:embedded:extrapolation}

\begin{figure}
    \centering
    \resizebox{0.7\textwidth}{!}{\includetikz{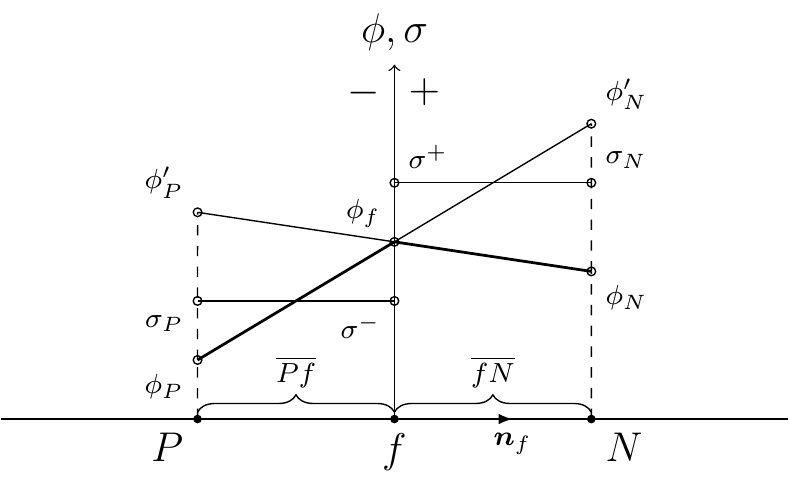}}
    \caption{One-sided extrapolation at a face of a material boundary. The thick black line shall illustrate the (non-linear) course of $\phi$ and the discontinuity in $\vec{n}_f \cdot \grad{\phi}$ from the owner cell ($P$) to the neighbour cell ($N$) across a face ($f$). The symbols ``+'' and ``-'' represent one side of the material boundary each. The values $\phi_{P,N}'$ are determined by extrapolation along $\phi_N$-$\phi_f$ and $\phi_P$-$\phi_f$ respectively.}
    \label{sfig:embedded:extrapolation}
  \end{figure}

  The face contributions to the sum in the Laplacian approximation according to \cref{eqn:fv:laplaciansigmaphi} are calculated based on central-differencing \cite{PHDTHESIS_Jasak_1996}. Leaving out the non-orthogonal correction and special treatment of skewed meshes for explanatory purposes, this reads:
  \begin{subequations}
    \label{eqn:fv:laplacian:facevalues}
    \begin{align}
      \left( \vec{S}_f \cdot \left(\sigma\grad{\phi}\right)_f \right)^P &\approx  L_f^P = \sigma_f \frac{\abs{\vec{S}_f}}{\abs{\vec{d}_f}} \left( \phi_N - \phi_P \right) \\
      \left( \vec{S}_f \cdot \left(\sigma\grad{\phi}\right)_f \right)^N &\approx  L_f^N = \sigma_f \frac{\abs{\vec{S}_f}}{\abs{\vec{d}_f}} \left( \phi_P - \phi_N \right) \text{.}
    \end{align}
  \end{subequations}
  where $\abs{\vec{d}_f} = \overline{PN} = \overline{Pf}+\overline{fN}$ is the cell-centre distance and the superscript $P$ means the face contribution to the owner CV and superscript $N$ means face contribution to the neighbour CV. Even though the central-differencing is again based on the hypothesis of linear varying fields, it does not need to be revised here if one follows the idea of \cite{ARTICLE_Vukcevic_Jasak_Gatin_2017} further. Instead a ``ghost''-value is used for the opposing cell across the face in order to respect the jump conditions of \cref{seqn:facejumpcond}:
  \begin{subequations}
    \label{eqn:fv:laplacian:jumpfacevalues}
    \begin{align}
      L_f^{'\,P} &= \sigma_P \frac{\abs{\vec{S}_f}}{\abs{\vec{d}_f}} \left( \phi_N' - \phi_P \right) \\
      L_f^{'\,N} &= \sigma_N \frac{\abs{\vec{S}_f}}{\abs{\vec{d}_f}} \left( \phi_P' - \phi_N \right) \text{.}
    \end{align}
  \end{subequations}
  The electrical conductivity at the face is taken from the corresponding side of the face as just explained. These ``ghost''-values $\phi_N'$ and $\phi_P'$ are derived by one-sided, linear extrapolations based on the face value of $\phi$:
  \begin{subequations}
    \label{seqn:extrapolationbase}
    \begin{align}
      \phi_N' &= \phi_f + \frac{\phi_f - \phi_P}{\overline{Pf}} \, \overline{fN} \\
      \phi_P' &= \phi_f + \frac{\phi_f - \phi_N}{\overline{fN}} \, \overline{Pf} \text{.}
    \end{align}
  \end{subequations}
  The situation across a face of the material boundary is sketched in \cref{sfig:embedded:extrapolation}, although the flux $F^{\shortpd{\vec{A}}{t}}_f = \shortpd{\vec{A}_f}{t} \cdot \vec{S}_f$ is not illustrated there. For both extrapolations in \cref{seqn:extrapolationbase}, second order accuracy is only maintained for completely unskewed meshes. Otherwise, a skew-correction may be necessary.

  Inserting the face value $\phi_f$ given by \cref{eqn:facephi} then yields:
  \begin{subequations}
    \label{seqn:extrapolation}
    \begin{align}
      \phi_N' &= \frac{\sigma_N}{\bar{\sigma}_f} \, \phi_N + \frac{\sigma_N - \sigma_P}{\bar{\sigma}_f} f_x     \left( \frac{\abs{\vec{d}_f}}{\abs{\vec{S}_f}} F^{\shortpd{\vec{A}}{t}}_f - \phi_P \right) \\
      \phi_P' &= \frac{\sigma_P}{\bar{\sigma}_f} \, \phi_P + \frac{\sigma_N - \sigma_P}{\bar{\sigma}_f} (1-f_x) \left( \frac{\abs{\vec{d}_f}}{\abs{\vec{S}_f}} F^{\shortpd{\vec{A}}{t}}_f + \phi_N \right) \text{.}
    \end{align}
  \end{subequations}
  By further substituting $\phi_N'$ and $\phi_P'$ from \cref{seqn:extrapolation} into \cref{eqn:fv:laplacian:jumpfacevalues}, we finally obtain the face contributions of the Laplacian discretisation for a material boundary face:
  \begin{subequations}
    \label{seqn:laplaciancontributions}
    \begin{align}
       L_f^{'\,P} &= \hat{\sigma}_f \frac{\abs{\vec{S}_f}}{\abs{\vec{d}_f}} \left( \phi_N - \phi_P \right) + \hat{\sigma}_f \left( 1 - \frac{\sigma_P}{\sigma_N} \right) f_x \, F^{\shortpd{\vec{A}}{t}}_f \label{seqn:laplaciancontributionsP} \\
       L_f^{'\,N} &= \hat{\sigma}_f \frac{\abs{\vec{S}_f}}{\abs{\vec{d}_f}} \left( \phi_P - \phi_N \right) - \hat{\sigma}_f \left( 1 - \frac{\sigma_N}{\sigma_P} \right) (1-f_x) \, F^{\shortpd{\vec{A}}{t}}_f \label{seqn:laplaciancontributionsN} \text{.}
    \end{align}
  \end{subequations}
  Here we have additionally introduced the harmonically interpolated electrical conductivity
  \begin{equation}
    \hat{\sigma}_f = \frac{\sigma_N \sigma_P}{\bar{\sigma}_f} = \left( \frac{f_x}{\sigma_N} + \frac{1 - f_x}{\sigma_P} \right)^{-1} \label{eqn:harmonicfacesigma} \text{,}
  \end{equation}
  which results from the derivation of the equations after combining the corresponding coefficients. We would like to stress at this point that for an electrostatic case ($\shortpd{\vec{A}}{t} = \vec{0}$), this harmonic interpolation of $\sigma$ is sufficient to obey the jump conditions (\ref{seqn:jumpcond}) in the Laplacian discretisation due to the vanishing face flux ($F^{\shortpd{\vec{A}}{t}}_f = 0$).

  Exactly as it is the case in \cite{ARTICLE_Vukcevic_Jasak_Gatin_2017}, a closer inspection of \cref{seqn:laplaciancontributions} reveals that an implicit discretisation of the Laplacian across material boundaries still leads to a symmetric matrix due to equal upper and lower off-diagonal coefficients. It is also still possible to reconstruct the diagonal contributions using negative off-diagonal coefficients. The last terms in \cref{seqn:laplaciancontributions} denote additional source terms that are not only anti-symmetric but also differ in magnitude depending on mesh geometry and the jump of $\sigma$ across the material boundary. Analogously to \cite{ARTICLE_Vukcevic_Jasak_Gatin_2017}, additional face flux from the source terms is here compensated by the jump in the gradient of $\phi$.


\section{Validation}
\label{sec:validation}

  In order to demonstrate the capabilities of our proposed method in the finite volume framework of \textit{foam-extend}, and to verify the corresponding implementation, this section will be devoted to two individual validation cases. The purpose of the first test case, which was motivated by an existing facility and from the existence of tailored magnetic fields in industrial applications, is to provide a general idea of how the numerical method from \cref{sec:efficiency} behaves depending on different mesh parameters. A second, more academical, test setup is used to validate the embedded discretisation from \cref{sec:embedded}.

\subsection{Rotating magnetic field}
\label{subsec:validation:multimag3D}

\begin{figure}
    \begin{subfigure}[c]{0.545\textwidth}
      \centering
      \resizebox{\linewidth}{!}{\includegraphics{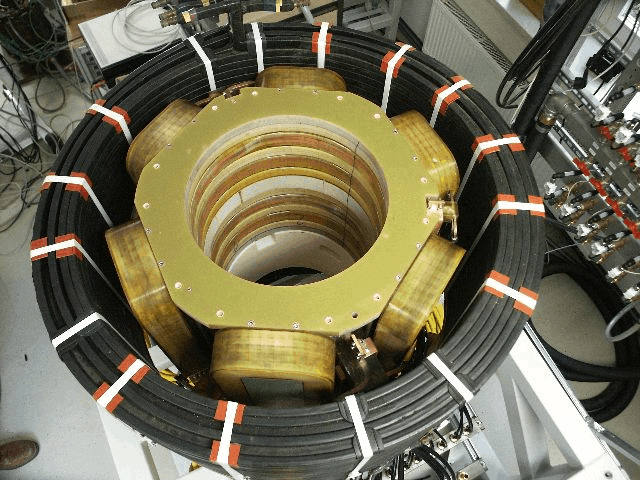}}
      \caption{The MULTIMAG facility at HZDR  \cite{ARTICLE_Pal_Cramer_Gundrum_Gerbeth_2009}.}
      \label{sfig:validation:multimag3D:multimag}
    \end{subfigure} \hspace*{1em}
    \begin{subfigure}[c]{0.41\textwidth}
      \centering
      \resizebox{\linewidth}{!}{\includegraphics{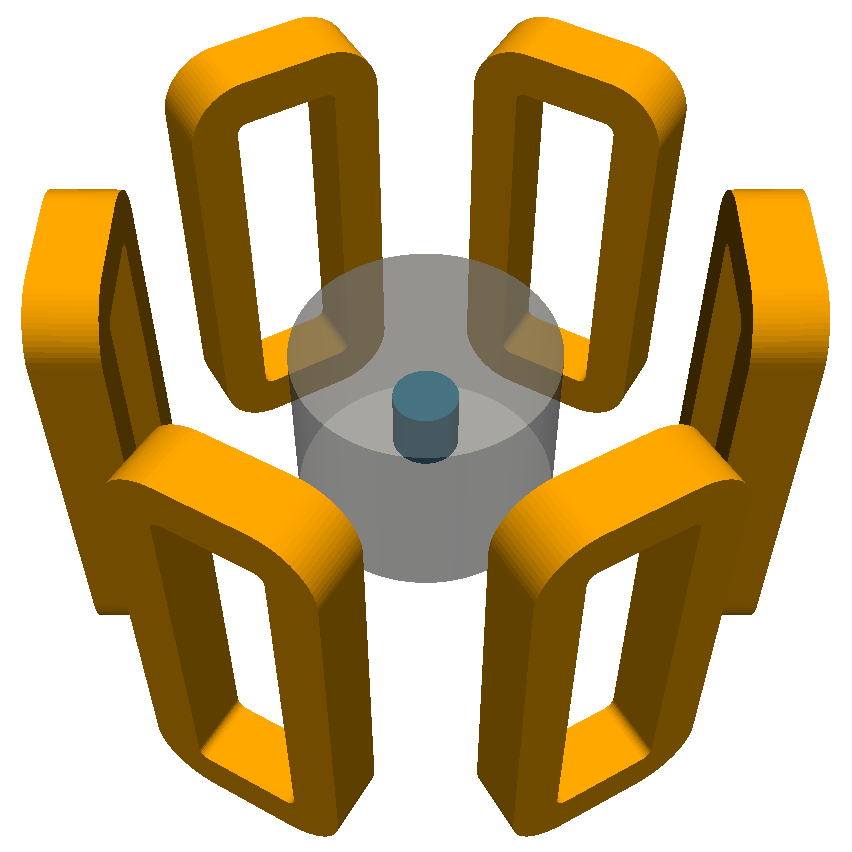}}
      \caption{Test case with conducting cylinder, non-conducting domain and 6 RMF coils.}
      \label{sfig:validation:multimag3D:sketch}
    \end{subfigure}
    \begin{subfigure}[c]{0.545\textwidth}
      \centering
      \resizebox{\linewidth}{!}{\includegraphics{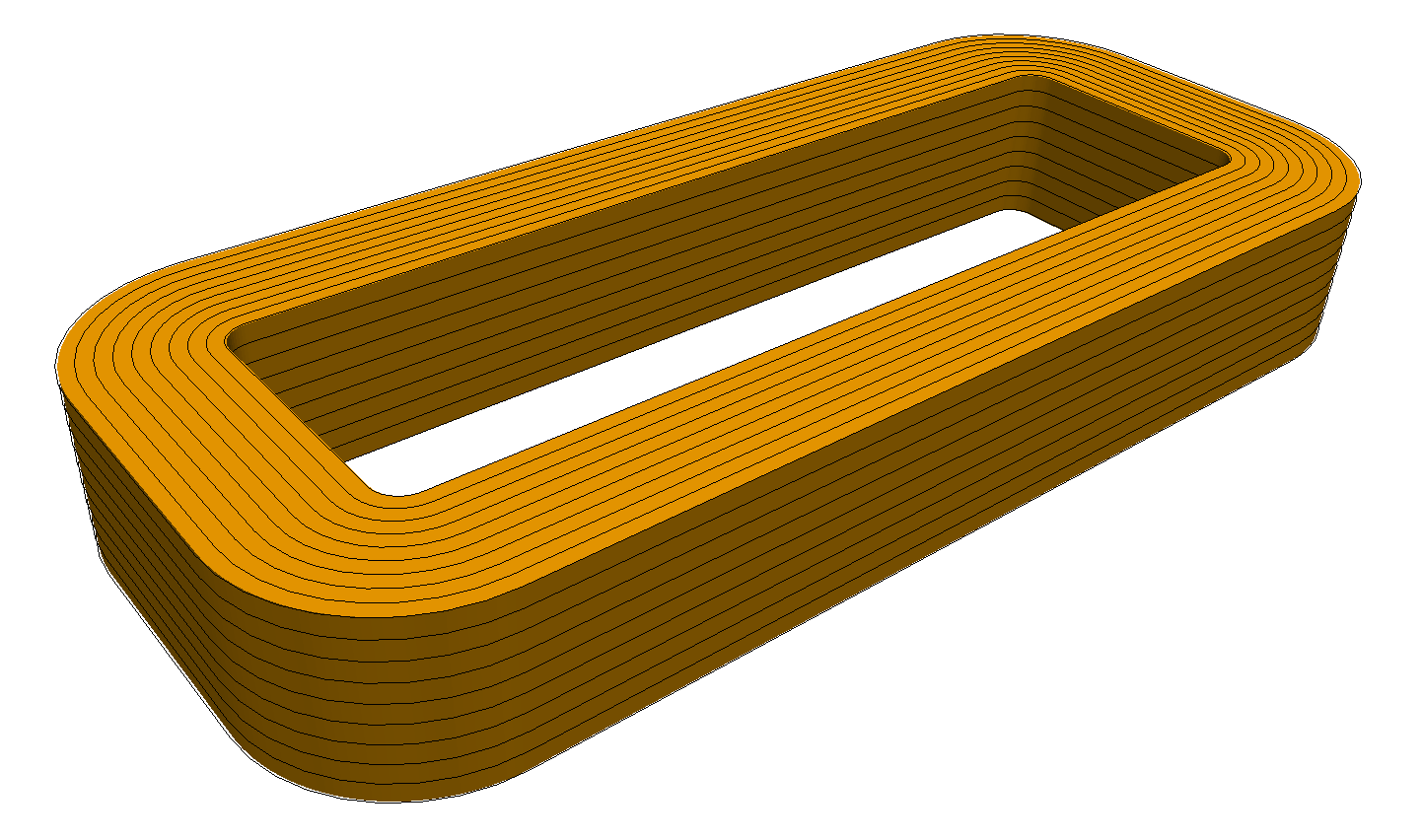}}
      \caption{Biot-Savart inductor with bundle of filaments.}
      \label{sfig:validation:multimag3D:coil}
    \end{subfigure} \hspace*{1em}
    \begin{subfigure}[c]{0.41\textwidth}
      \centering
      \resizebox{0.72\linewidth}{!}{\includegraphics{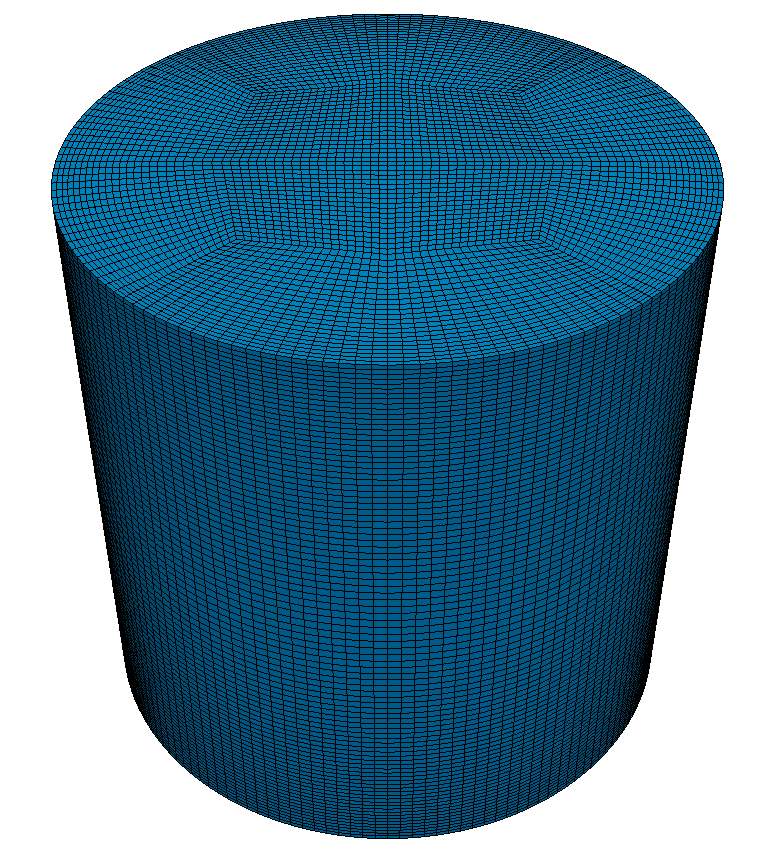}}
      \caption{Conducting cylinder with finite number of hexahedral cells.}
      \label{sfig:validation:multimag3D:cylinder}
    \end{subfigure}
    \caption{Rotating magnetic field (RMF) test case: Motivation from scientific facility \subref{sfig:validation:multimag3D:multimag}, numerical model \subref{sfig:validation:multimag3D:sketch}, inductor discretisation based on an edge mesh \subref{sfig:validation:multimag3D:coil} and conductor discretisation based on a finite volume mesh \subref{sfig:validation:multimag3D:cylinder}.}
    \label{fig:validation:multimag3D:multimag}
  \end{figure}

  The first test concerns an eddy-current problem of multiple inductor coils with different phase shifts $\alpha_\mathrm{0}$ according to \cref{eqn:current0,seqn:biot-svart:phase}. We have selected a coil setup similar to the RMF-coils (RMF: rotating magnetic field) of the MULTIMAG (MULTI purpose MAGnetic fields) facility \cite{ARTICLE_Pal_Cramer_Gundrum_Gerbeth_2009} at the HZDR (Helmholtz-Zentrum Dresden-Rossendorf). A picture of this facility is shown in \cref{sfig:validation:multimag3D:multimag}, whereas the derived numerical model is illustrated in \cref{sfig:validation:multimag3D:sketch}.

  For a finite, electrically conducting cylinder (cf. \cref{sfig:validation:multimag3D:sketch} \& \cref{sfig:validation:multimag3D:cylinder}) with a height of $2H$ and radius $R$, which is axially aligned with the RMF, an analytical solution for the corresponding time-averaged Lorentz-force (\ref{eqn:harmonic:lorentz}) is available \cite{ARTICLE_Gorbachev_Nikitin_Ustinov_1974, ARTICLE_Grants_Gerbeth_2001, ARTICLE_Pal_Cramer_Gundrum_Gerbeth_2009} for small shielding parameters:
  \begin{equation}
    S = \mu_\mathrm{0}\sigma\omega_\mathrm{0}R^2 < 1 \label{eqn:validation:multimag3D:shieldingparameter} \text{.}
  \end{equation}
  Provided that \cref{eqn:magneticreynolds} and the low-frequency approximation \cite{BOOK_Davidson_An_Introduction_To_Magnetohydrodynamics}
  \begin{equation}
    \mathrm{Re}_\mathrm{m} \ll S < 1 \label{eqn:validation:multimag3D:lowfreq}
  \end{equation}
  both hold, the time-averaged Lorentz-force inside the cylinder contains only an azimuthal component, which can be written in cylindrical coordinates ($r$, $\varphi$, $z$) as
  \begin{equation}
    \avg{\vec{F}}{t} = \frac{\sigma\omega_\mathrm{0}B_\mathrm{0}^2}{2} ~ R ~ s(r, z) ~ \vec{e}_\mathrm{\varphi} \label{eqn:validation:multimag3D:analytical} \text{,}
  \end{equation}
  where $B_\mathrm{0}=\norm{\vec{B}(r=0, z=0)}$ is the magnitude of the magnetic field in the centroid of the cylinder. The shape function $s(r, z)$ reads:
  \begin{equation}
    s(r, z) = \frac{r}{R} - \Sum{k=1}{\infty}{c_k J_1 \left(\lambda_k \frac{r}{R} \right) \cosh\left( \lambda_k \frac{z}{R} \right)} \label{eqn:validation:multimag3D:shapefunction} \text{.}
  \end{equation}
  Based on the Bessel functions of the first kind $J_1$ and order and the roots $\lambda_k$ of its first derivative $J'_1$, the coefficients in \cref{eqn:validation:multimag3D:shapefunction} are defined as:
  \begin{equation}
    c_k = \frac{2}{(\lambda_k^2 - 1) J_1(\lambda_k) \cosh\left( \lambda_k \frac{H}{R} \right)} \label{eqn:validation:multimag3D:shapecoeffs} \text{.}
  \end{equation}

  We have performed numerical simulations for an aspect ratio of $H/R=1$ with $R=30~\unit{mm}$, an angular frequency of $\omega_\mathrm{0}=2\pi~50~\unit{Hz}$ and $B_\mathrm{0}=0.4216 \times 10^{-3}~\unit{T}$. The value of $B_\mathrm{0}$ originates from an underlying magnetic Taylor number of $\mathrm{Ta}=\sigma\omega_\mathrm{0}B_\mathrm{0}^2 R^4 / (2\rho\nu^2)=1\times 10^{5}$ for the material properties ($\rho=6353~\unitfrac{kg}{m^3}$, $\nu=3.436\times 10^{-7}~\unitfrac{m^2}{s}$, $\sigma=3.289\times 10^6~\unitfrac{S}{m}$) of the liquid-metal alloy Gallium-Indium-Tin (GaInSn) at room temperature ($T_\mathrm{0}=20\degree \unit{C}$).

  In the numerical model as shown in \cref{sfig:validation:multimag3D:sketch}, the conducting cylinder is surrounded by a non-conducing region which extends up to a distance of $R_\mathrm{\infty}=4R$ in axial and radial direction from its centroid. That is, the volume of the non-conducting region is seven times larger than the volume of the conducting domain.

  The reader may have recognised that the non-conducting region does not contain the coils. This is possible due the utilisation of Biot-Savart's law as presented in \cref{subsec:biot-svart}. According to \cref{sfig:domain:impressedmagpot:impressedmagpot:gammaC}, boundary conditions for the impressed magnetic vector potential $\vec{A}_\mathrm{0}$ in $\Omega_\mathrm{C}$ are calculated from \cref{eqn:biot-savart}. If a coil is modelled on the basis of \cref{eqn:biot-svart:edge} and represented by a finite number of piecewise linear edges, it does not need to be part of the finite volume mesh. \Cref{sfig:validation:multimag3D:coil} shows the separate edge-discretisation of one RMF Biot-Savart inductor with a bundle of such filaments.

  \begin{figure}
    \begin{subfigure}[c]{0.49\textwidth}
      \centering
      \resizebox{\linewidth}{!}{\includegraphics{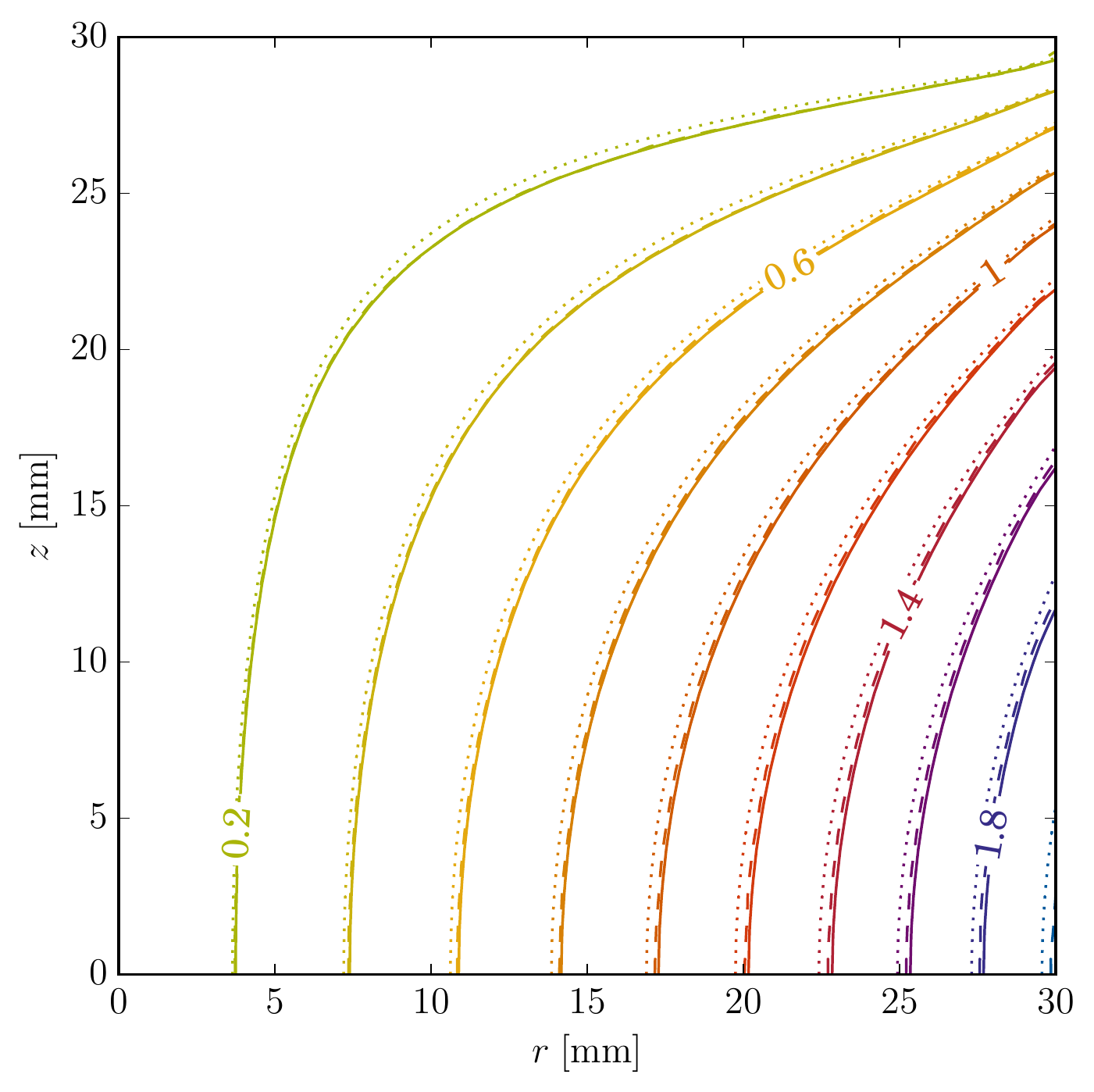}}
      \caption{Azimuthal component: $\avg{\vec{F}}{t} \cdot \vec{e}_\mathrm{\varphi} ~[\unitfrac{N}{m^3}]$}
      \label{sfig:validation:multimag3D:comparefa}
    \end{subfigure}
    \begin{subfigure}[c]{0.49\textwidth}
      \centering
      \resizebox{\linewidth}{!}{\includegraphics{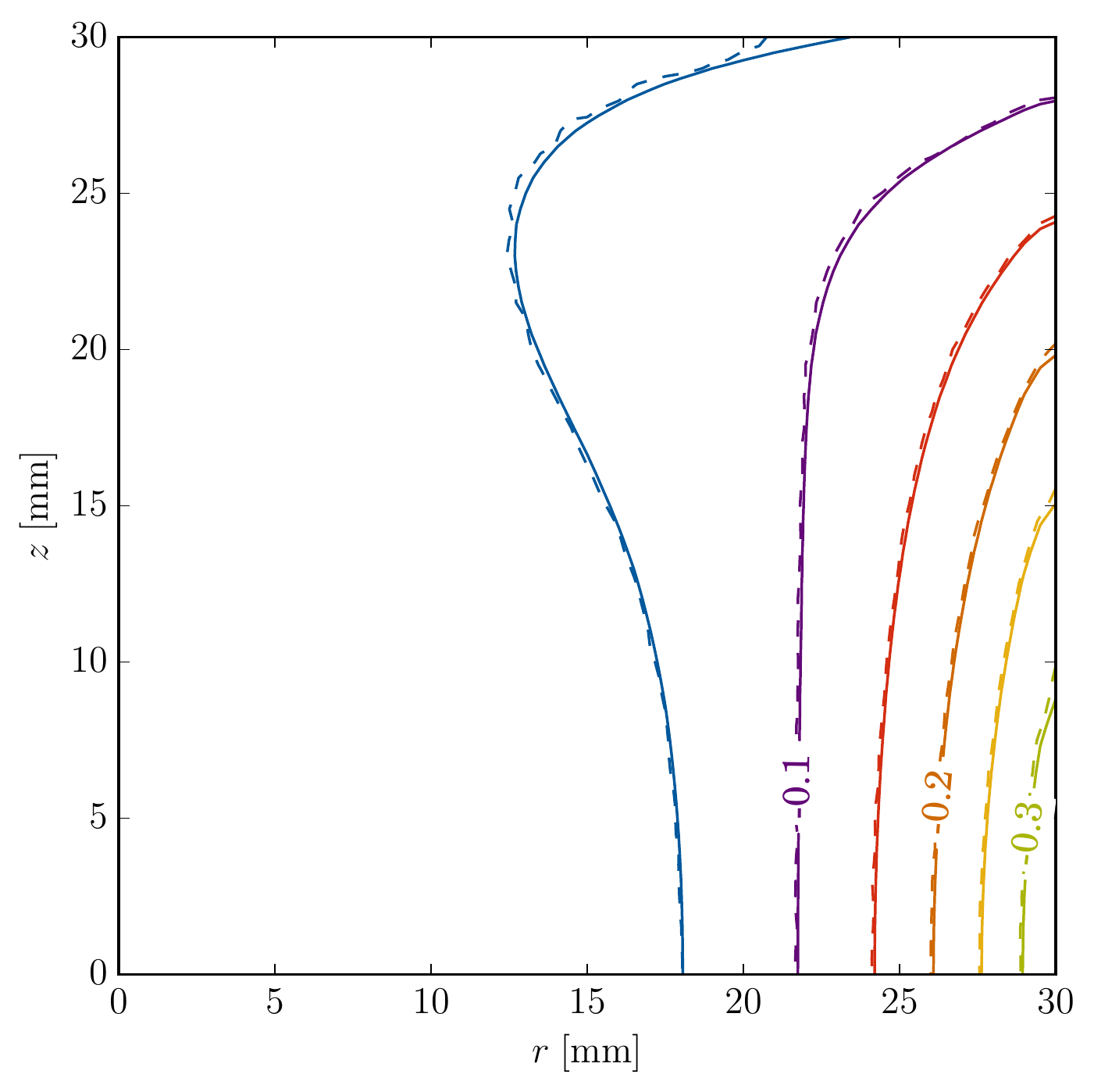}}
      \caption{Radial component: $\avg{\vec{F}}{t} \cdot \vec{e}_\mathrm{r} ~[\unitfrac{N}{m^3}]$}
      \label{sfig:validation:multimag3D:comparefr}
    \end{subfigure}
    \begin{subfigure}[c]{0.49\textwidth}
      \centering
      \resizebox{\linewidth}{!}{\includegraphics{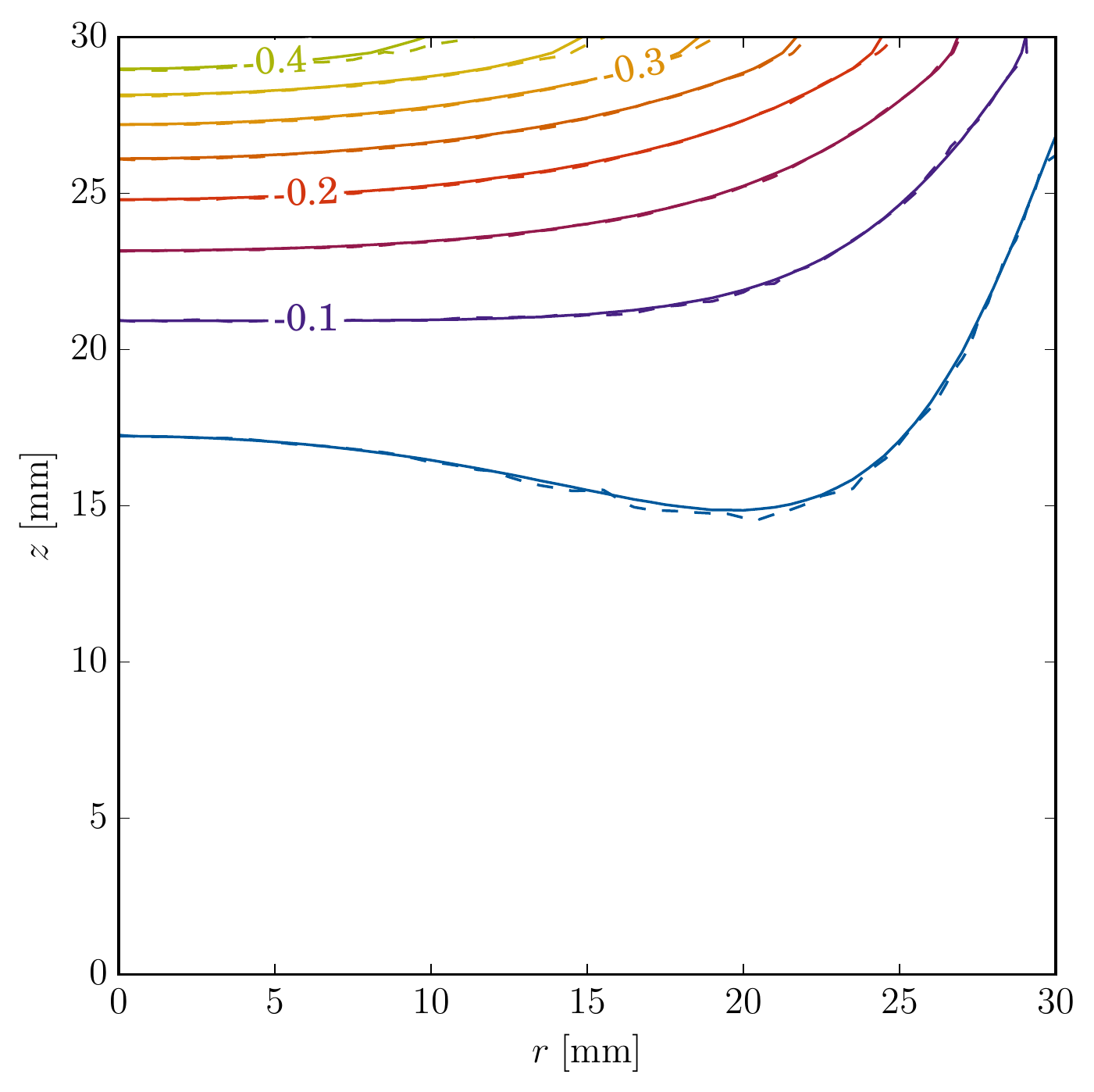}}
      \caption{Axial component: $\avg{\vec{F}}{t} \cdot \vec{e}_\mathrm{z} ~[\unitfrac{N}{m^3}]$}
      \label{sfig:validation:multimag3D:comparefz}
    \end{subfigure} \hspace*{0.1em}
    \begin{subfigure}[c]{0.49\textwidth}
      \centering
      \resizebox{\linewidth}{!}{\includegraphics{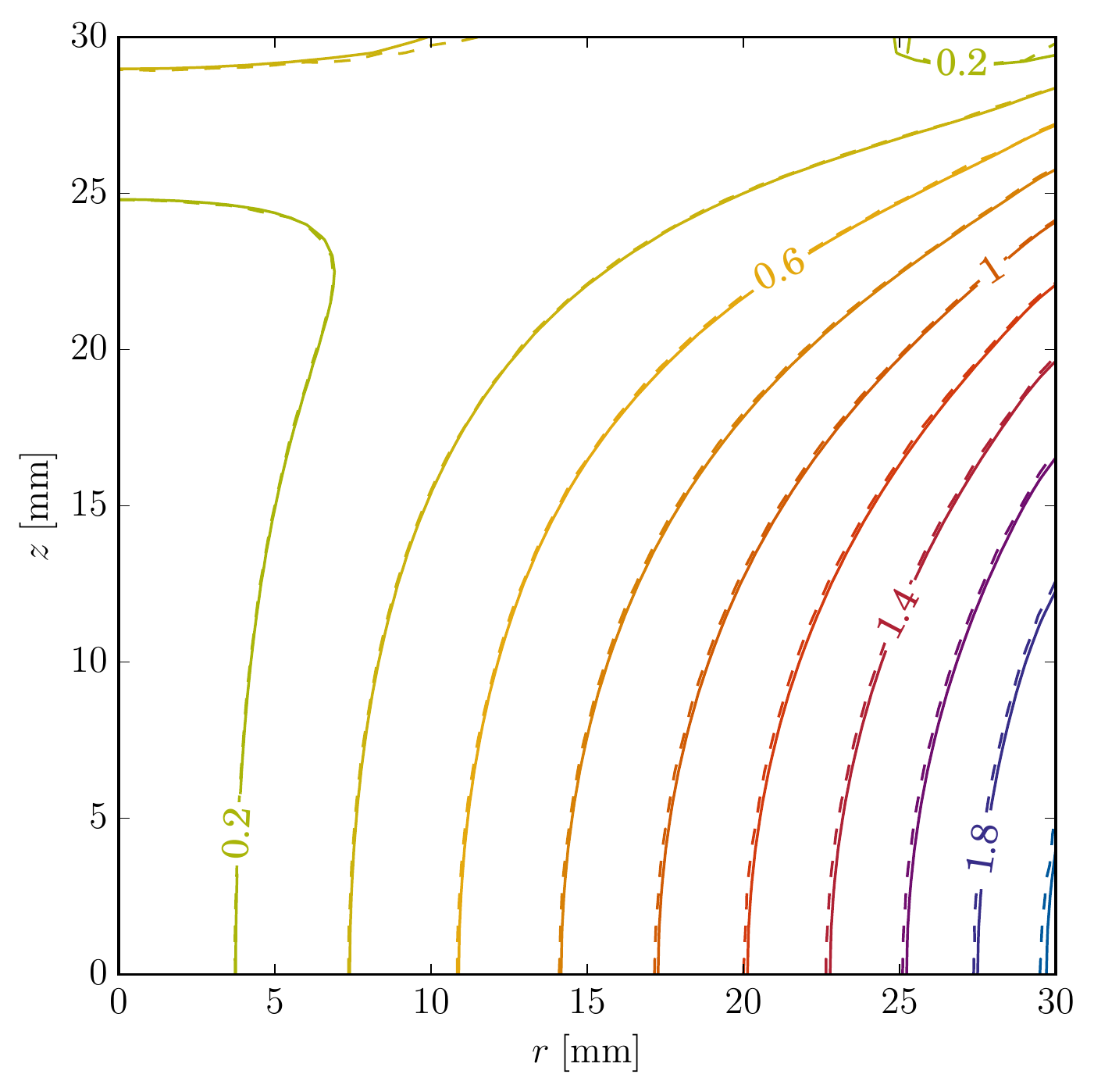}}
      \caption{Magnitude: $\norm{\avg{\vec{F}}{t}} ~[\unitfrac{N}{m^3}]$}
      \label{sfig:validation:multimag3D:comparef}
    \end{subfigure}
    \caption{Comparison of results (without low-frequency approximation) from a numerical simulation of our implementation in \textit{foam-extend} (solid) with results from the commercial software \textit{Cobham Opera 3D} \cite{MISC_Opera3D} (dashed). Figure \subref{sfig:validation:multimag3D:comparefa} additionally shows the analytical solution (dotted) from \cref{eqn:validation:multimag3D:analytical} based on the first $40$ addends of the infinite series of $s(r,z)$.}
    \label{fig:validation:multimag3D:comparef}
  \end{figure}

  All four $z$-$r$ contour plots of \cref{fig:validation:multimag3D:comparef} depict a comparison of results for the time-averaged Lorentz-force from a numerical simulation of our implementation in \textit{foam-extend} (solid lines) with results from the commercial FEM-software \textit{Cobham Opera 3D} \cite{MISC_Opera3D} (dashed lines). Both data sets are the outcome of formulations without low-frequency approximation. The underlying mesh for our solution consists of $N\approx 7\times 10^5$ cells, whereof $N_\mathrm{C}\approx 1.2\times 10^5$ were used for the conducting region. These parameters correspond to the highlighted mesh ($1.000$) in \cref{tab:validation:multimag3D:meshes}. Using one single CPU-core (\textit{Intel i5-3570}), the simulation time for this mesh, given a global residual of $1\times 10^{-8}$, amounts to $\approx 500~\unit{s}$. For the reference solution in \textit{Cobham Opera 3D}, we have used a finer mesh consisting of $\approx 5\times 10^6$ linear and $\approx 1.5\times 10^5$ quadratic finite elements, on which convergence with the same residual was achieved after $\approx 6600~\unit{s}$ with the help of four CPU-cores of the same processor. For all components (\crefrange{sfig:validation:multimag3D:comparefa}{sfig:validation:multimag3D:comparefz}) and the magnitude (\cref{sfig:validation:multimag3D:comparef}) of $\avg{\vec{F}}{t}$, we found a very good agreement of both results.

  The plot in \cref{sfig:validation:multimag3D:comparefa} additionally relates the numerical results to the analytical solution (dotted line). There are two reasons which are responsible for the slight, but obvious deviation between numerical and analytical data of the azimuthal component of $\avg{\vec{F}}{t}$: Firstly, \cref{eqn:validation:multimag3D:analytical} has been derived using the low-frequency approximation together with the assumption of a purely axial magnetic vector potential. Secondly, the shape function in \cref{eqn:validation:multimag3D:shapefunction} needs to be approximated using only a finite number (in our case $40$) addends of its infinite series.

  \begin{figure}
    \centering
    \resizebox{0.49\linewidth}{!}{\includegraphics{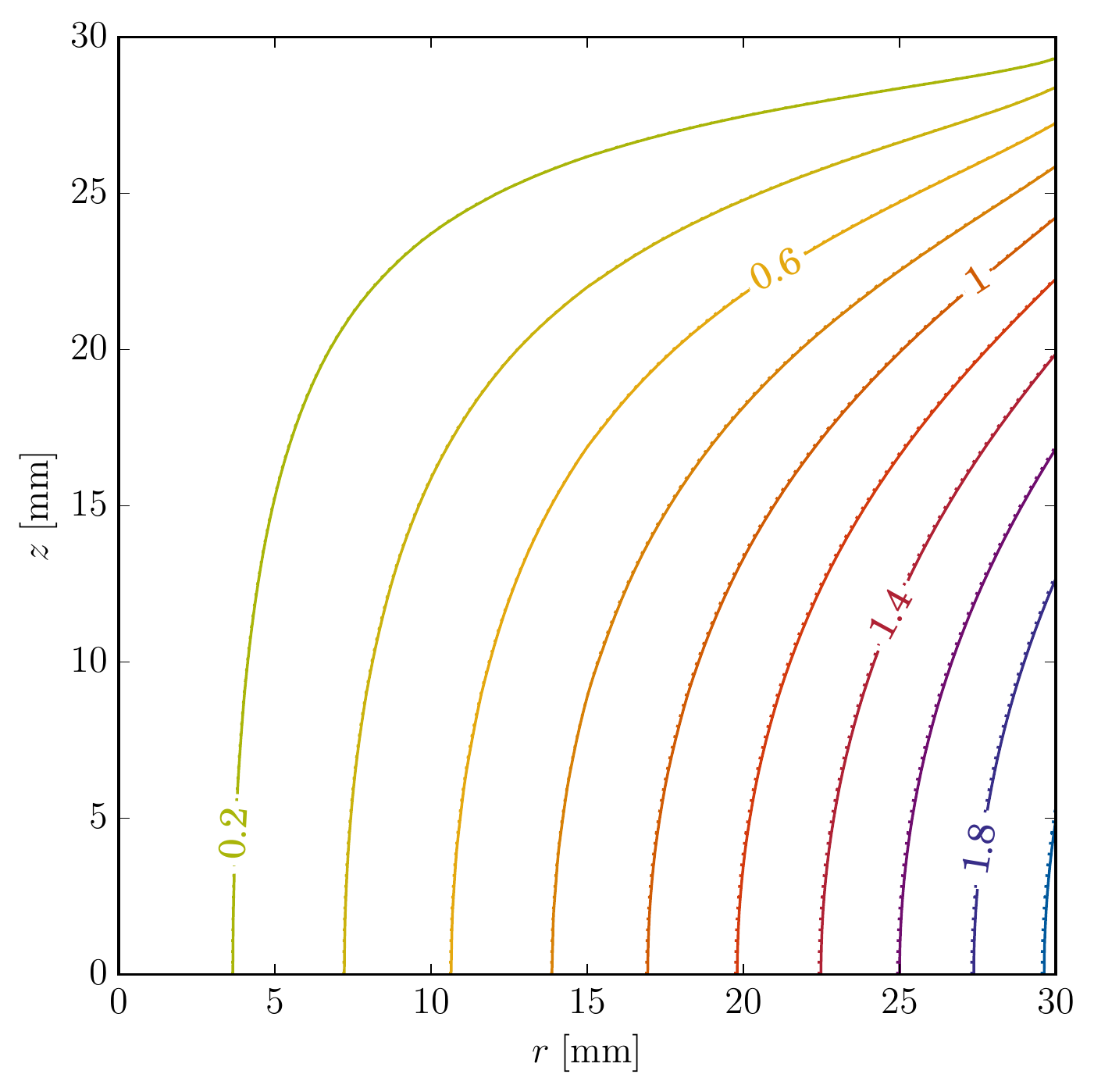}}
    \caption{Comparison of the analytical solution (dotted) of the azimuthal part $\avg{\vec{F}}{t} \cdot \vec{e}_\mathrm{\varphi} ~[\unitfrac{N}{m^3}]$ of the time-averaged Lorentz-force from \cref{eqn:validation:multimag3D:analytical} with results of a numerical simulation (solid) based on the low-frequency approximation from \cref{eqn:validation:multimag3D:lowfreq} and a magnetic vector potential with only an axial component (infinitely high coils).}
    \label{fig:validation:multimag3D:comparefalowf}
  \end{figure}

  In our code, the low-frequency assumption can be modelled numerically by assuming the reduced magnetic vector potential, which was introduced in \cref{eqn:magpotsplit}, to vanish completely ($\vec{A}'\equiv\vec{0}$). As a consequence, \cref{eqn:reducedmagpot:omega} may be dropped. It is therefore only necessary to solve for the electric scalar potential in $\Omega_\mathrm{C}$. The boundary condition from \cref{eqn:elepot:bc:gammaC} for $\phi$ on $\Gamma_\mathrm{C}$ simplifies to $\vec{n}_\mathrm{C} \cdot \grad{\phi} = -\vec{n}_\mathrm{C} \cdot \shortpd{\vec{A}_\mathrm{0}}{t}$. An almost perfect axial (impressed) magnetic vector potential $\vec{A}_\mathrm{0}$ was realised by simply setting the length of the vertical filaments of our Biot-Savart inductors (cf. \cref{sfig:validation:multimag3D:coil}) much larger than the height of the conducting cylinder. With the assumptions of \cref{eqn:validation:multimag3D:analytical} numerically met, a very good agreement with our numerical solution has finally been achieved. This is illustrated in \cref{fig:validation:multimag3D:comparefalowf}.

  As explained in \cite{PHDTHESIS_Jasak_1996}, spatial discretisation of the FVM in \textit{OpenFOAM} is ideally second order accurate, if linear interpolation and central differencing is used. This accuracy is however limited due to mesh skewness and non-orthogonality. In case of the O-grid of the cylinder-mesh (cf. \cref{sfig:validation:multimag3D:coil}), the bulk mesh-quality is very good, but it is not perfect from a local point of view. To analyse the grid-dependency of the error, we will hereinafter refer to the following three error-norms:
  \begin{subequations}
    \label{seqn:errornorms}
    \begin{align}
      \norm{\mathcal{E}}_\infty &= \frac{\mathrm{max}_k \, \abs{\overbar{\mathcal{Q}}_k - \mathcal{Q}_k}}{\mathrm{max}_k \,\abs{\overbar{\mathcal{Q}}_k}} \label{seqn:errornorms:infty} \\
      \norm{\mathcal{E}}_1      &= \frac{\sum_{k=1}^{N} \, \abs{\overbar{\mathcal{Q}}_k - \mathcal{Q}_k}}{\sum_{k=1}^{N}\,\abs{\overbar{\mathcal{Q}}_k}} \label{seqn:errornorms:1} \\
      \norm{\mathcal{E}}_2      &= \left(\frac{\sum_{k=1}^{N} \, \abs{\overbar{\mathcal{Q}}_k - \mathcal{Q}_k}^2}{\sum_{k=1}^{N}\,\abs{\overbar{\mathcal{Q}}_k}^2} \right)^{1/2} \label{seqn:errornorms:2} \text{.}
    \end{align}
  \end{subequations}
  Therein, $N$ is the number of sample points, $k$ a sample point index, $\mathcal{Q}_k$ is a local value of an approximation of a quantity of interest (e.g. a numerical solution of $\avg{\vec{F}}{t}$ at sample location $k$), $\overbar{\mathcal{Q}}_k$ is the reference value of $\mathcal{Q}_k$ (e.g. the most accurate value available at $k$) and $\overbar{\mathcal{Q}}_k-\mathcal{Q}_k=\mathcal{E}_k$ is regarded as the absolute error of an approximation of a certain quantity with respect to a reference solution of that quantity.

  \begin{table}
    \caption{Numerical meshes with different cell-sizes: The first column contains mesh scaling factors, $\triangle$ is a relative cell-size and $\triangle/\triangle_\mathrm{max}$ the corresponding normalised cell-size. The total number of cells for each mesh is represented by $N$ and the number of cells in the conducting region by $N_\mathrm{C}$. In the last column, the computational time for one CPU-core is listed. The highlighted mesh (1.000) corresponds to the default from which refinement/coarsening was performed in both directions.}
    \label{tab:validation:multimag3D:meshes}
    \centering
    \small
    \renewcommand{\arraystretch}{1.1}
    \begin{tabularx}{\textwidth}{X X X l l l l}
      \hline
      Mesh &$\triangle$ &$\triangle/\triangle_\mathrm{max}$ &$N$ &$N_\mathrm{C}$ &$N/N_\mathrm{C}$ &$t_\mathrm{CPU}~[\unit{s}]$ \\
      \hline
      $0.125$ &$8.00$ &$1.0000$ &$1792$ &$384$ &$4.67$ &$1$ \\
      $0.250$ &$4.00$ &$0.5000$ &$11832$ &$2112$ &$5.60$ &$5$ \\
      $0.375$ &$2.67$ &$0.3333$ &$43000$ &$8160$ &$5.27$ &$15$ \\
      $0.500$ &$2.00$ &$0.2500$ &$91872$ &$15840$ &$5.80$ &$40$ \\
      $0.750$ &$1.33$ &$0.1667$ &$367200$ &$62560$ &$5.87$ &$300$ \\
      \rowcolor{hzdr-gray5}
      $1.000$ &$1.00$ &$0.1250$ &$698880$ &$120960$ &$5.77$ &$500$ \\
      $1.500$ &$0.67$ &$0.0833$ &$2629600$ &$454480$ &$5.79$ &$2700$ \\
      $2.000$ &$0.50$ &$0.0625$ &$5684352$ &$967680$ &$5.87$ &$10800$ \\
      $2.500$ &$0.40$ &$0.0500$ &$11325600$ &$1908000$ &$5.94$ &$31200$ \\
      \hline
    \end{tabularx}
  \end{table}

  \begin{figure}
    \centering
    \resizebox{0.69\linewidth}{!}{\includegraphics{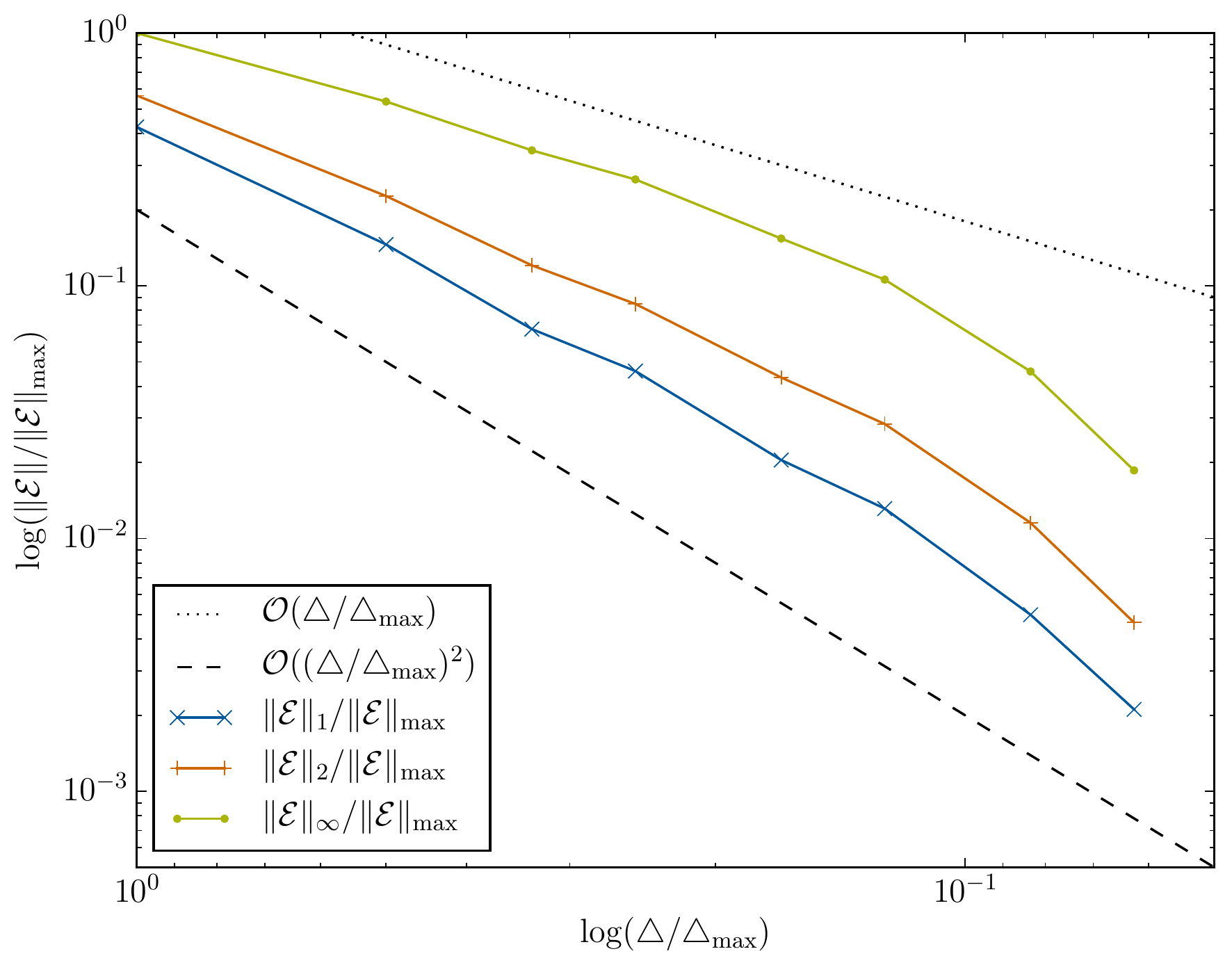}}
    \caption{Convergence rate of the normalised error norms from \cref{seqn:errornorms} for the time-averaged Lorentz-force ($\mathcal{Q}=\avg{\vec{F}}{t}$) and different normalised mesh cell-sizes according to \cref{tab:validation:multimag3D:meshes}. The finest mesh ($2.500$) has served as reference solution for the error-estimation.}
    \label{fig:validation:multimag3D:convergence}
  \end{figure}

  For all mesh variations from \cref{tab:validation:multimag3D:meshes} and its relative cell-sizes $\triangle$, \cref{fig:validation:multimag3D:convergence} shows the course of the corresponding error norms from \cref{seqn:errornorms} with respect to the best solution with the finest mesh ($2.500$). A normalisation has been applied which is based on the global maximum value $\norm{\mathcal{E}}_{max}$ that any norm may take on (not to be confused with $\norm{\mathcal{E}}_{\infty}$), and the maximum of all cell-sizes $\triangle_\mathrm{max}$. As expected, we found a convergence rate between first (dotted line) and second order (dashed line). The infinity-norm ($\norm{\mathcal{E}}_\infty$) shows the worst behaviour, as it represents a global maximum of the error. In contrast to that, the $L^2$-Norm shows the best behaviour as here the error is quadratically weighted and describes a mean error. This exactly reflects the mesh quality as explained above.

  Besides discretisation errors, the quality of the numerical results also depends on the size of the non-conducting region, as the domain of Maxwell's equations (\ref{seqn:maxwell}) is actually unbounded. The truncation of $\Omega_\mathrm{0}$ at $\Gamma_\mathrm{\infty}$ introduces a related truncation error. In order to investigate the impact of this error, it is helpful to regard all currents of the finite volume discretisation as a general, localised current-distribution of size $R_J$. According to \cite[Section 5.6]{BOOK_Jackson_Classical_Electrodynamics_1975}, we may approximate the strength of the magnetic field $B(r)$ at some distant radius $r > R_J$ with
  \begin{equation}
    B(r) \approx \frac{J}{r^3} \label{eqn:validation:multimag3D:dipole} \text{,}
  \end{equation}
  where $J$ is a coefficient which represents the influence of the first magnetic moment of the localised current-distribution. Using this simple approximation, the ratio of the magnetic field strengths $B_\infty = B(R_\infty)$ and $B_J = B(R_J)$ reads: $B_\infty/B_J \approx (R_\infty/R_J)^3$. By implication, we can approximately express the ratio of the radii in terms of the ratio of magnetic field strengths:
  \begin{equation}
    \frac{R_\infty}{R_J} \approx \left(\frac{B_\infty}{B_J}\right)^\frac{1}{3} \label{eqn:validation:multimag3D:diploeratior} \text{.}
  \end{equation}
  \Cref{fig:validation:multimag3D:ncconvergence} demonstrates that this relation is a very good approximation to estimate the relative error of the Lorentz-force ($\mathcal{Q}=\avg{\vec{F}}{t}$) with respect to $R_\infty$ according to the different meshes from \cref{tab:validation:multimag3D:ncmeshes}. To be specific, for a relative error of e.g. $1\%$ we need a size of $R_\infty \approx 4.6 R_J$ and for a relative error of $0.1\%$ we need a size of $R_\infty \approx 10.0 R_J$. With this rule-of-thumb, we again want to emphasise the advantage of using Biot-Savart inductors, where $R_J$ is defined by the size $R$ of the conducting region only. In case of discretised inductors, $R_J$ is given by the size of the coils and is thus significantly larger.

  \begin{table}
    \caption{Numerical meshes with different absolute size of the non-conducting region: The first column contains mesh scaling factors, $R$ is the radius of the conducting cylinder and $R_\mathrm{\infty}$ the radius of the non-conducting region. The total number of cells for each mesh is represented by $N$ and the number of cells in the conducting region by $N_\mathrm{C}$. In the last column, the computational time for one CPU-core is listed. The highlighted mesh (1.000) corresponds to the default from which refinement/coarsening was performed.}
    \label{tab:validation:multimag3D:ncmeshes}
    \centering
    \small
    \renewcommand{\arraystretch}{1.1}
    \begin{tabularx}{\textwidth}{X X l l l l}
      \hline
      Mesh &$R_\mathrm{\infty}/R$ &$N$ &$N_\mathrm{C}$ &$N/N_\mathrm{C}$ &$t_\mathrm{CPU}~[\unit{s}]$ \\
      \hline
      $0.300$ &$1.2$ &$251904$ &$120960$ &$2.08$ &$100$ \\
      $0.375$ &$1.5$ &$342912$ &$120960$ &$2.83$ &$140$ \\
      $0.500$ &$2.0$ &$447744$ &$120960$ &$3.70$ &$210$ \\
      $0.750$ &$3.0$ &$587520$ &$120960$ &$4.85$ &$320$ \\
      \rowcolor{hzdr-gray5}
      $1.000$ &$4.0$ &$698880$ &$120960$ &$5.78$ &$400$ \\
      $2.000$ &$8.0$ &$950400$ &$120960$ &$7.86$ &$660$ \\
      $4.000$ &$16.0$ &$1240320$ &$120960$ &$10.25$ &$1000$ \\
      $8.000$ &$32.0$ &$1534080$ &$120960$ &$12.68$ &$1500$ \\
      $16.000$ &$64.0$ &$1896960$ &$120960$ &$15.68$ &$2100$ \\
      $32.000$ &$128.0$ &$2298240$ &$120960$ &$19.0$ &$7800$ \\
      \hline
    \end{tabularx}
  \end{table}

  \begin{figure}
    \centering
    \small
    \resizebox{0.69\linewidth}{!}{\includegraphics{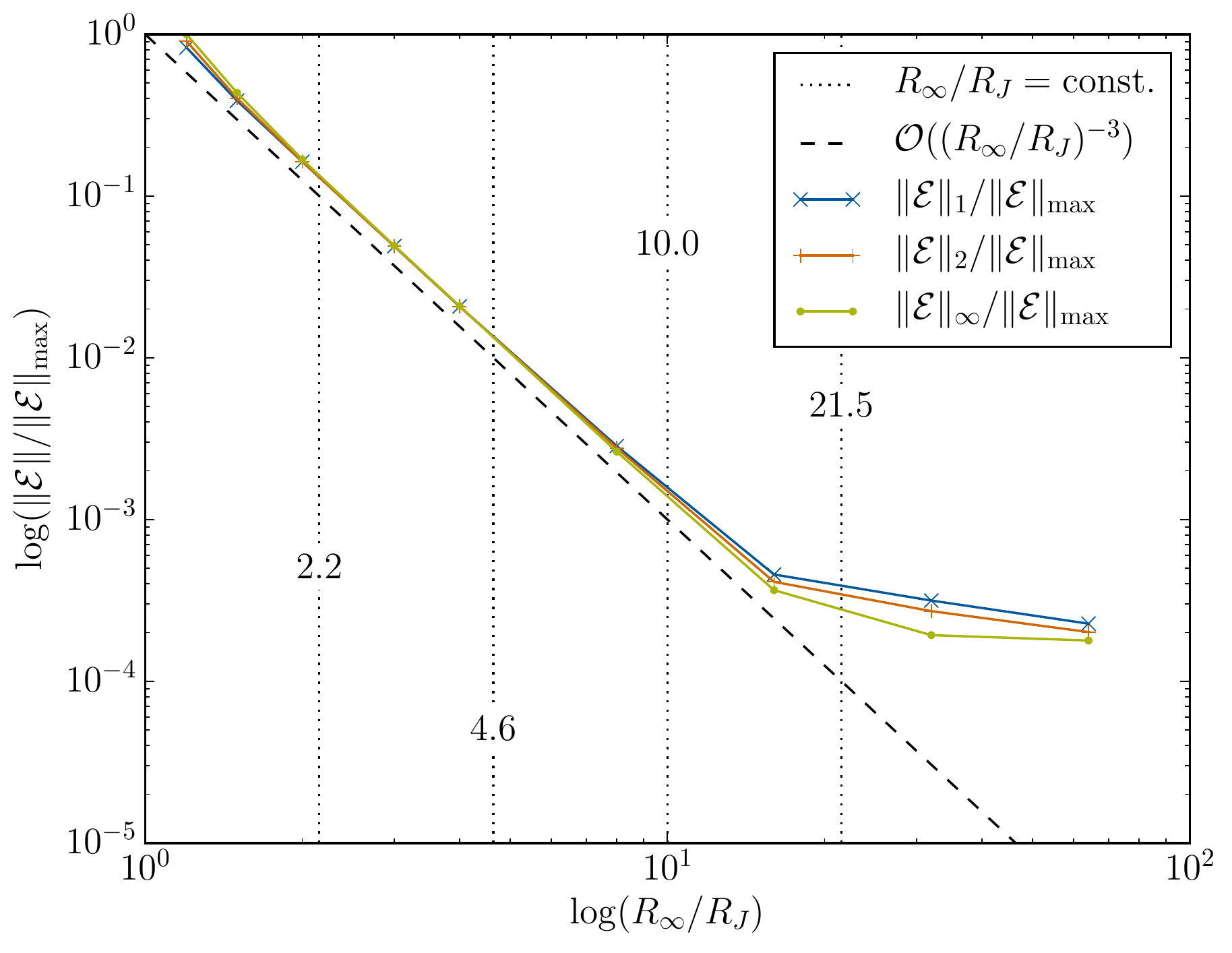}}
    \hspace*{0.19em}
    \renewcommand{\arraystretch}{1.1}
    \begin{tabularx}{0.26\linewidth}[b]{X X}
      \hline
      $B_\mathrm{\infty}/B_J$ &$R_\mathrm{\infty}/R_J$ \\
      \hline
      $1\times10^{-1}$ &$2.2$ \\
      $1\times10^{-2}$ &$4.6$ \\
      $1\times10^{-3}$ &$10.0$ \\
      $1\times10^{-4}$ &$21.5$ \\
      \hline
      \rule{0pt}{4.8ex}
    \end{tabularx}
    \captionlistentry[table]{}
    \captionsetup{labelformat=andtable}
    \caption{Convergence rate of the normalised error norms from \cref{seqn:errornorms} for the time-averaged Lorentz-force ($\mathcal{Q}=\avg{\vec{F}}{t}$) and different absolute sizes of the non-conducting region according to \cref{tab:validation:multimag3D:ncmeshes}. The largest mesh ($32.000$) has served as reference solution for the error-estimation. The norms are also compared to the ratio (dotted) from \cref{eqn:validation:multimag3D:diploeratior}. The table shows some selected values for $R_\mathrm{\infty}/R_J$, which are also marked in the plot.}
    \label{fig:validation:multimag3D:ncconvergence}
  \end{figure}

\subsection{Material boundary}
\label{subsec:validation:rgsSimpleBlock3D}

  A second validation case concerns an eddy-current problem of three inductor coils with equal phase shift $\alpha_\mathrm{0}=0$. The basic geometry for this setup was derived from our model of the RGS process \cite{ARTICLE_Beckstein_Galindo_Gerbeth_2015, INPROCEEDINGS_Beckstein_Gerbeth_Galindo_2016}. The inductor coils exactly match the ones from the RGS model, while only the bulk fluid-region with material properties of liquid silicon serves us as conducting domain. As a somewhat academical extension, we have reduced the electrical conductivity $\sigma$ in one half of the conductor by a factor of 12 in order to simulate a material change and validate our embedded discretisation from \cref{sec:embedded}.

  \begin{figure}
    \begin{subfigure}[c]{0.545\textwidth}
      \centering
      \resizebox{\linewidth}{!}{\includegraphics{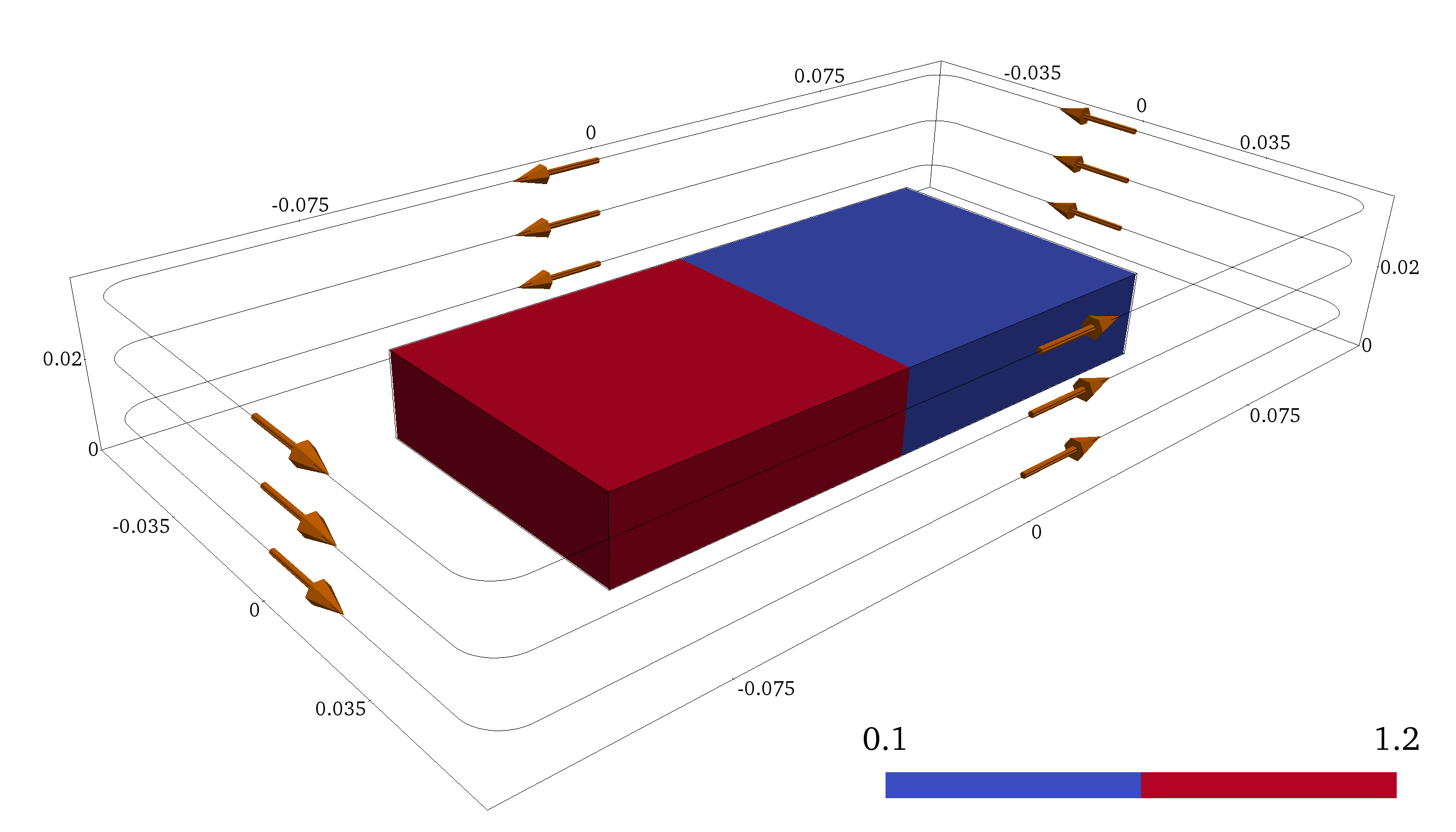}}
      \caption{Geometry and dimensions (in $[\unit{m}]$) of coils and the fluid region according to our model \cite{ARTICLE_Beckstein_Galindo_Gerbeth_2015, INPROCEEDINGS_Beckstein_Gerbeth_Galindo_2016} of the RGS process. The electrical conductivity $\sigma$ (in $[\times 10^{6}~\unitfrac{S}{m}]$) has been artificially reduced in one half of the conductor.}
      \label{sfig:validation:rgsSimpleBlock3D:sigma}
    \end{subfigure} \hspace*{1em} \vspace*{0.5em}
    \begin{subfigure}[c]{0.41\textwidth}
      \centering
      \resizebox{\linewidth}{!}{\includegraphics{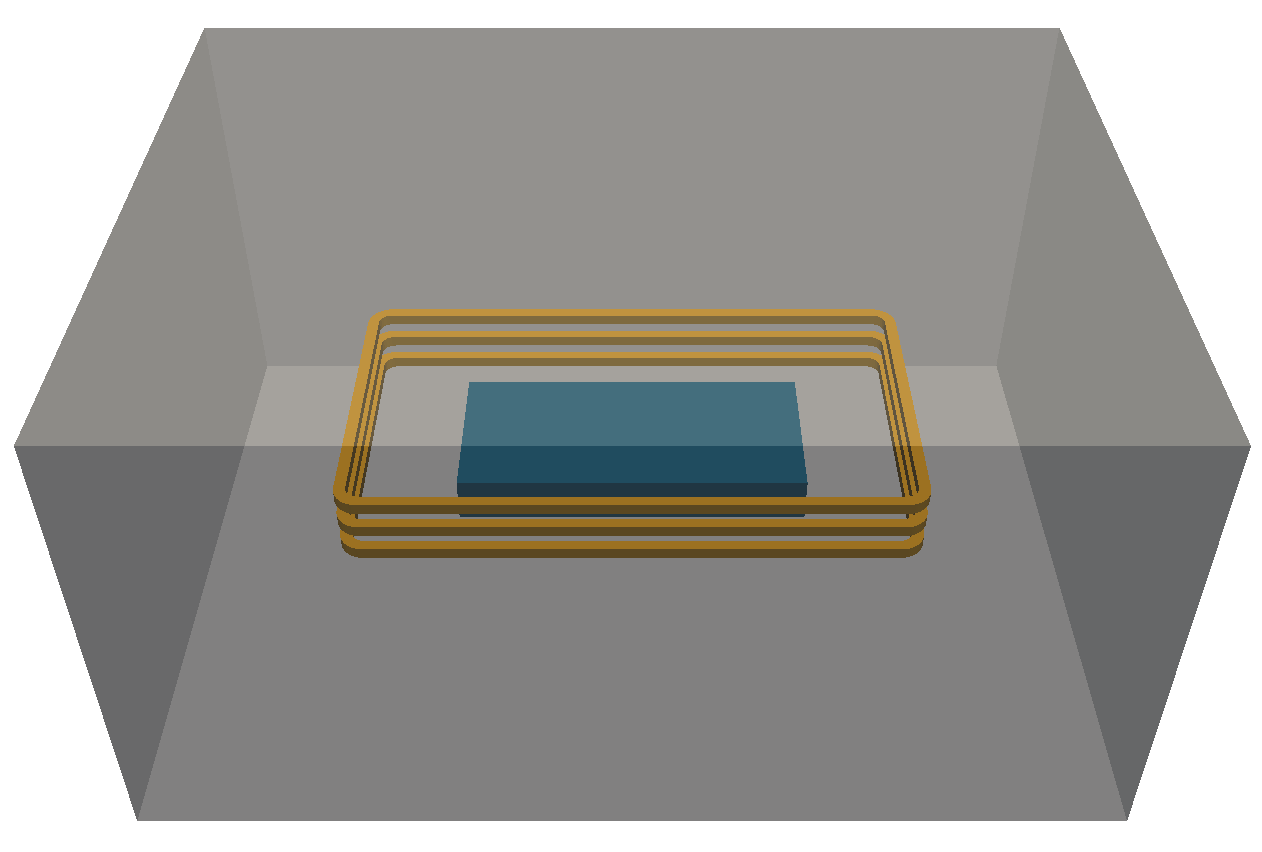}}
      \caption{Test case with conducting block, non-conducting domain and 3 excitation coils.}
      \label{sfig:validation:rgsSimpleBlock3D:sketch}
    \end{subfigure}
    \begin{subfigure}[c]{0.545\textwidth}
      \centering
      \resizebox{\linewidth}{!}{\includegraphics{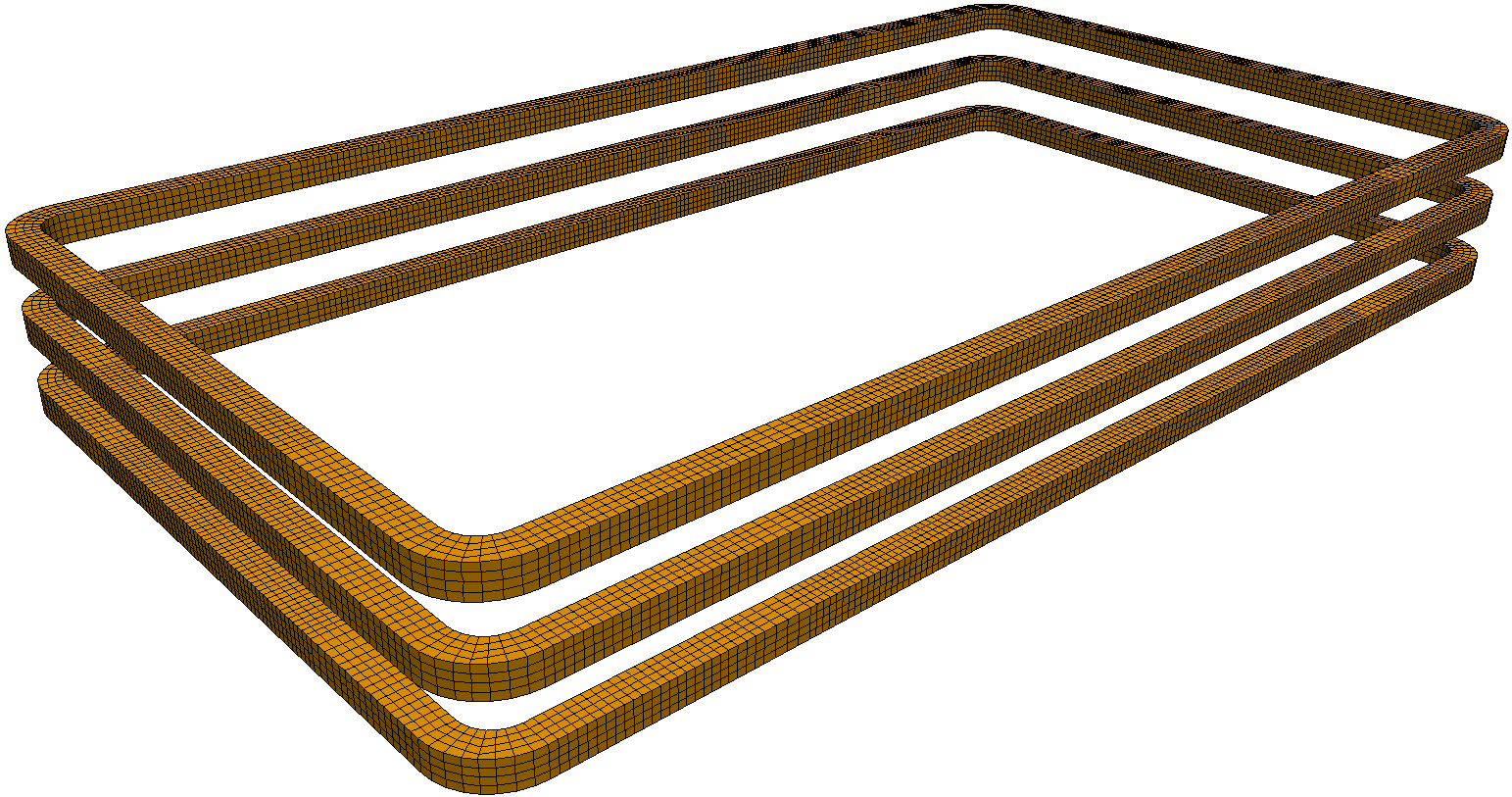}}
      \caption{Inductors with hexahedral mesh.}
      \label{sfig:validation:rgsSimpleBlock3D:coil}
    \end{subfigure} \hspace*{1em}
    \begin{subfigure}[c]{0.41\textwidth}
      \centering
      \resizebox{\linewidth}{!}{\includegraphics{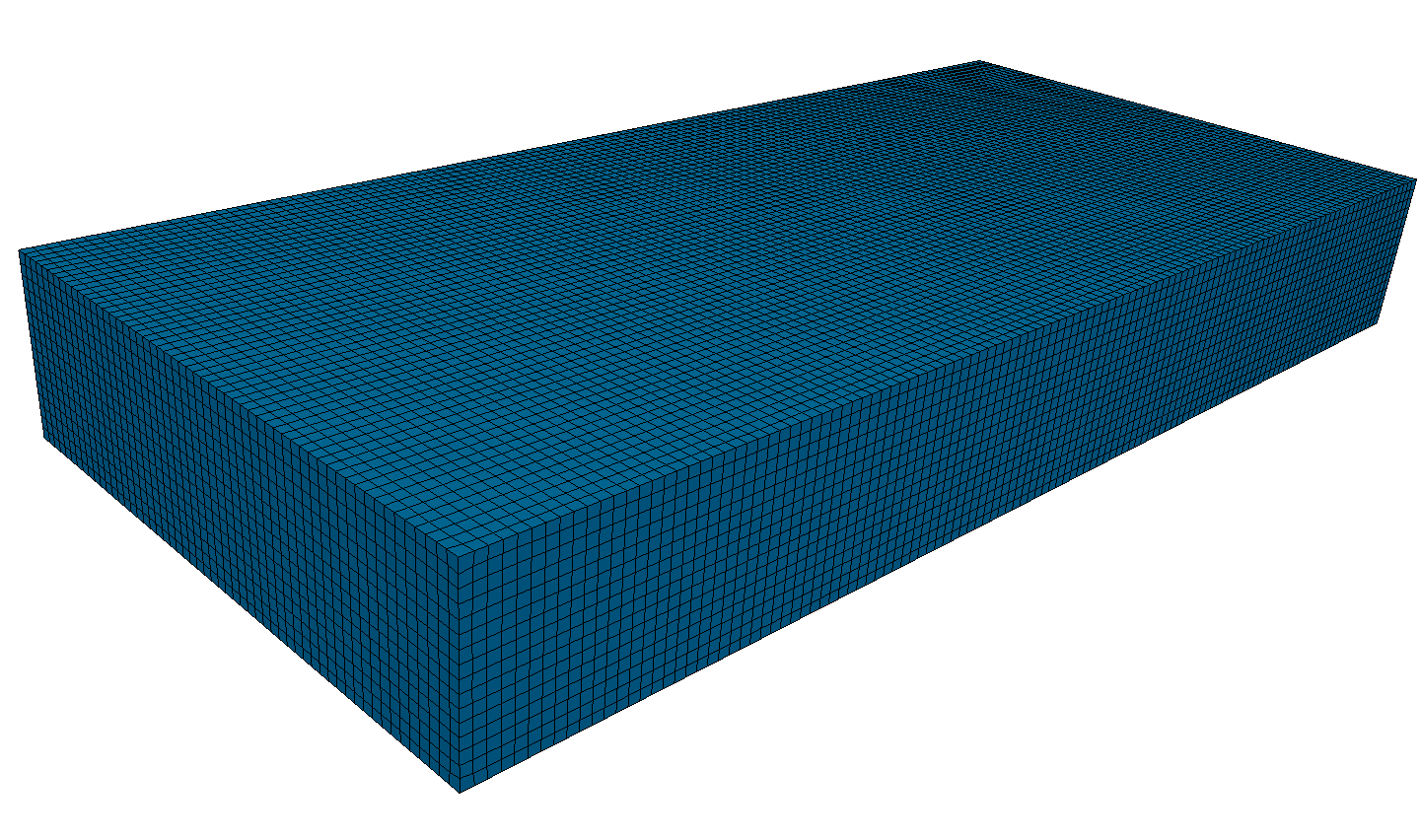}}
      \caption{Conducting block with hexahedral finite volume mesh.}
      \label{sfig:validation:rgsSimpleBlock3D:block}
    \end{subfigure}
    \caption{Material boundary test case: Motivation from the RGS \subref{sfig:validation:rgsSimpleBlock3D:sigma}, numerical model \subref{sfig:validation:rgsSimpleBlock3D:sketch} and inductor/conductor discretisation based on a finite volume mesh \subref{sfig:validation:rgsSimpleBlock3D:coil}/\subref{sfig:validation:rgsSimpleBlock3D:block}.}
    \label{fig:validation:rgsSimpleBlock3D:rgsSimpleBlock}
  \end{figure}

  In \cref{fig:validation:rgsSimpleBlock3D:rgsSimpleBlock}, an overview of the test case is given. \Cref{sfig:validation:rgsSimpleBlock3D:sigma} shows the basic setup. The conducting region measures $70~\unit{mm}$ in length ($x$-direction), $150~\unit{mm}$ in width ($y$-direction) and $20~\unit{mm}$ in height ($z$-direction). The jump discontinuity of $\sigma$ occurs at the centre in $y$-direction. An electrical conductivity of liquid silicon of $\sigma^+=1.2 \times 10^{6}~\unitfrac{S}{m}$ was used in one half and a value of $\sigma^-=1/12 ~\sigma^+ = 1 \times 10^{5}~\unitfrac{S}{m}$ in the other half.

  For this validation case, the inductor coil assembly with a span of $130~\unit{mm}$ in length, $240~\unit{mm}$ in width, corner radius of $8~\unit{mm}$, offset of $9~\unit{mm}$, an individual cross-section of $5~\unit{mm} \times 5~\unit{mm}$ and with its centroid being $13~\unit{mm}$ above the conductor's centroid, is situated within the non-conducting domain. It is fully discretised and part of the finite volume mesh which is illustrated in \cref{sfig:validation:rgsSimpleBlock3D:sketch}. The size of the non-conducting region is not to scale and actually dimensioned such that a ratio of $B_\mathrm{\infty}/B \approx 0.01$ was achieved.

\begin{figure}
    \begin{subfigure}[c]{\linewidth}
      \centering
      \resizebox{\linewidth}{!}{\includegraphics{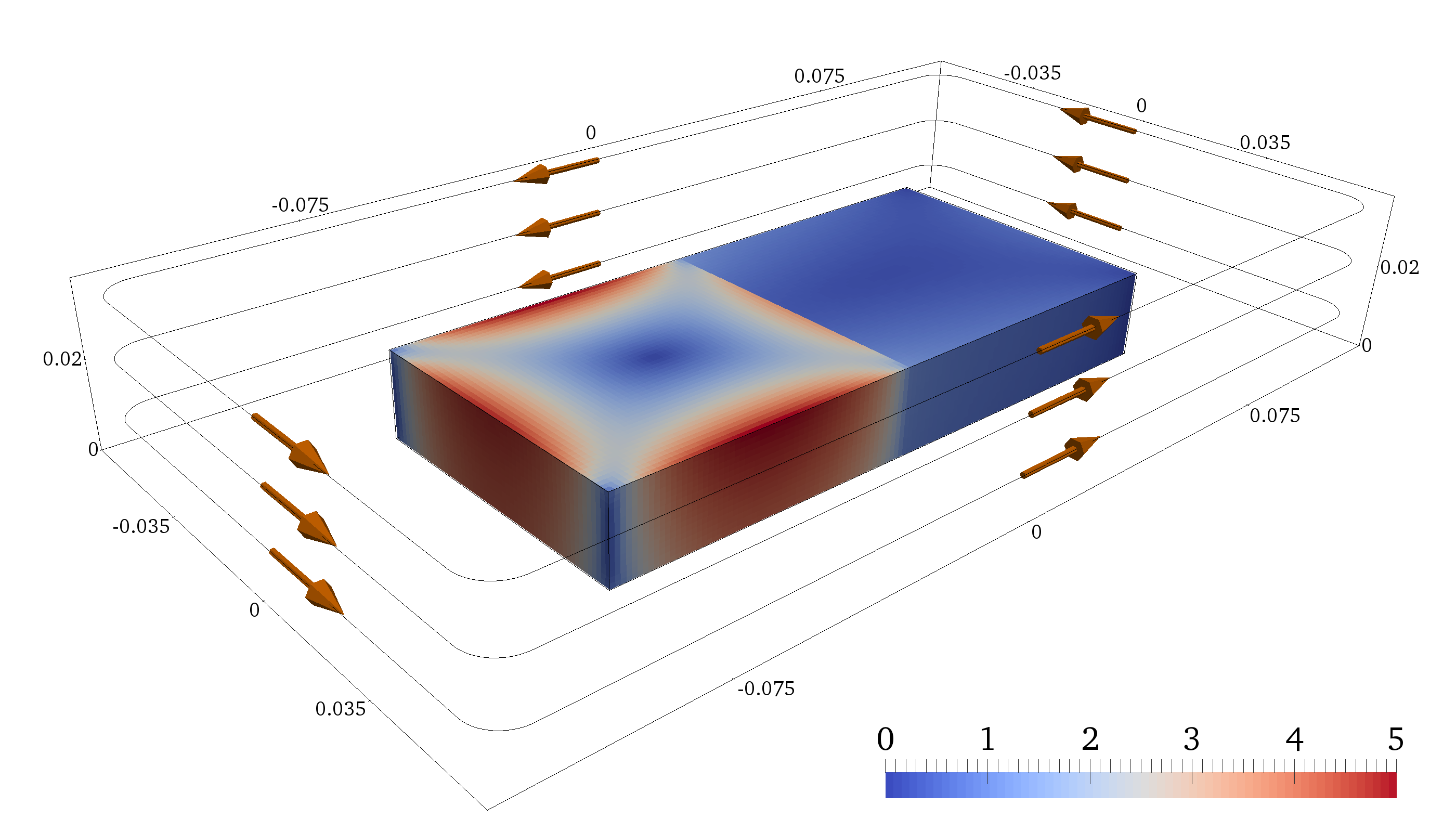}}
      \caption{Induced current density $\norm{\vec{j}'}$ (in $[\times 10^6 \unitfrac{A}{m^2}]$) for a coil frequency of $\omega_\mathrm{0}=1~\unit{kHz}$.}
      \label{sfig:validation:rgsSimpleBox3D:f001000Hz_j}
    \end{subfigure}
    \begin{subfigure}[c]{\linewidth}
      \centering
      \resizebox{\linewidth}{!}{\includegraphics{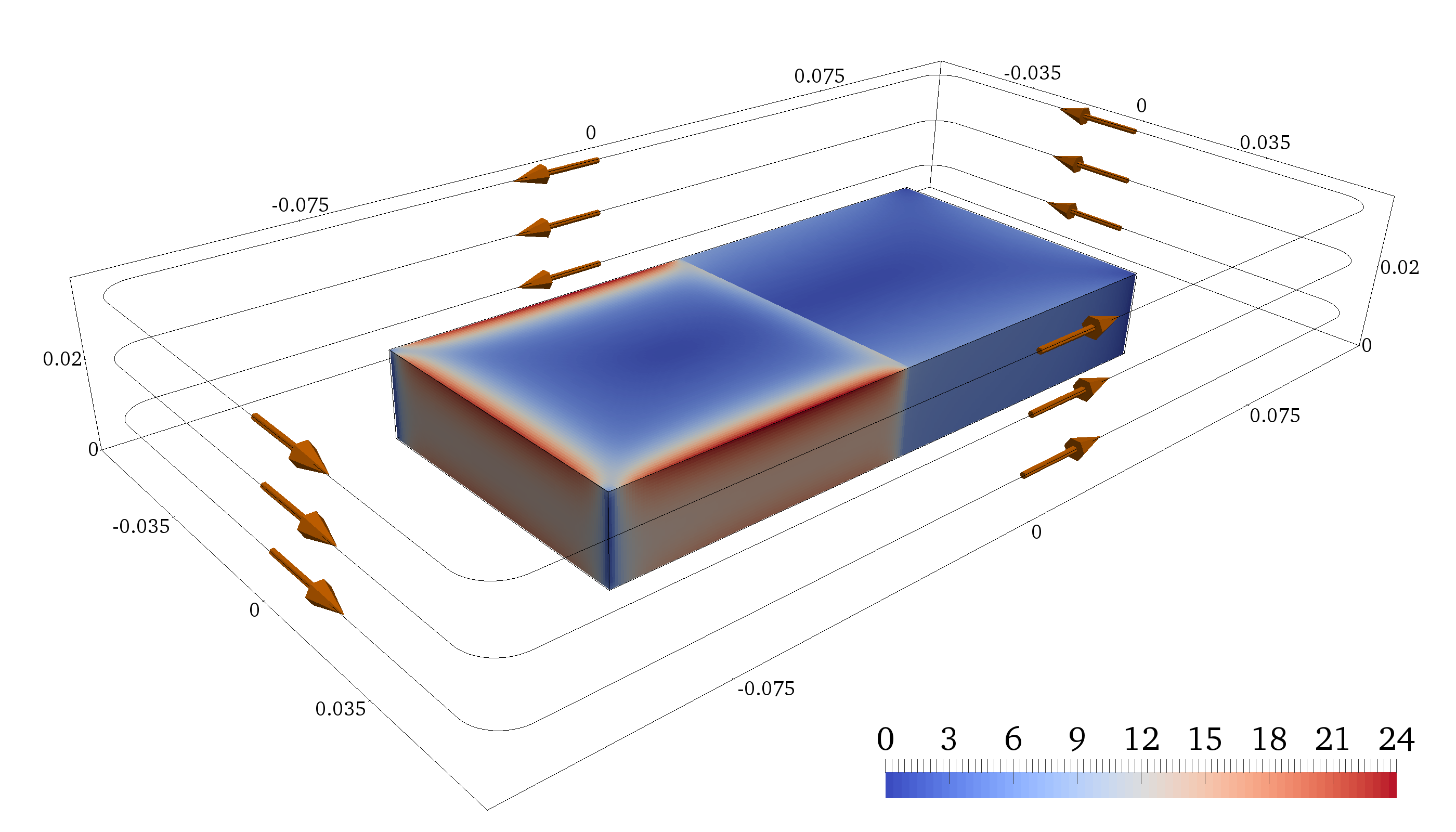}}
      \caption{Induced current density $\norm{\vec{j}'}$ (in $[\times 10^6 \unitfrac{A}{m^2}]$) for a coil frequency of $\omega_\mathrm{0}=10~\unit{kHz}$.}
      \label{sfig:validation:rgsSimpleBox3D:f010000Hz_j}
    \end{subfigure}
    \caption{Results for the amplitude of the induced current density $\vec{j}'$ from simulations with jump discontinuity in $\sigma$ for two different frequencies.}
    \label{fig:validation:rgsSimpleBlock3D:j}
  \end{figure}

  Both contour plots in \cref{fig:validation:rgsSimpleBlock3D:j} show numerical results for the induced current density from simulations with two different frequencies of $1~\unit{kHz}$ and $10~\unit{kHz}$. Thanks to our embedded discretisation, the flux continuity across the material boundary is preserved, while the jump in tangential current density is sharply resolved. The corresponding mesh consists of $N\approx 5\times 10^5$ cells, whereof $N\approx 1\times 10^5$ cells were used for the conducting region.

\begin{figure}
    \begin{subfigure}[c]{\linewidth}
      \centering
      \resizebox{\linewidth}{!}{\includegraphics{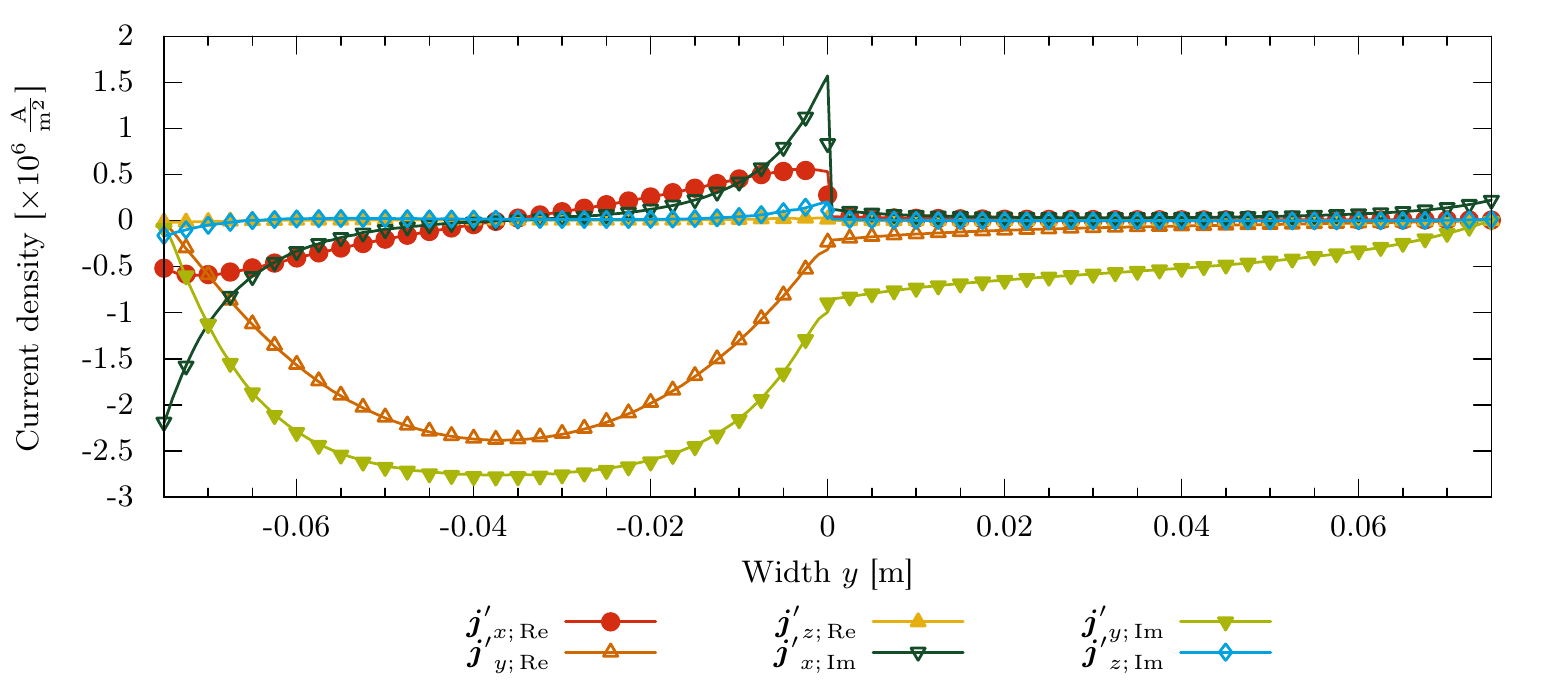}}
      \caption{Course of the induced current density for a coil frequency of $\omega_\mathrm{0}=1~\unit{kHz}$.}
      \label{fig:validation:rgsSimpleBox3D:ofmesh2_f001000Hz_line_y2_rgsSimpleBlock3D_f001000_j}
    \end{subfigure}
    \begin{subfigure}[c]{\linewidth}
      \centering
      \resizebox{\linewidth}{!}{\includegraphics{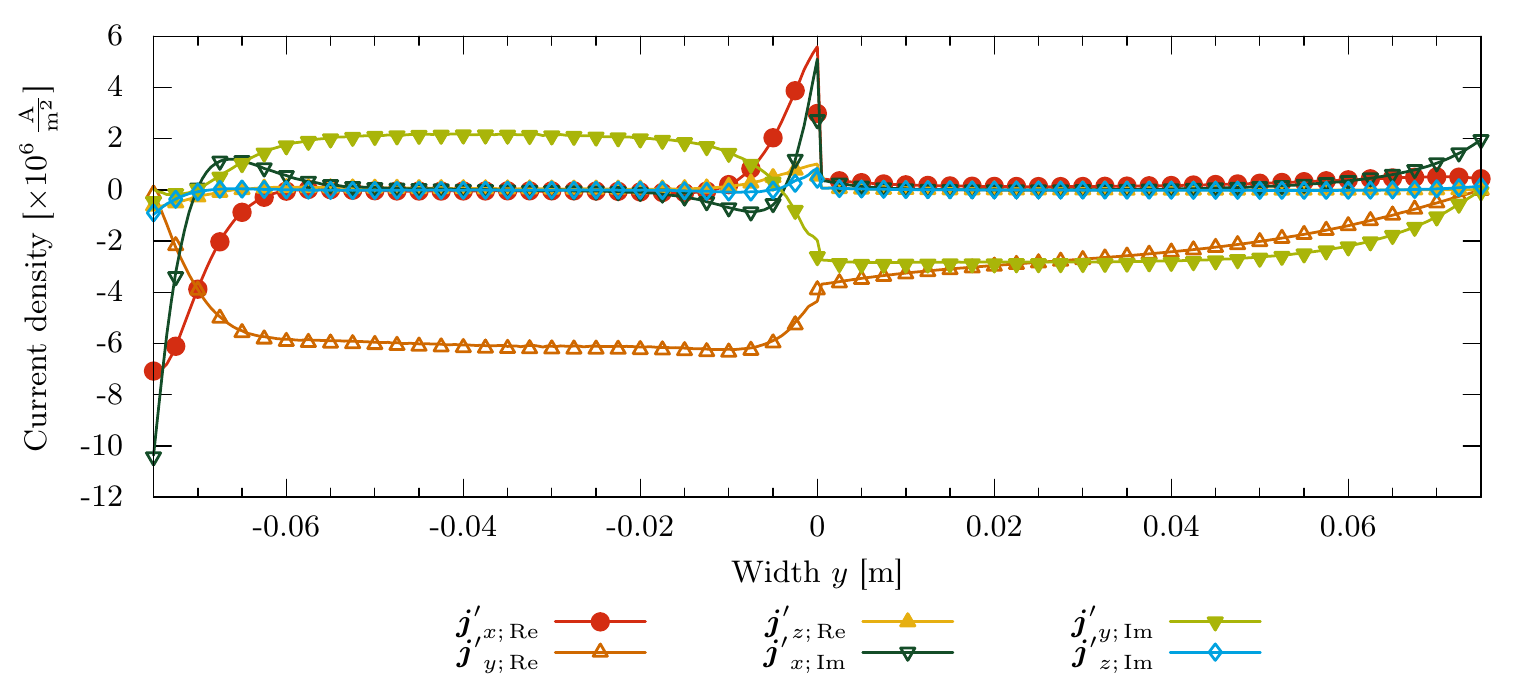}}
      \caption{Course of the induced current density for a coil frequency of $\omega_\mathrm{0}=10~\unit{kHz}$.}
      \label{fig:validation:rgsSimpleBox3D:ofmesh2_f001000Hz_line_y2_rgsSimpleBlock3D_f010000_j}
    \end{subfigure}
    \caption{Comparison of results of all components of the complex phasor of $\vec{j}'$ across the jump in $\sigma$ from our implementation in \textit{foam-extend} (solid lines) with results from the commercial FEM-software \textit{Cobham Opera 3D} \cite{MISC_Opera3D} (markers) for two different frequencies.}
    \label{fig:validation:rgsSimpleBox3D:ofmesh2_f001000Hz_line_y2_rgsSimpleBlock3D_j}
  \end{figure}

  By comparing the results from our simulations with results from the commercial FEM-software \textit{Cobham Opera 3D} \cite{MISC_Opera3D} and a corresponding mesh with $\approx 7 \times 10^6$ finite elements, we revealed a perfect agreement of any local data. \Cref{fig:validation:rgsSimpleBox3D:ofmesh2_f001000Hz_line_y2_rgsSimpleBlock3D_j} shows the local course of all components of the complex phasor from \cref{eqn:phasor} of $\vec{j}'$ across the jump in $\sigma$. The sampled data has been taken from a line along the $y$-direction through the centre of the conducting region (in terms of height), $5 ~\unit{mm}$ below the conductor surface in $x$-direction. The plots clearly present the continuous components in $y$-direction and occurring jumps in $x$-direction and $z$-direction.

\section{Conclusion}
\label{sec:conclusion}

  A new multi-mesh concept for the solution of 3D eddy-current problems in the finite volume framework of \textit{foam-extend} (\textit{OpenFOAM}) was presented in detail. The corresponding formulation is based on the magnetic vector potential and the electric scalar potential. A semi-coupled solution of both potentials is memory efficient and, due to the usage of block-coupled matrices, numerically robust also for higher frequencies. Even though it involves a non-conducting domain around the region of interest, the former is only used for the solution of the magnetic vector potential. By means of edge-based inductors in combination with splitting the magnetic vector potential into an impressed and reduced part, both the size of the non-conducting region and the computational costs of Biot-Savart's law were minimised. We have furthermore derived and included a special discretisation scheme to correctly resolve jump discontinuities in the electrical conductivity within the conducting region at material boundaries. An extensive validation for all aspects of the numerical approach was accomplished and demonstrated. The method is fully parallelised and may be readily coupled to models of other physical phenomena within the library of \textit{foam-extend}. Just recently we have investigated the three-dimensional behaviour of a free-surface flow under the influence of magnetic fields in the Ribbon Growth on Substrate process \cite{ARTICLE_Beckstein_Galindo_Gerbeth_2017}. Related numerical simulations have been performed by means of a successful and efficient integration of hydrodynamic and the presented electromagnetic methods within one simulation tool.


\section*{Acknowledgements}
\label{sec:acknowledgements}

  This work was funded by the German Helmholtz Association in frame of the Alliance Liquid Metal Technologies (LIMTECH), Project C2. Their financial support for research and development is gratefully acknowledged. Special thanks go to Prof. Hrvoje Jasak, Prof. Željko Tuković, Dr. Henrik Rusche and all other people involved in the NUMAP-FOAM Summerschool 2015 at the Faculty of Mechanical Engineering and Naval Architecture (FAMENA) in Zagreb for helpful lessons, fruitful debates and shared code.





\section*{References}

\bibliographystyle{elsarticle-num-names}
\hbadness 10000
\bibliography{paper}

\begin{thebibliography}{41}
\providecommand{\natexlab}[1]{#1}
\providecommand{\url}[1]{\texttt{#1}}
\providecommand{\urlprefix}{URL }
\expandafter\ifx\csname urlstyle\endcsname\relax
  \providecommand{\doi}[1]{doi:\discretionary{}{}{}#1}\else
  \providecommand{\doi}[1]{doi:\discretionary{}{}{}\begingroup
  \urlstyle{rm}\url{#1}\endgroup}\fi
\providecommand{\bibinfo}[2]{#2}

\bibitem[{Carpenter(1977)}]{ARTICLE_Carpenter_1977}
\bibinfo{author}{C.~J. Carpenter}, \bibinfo{title}{Comparison of alternative
  formulations of 3-dimensional magnetic-field and eddy-current problems at
  power frequencies}, \bibinfo{journal}{Proceedings of the Institution of
  Electrical Engineers} \bibinfo{volume}{124}~(\bibinfo{number}{11})
  (\bibinfo{year}{1977}) \bibinfo{pages}{1026--1034}.

\bibitem[{Biro(1999)}]{ARTICLE_Biro_1999}
\bibinfo{author}{O.~Biro}, \bibinfo{title}{Edge element formulations of eddy
  current problems}, \bibinfo{journal}{Computer Methods in Applied Mechanics
  and Engineering} \bibinfo{volume}{169} (\bibinfo{year}{1999})
  \bibinfo{pages}{391--405}.

\bibitem[{Biro and Preis(2000)}]{ARTICLE_Biro_Preis_2000}
\bibinfo{author}{O.~Biro}, \bibinfo{author}{K.~Preis}, \bibinfo{title}{An edge
  finite element eddy current formulation using a reduced magnetic and a
  current vector potential}, \bibinfo{journal}{IEEE Transactions on Magnetics}
  \bibinfo{volume}{36}~(\bibinfo{number}{5}) (\bibinfo{year}{2000})
  \bibinfo{pages}{3128--3130}.

\bibitem[{Xu and Simkin(2004)}]{ARTICLE_Xu_Simkin_2004}
\bibinfo{author}{E.~X. Xu}, \bibinfo{author}{J.~Simkin}, \bibinfo{title}{Total
  and reduced magnetic vector potentials and electrical scalar potential for
  eddy current calculation}, \bibinfo{journal}{IEEE Transactions on Magnetics}
  \bibinfo{volume}{40}~(\bibinfo{number}{2}) (\bibinfo{year}{2004})
  \bibinfo{pages}{938--940}.

\bibitem[{Biro and Valli(2007)}]{ARTICLE_Biro_Valli_2007}
\bibinfo{author}{O.~Biro}, \bibinfo{author}{A.~Valli}, \bibinfo{title}{The
  Coulomb gauged vector potential formulation for the eddy-current problem in
  general geometry: Well-posedness and numerical approximation},
  \bibinfo{journal}{Computer Methods in Applied Mechanics and Engineering}
  \bibinfo{volume}{196} (\bibinfo{year}{2007}) \bibinfo{pages}{1890--1904}.

\bibitem[{Stratton(1941)}]{BOOK_Stratton_Electromagnetic_Theory_1941}
\bibinfo{author}{J.~A. Stratton}, \bibinfo{title}{Electromagnetic Theory},
  \bibinfo{publisher}{McGraw-Hill}, \bibinfo{year}{1941}.

\bibitem[{Ferziger and Peri{\'
  c}(2002)}]{BOOK_FerzigerPeric_Computational_Methods_2002}
\bibinfo{author}{J.~H. Ferziger}, \bibinfo{author}{M.~Peri{\' c}},
  \bibinfo{title}{Computational Methods for Fluid Dynamics},
  \bibinfo{publisher}{Springer}, \bibinfo{edition}{3} edn.,
  \bibinfo{year}{2002}.

\bibitem[{Spitans et~al.(2014)Spitans, Baake, Nacke, and
  Jakovics}]{ARTICLE_Spitans_Baake_Nacke_Jakovics_2014}
\bibinfo{author}{S.~Spitans}, \bibinfo{author}{E.~Baake},
  \bibinfo{author}{B.~Nacke}, \bibinfo{author}{A.~Jakovics}, \bibinfo{title}{A
  numerical model for coupled free surface and liquid metal flow calculation in
  electromagnetic field}, \bibinfo{journal}{International Journal of Applied
  Electromagnetics and Mechanics} \bibinfo{volume}{44} (\bibinfo{year}{2014})
  \bibinfo{pages}{171--182}.

\bibitem[{Beckstein et~al.(2015)Beckstein, Galindo, and
  Gerbeth}]{ARTICLE_Beckstein_Galindo_Gerbeth_2015}
\bibinfo{author}{P.~Beckstein}, \bibinfo{author}{V.~Galindo},
  \bibinfo{author}{G.~Gerbeth}, \bibinfo{title}{Electromagnetic flow control in
  the Ribbon Growth on Substrate ({RGS}) process},
  \bibinfo{journal}{Magnetohydrodynamics}
  \bibinfo{volume}{51}~(\bibinfo{number}{2}) (\bibinfo{year}{2015})
  \bibinfo{pages}{385--396}.

\bibitem[{Beckstein et~al.(2016)Beckstein, Galindo, and
  Gerbeth}]{INPROCEEDINGS_Beckstein_Gerbeth_Galindo_2016}
\bibinfo{author}{P.~Beckstein}, \bibinfo{author}{V.~Galindo},
  \bibinfo{author}{G.~Gerbeth}, \bibinfo{title}{Modelling free-surface dynamics
  in the Ribbon Growth on Substrate process ({RGS})}, in:
  \bibinfo{booktitle}{Proceedings of the 10\textsuperscript{th} International
  PAMIR Conference on Fundamental and Applied MHD}, \bibinfo{address}{Cagliari,
  Italy}, \bibinfo{pages}{257--261}, \bibinfo{year}{2016}.

\bibitem[{Djambazov et~al.(2015)Djambazov, Bojarevics, Pericleous, and
  Croft}]{ARTICLE_Djambazov_Bojarevics_Pericleous_Croft_2015}
\bibinfo{author}{G.~Djambazov}, \bibinfo{author}{V.~Bojarevics},
  \bibinfo{author}{K.~Pericleous}, \bibinfo{author}{N.~Croft},
  \bibinfo{title}{Finite volume solutions for electromagnetic induction
  processing}, \bibinfo{journal}{Applied Mathematical Modelling}
  \bibinfo{volume}{39}~(\bibinfo{number}{16}) (\bibinfo{year}{2015})
  \bibinfo{pages}{4733--4745}.

\bibitem[{Haber et~al.(2000)Haber, Ascher, Aruliah, and
  Oldenburg}]{ARTICLE_Haber_Ascher_Aruliah_Oldenburg_2000}
\bibinfo{author}{E.~Haber}, \bibinfo{author}{U.~Ascher},
  \bibinfo{author}{D.~Aruliah}, \bibinfo{author}{D.~Oldenburg},
  \bibinfo{title}{Fast Simulation of 3D Electromagnetic Problems Using
  Potentials}, \bibinfo{journal}{Journal of Computational Physics}
  \bibinfo{volume}{163}~(\bibinfo{number}{1}) (\bibinfo{year}{2000})
  \bibinfo{pages}{150--171}.

\bibitem[{Aruliah et~al.(2001)Aruliah, Ascher, Haber, and
  Oldenburg}]{ARTICLE_Aruliah_Ascher_Haber_Oldenburg_2001}
\bibinfo{author}{D.~A. Aruliah}, \bibinfo{author}{U.~M. Ascher},
  \bibinfo{author}{E.~Haber}, \bibinfo{author}{D.~Oldenburg}, \bibinfo{title}{A
  Method for the forward modelling of 3-D electromagnetic quasi-static
  problems}, \bibinfo{journal}{Mathematical Models and Methods in Applied
  Sciences} \bibinfo{volume}{11}~(\bibinfo{number}{1}) (\bibinfo{year}{2001})
  \bibinfo{pages}{1--21}.

\bibitem[{MIS(2016{\natexlab{a}})}]{MISC_foam-extend}
\bibinfo{title}{The foam-extend project},
  \urlprefix\url{http://www.foam-extend.org},
  \bibinfo{year}{2016}{\natexlab{a}}.

\bibitem[{Jasak(1996)}]{PHDTHESIS_Jasak_1996}
\bibinfo{author}{H.~Jasak}, \bibinfo{title}{Error Analysis and Estimation for
  the Finite Volume Method with Applications to Fluid Flows}, Ph.D. thesis,
  \bibinfo{school}{Imperial College London}, \bibinfo{year}{1996}.

\bibitem[{Weller et~al.(1998)Weller, Tabor, Jasak, and
  Fureby}]{ARTICLE_Weller_Tabor_Jasak_Fureby_1998}
\bibinfo{author}{H.~Weller}, \bibinfo{author}{G.~Tabor},
  \bibinfo{author}{H.~Jasak}, \bibinfo{author}{C.~Fureby}, \bibinfo{title}{A
  tensorial approach to computational continuum mechanics using object oriented
  techniques}, \bibinfo{journal}{Computers in Physics}
  \bibinfo{volume}{12}~(\bibinfo{number}{6}) (\bibinfo{year}{1998})
  \bibinfo{pages}{620--631}.

\bibitem[{MIS(2016{\natexlab{b}})}]{MISC_OpenFOAM}
\bibinfo{title}{OpenFOAM - Free Open Source CFD},
  \urlprefix\url{http://www.openfoam.org}, \bibinfo{year}{2016}{\natexlab{b}}.

\bibitem[{Darwish et~al.(2009)Darwish, Sraj, and
  Moukalled}]{ARTICLE_Darwish_Sraj_Moukalled_2009}
\bibinfo{author}{M.~Darwish}, \bibinfo{author}{I.~Sraj},
  \bibinfo{author}{F.~Moukalled}, \bibinfo{title}{A coupled finite volume
  solver for the solution of incompressible flows on unstructured grids},
  \bibinfo{journal}{Journal of Computational Physics} \bibinfo{volume}{228}
  (\bibinfo{year}{2009}) \bibinfo{pages}{180--201}.

\bibitem[{Jareteg et~al.(2014)Jareteg, Vuk{\v c}evi{\' c}, and
  Jasak}]{INPROCEEDINGS_Jareteg_Vukcevic_Jasak_2014}
\bibinfo{author}{K.~Jareteg}, \bibinfo{author}{V.~Vuk{\v c}evi{\' c}},
  \bibinfo{author}{H.~Jasak}, \bibinfo{title}{pUCoupledFoam - An open source
  coupled incompressible pressure-velocity solver based on foam-extend}, in:
  \bibinfo{booktitle}{9th OpenFOAM Workshop}, \bibinfo{year}{2014}.

\bibitem[{Patankar and Spalding(1972)}]{ARTICLE_Patankar_Spalding_1972}
\bibinfo{author}{S.~Patankar}, \bibinfo{author}{D.~Spalding}, \bibinfo{title}{A
  calculation procedure for heat, mass and momentum transfer in
  three-dimensional parabolic flows}, \bibinfo{journal}{International Journal
  of Heat and Mass Transfer} \bibinfo{volume}{15}~(\bibinfo{number}{10})
  (\bibinfo{year}{1972}) \bibinfo{pages}{1787--1806}.

\bibitem[{Binns et~al.(1992)Binns, Lawrenson, and
  Trowbridge}]{BOOK_BinnsLawrensonTrowbridge_Electric_Magnetic_Fields_1992}
\bibinfo{author}{K.~J. Binns}, \bibinfo{author}{P.~J. Lawrenson},
  \bibinfo{author}{C.~W. Trowbridge}, \bibinfo{title}{The Analytical and
  Numerical Solution of Electric and Magnetic Fields},
  \bibinfo{publisher}{Wiley}, \bibinfo{year}{1992}.

\bibitem[{MIS(2016{\natexlab{c}})}]{MISC_Opera3D}
\bibinfo{title}{Opera-3D Design Software}, \urlprefix\url{http://operafea.com},
  \bibinfo{year}{2016}{\natexlab{c}}.

\bibitem[{Vuk{\v c}evi{\' c}(2016)}]{PHDTHESIS_Vukcevic_2016}
\bibinfo{author}{V.~Vuk{\v c}evi{\' c}}, \bibinfo{title}{Numerical Modelling of
  Coupled Potential and Viscous Flow for Marine Applications}, Ph.D. thesis,
  \bibinfo{school}{University of Zagreb}, \bibinfo{year}{2016}.

\bibitem[{Vuk{\v c}evi{\' c} et~al.(2017)Vuk{\v c}evi{\' c}, Jasak, and
  Gatin}]{ARTICLE_Vukcevic_Jasak_Gatin_2017}
\bibinfo{author}{V.~Vuk{\v c}evi{\' c}}, \bibinfo{author}{H.~Jasak},
  \bibinfo{author}{I.~Gatin}, \bibinfo{title}{Implementation of the Ghost Fluid
  Method for Free Surface Flows in Polyhedral Finite Volume Framework},
  \bibinfo{journal}{Computers \& Fluids}  (\bibinfo{year}{2017})
  \bibinfo{pages}{In Press}.

\bibitem[{Fedkiw et~al.(1999)Fedkiw, Aslam, and
  Xu}]{ARTICLE_Fedkiw_Aslam_Xu_1999}
\bibinfo{author}{R.~P. Fedkiw}, \bibinfo{author}{T.~Aslam},
  \bibinfo{author}{S.~Xu}, \bibinfo{title}{The Ghost Fluid Method for
  Deflagration and Detonation Discontinuities}, \bibinfo{journal}{Journal of
  Computational Physics} \bibinfo{volume}{154}~(\bibinfo{number}{2})
  (\bibinfo{year}{1999}) \bibinfo{pages}{393--427}.

\bibitem[{Huang et~al.(2007)Huang, Carrica, and
  Stern}]{ARTICLE_Huang_Carrica_Stern_2007}
\bibinfo{author}{J.~Huang}, \bibinfo{author}{P.~M. Carrica},
  \bibinfo{author}{F.~Stern}, \bibinfo{title}{Coupled ghost fluid/two-phase
  level set method for curvilinear body-fitted grids},
  \bibinfo{journal}{International Journal for Numerical Methods in Fluids}
  \bibinfo{volume}{44} (\bibinfo{year}{2007}) \bibinfo{pages}{867--897}.

\bibitem[{Desjardins et~al.(2008)Desjardins, Moureau, and
  Pitsch}]{ARTICLE_Desjardins_Moureau_Pitsch_2008}
\bibinfo{author}{O.~Desjardins}, \bibinfo{author}{V.~Moureau},
  \bibinfo{author}{H.~Pitsch}, \bibinfo{title}{An accurate conservative level
  set/ghost fluid method for simulating turbulent atomization},
  \bibinfo{journal}{Journal of Computational Physics}
  \bibinfo{volume}{227}~(\bibinfo{number}{18}) (\bibinfo{year}{2008})
  \bibinfo{pages}{8395--8416}.

\bibitem[{Lalanne et~al.(2015)Lalanne, Villegas, Tanguy, and
  Risso}]{ARTICLE_Lalanne_Villegas_Tanguy_Risso_2015}
\bibinfo{author}{B.~Lalanne}, \bibinfo{author}{L.~R. Villegas},
  \bibinfo{author}{S.~Tanguy}, \bibinfo{author}{F.~Risso}, \bibinfo{title}{On
  the computation of viscous terms for incompressible two-phase flows with
  Level Set/Ghost Fluid Method}, \bibinfo{journal}{Journal of Computational
  Physics} \bibinfo{volume}{301} (\bibinfo{year}{2015})
  \bibinfo{pages}{289--307}.

\bibitem[{Johansen and Colella(1998)}]{ARTICLE_Johansen_Colella_1998}
\bibinfo{author}{H.~Johansen}, \bibinfo{author}{P.~Colella}, \bibinfo{title}{A
  Cartesian Grid Embedded Boundary Method for Poisson’s Equation on Irregular
  Domains}, \bibinfo{journal}{Journal of Computational Physics}
  \bibinfo{volume}{147} (\bibinfo{year}{1998}) \bibinfo{pages}{60--85}.

\bibitem[{Crockett et~al.(2010)Crockett, Colella, and
  Graves}]{ARTICLE_Crockett_Colella_Graves_2010}
\bibinfo{author}{R.~K. Crockett}, \bibinfo{author}{P.~Colella},
  \bibinfo{author}{D.~T. Graves}, \bibinfo{title}{A Cartesian grid embedded
  boundary method for solving the Poisson and heat equations with discontinuous
  coefficients in three dimensions}, \bibinfo{journal}{Journal of Computational
  Physics} \bibinfo{volume}{230} (\bibinfo{year}{2010})
  \bibinfo{pages}{613--628}.

\bibitem[{Wang et~al.(2013)Wang, Glimm, Samulyak, Jiao, and
  Diao}]{ARTICLE_Wang_Glimm_Samulyak_Jiao_Diao_2013}
\bibinfo{author}{S.~Wang}, \bibinfo{author}{J.~Glimm},
  \bibinfo{author}{R.~Samulyak}, \bibinfo{author}{X.~Jiao},
  \bibinfo{author}{C.~Diao}, \bibinfo{title}{An Embedded Boundary Method for
  Two Phase Incompressible Flow}, \bibinfo{journal}{ArXiv e-prints} .

\bibitem[{Moreau(1990)}]{BOOK_Moreau_Magnetohydrodynamics_1990}
\bibinfo{author}{R.~Moreau}, \bibinfo{title}{Magnetohydrodynamics},
  \bibinfo{publisher}{Kluwer}, \bibinfo{year}{1990}.

\bibitem[{Davidson(2001)}]{BOOK_Davidson_An_Introduction_To_Magnetohydrodynamics}
\bibinfo{author}{P.~A. Davidson}, \bibinfo{title}{An Introduction to
  Magnetohydrodynamics}, \bibinfo{publisher}{Cambridge University Press},
  \bibinfo{year}{2001}.

\bibitem[{Weber et~al.(2013)Weber, Galindo, Stefani, Weier, and
  Wondrak}]{ARTICLE_Weber_Galindo_Stefani_Weier_Wondrak_2013}
\bibinfo{author}{N.~Weber}, \bibinfo{author}{V.~Galindo},
  \bibinfo{author}{F.~Stefani}, \bibinfo{author}{T.~Weier},
  \bibinfo{author}{T.~Wondrak}, \bibinfo{title}{Numerical simulation of the
  {T}ayler instability in liquid metals}, \bibinfo{journal}{New Journal of
  Physics} \bibinfo{volume}{15} (\bibinfo{year}{2013}) \bibinfo{pages}{043034}.

\bibitem[{Versteeg and
  Malalasekera(2007)}]{BOOK_VersteegMalalasekera_An_Introduction_To_Computational_Fluid_Dynamics_2007}
\bibinfo{author}{H.~K. Versteeg}, \bibinfo{author}{W.~Malalasekera},
  \bibinfo{title}{An introduction to Computational Fluid Dynamics: The finite
  volume method}, \bibinfo{publisher}{Prentice Hall}, \bibinfo{edition}{2}
  edn., \bibinfo{year}{2007}.

\bibitem[{Tukovi{\' c} and Jasak(2012)}]{ARTICLE_Tukovic_Jasak_2012}
\bibinfo{author}{{\v Z}.~Tukovi{\' c}}, \bibinfo{author}{H.~Jasak},
  \bibinfo{title}{A moving mesh finite volume interface tracking method for
  surface tension dominated interfacial fluid flow},
  \bibinfo{journal}{Computers \& Fluids} \bibinfo{volume}{55}
  (\bibinfo{year}{2012}) \bibinfo{pages}{70--84}.

\bibitem[{Pal et~al.(2009)Pal, Cramer, Gundrum, and
  Gerbeth}]{ARTICLE_Pal_Cramer_Gundrum_Gerbeth_2009}
\bibinfo{author}{J.~Pal}, \bibinfo{author}{A.~Cramer},
  \bibinfo{author}{T.~Gundrum}, \bibinfo{author}{G.~Gerbeth},
  \bibinfo{title}{{MULTIMAG} - A {MULTI}purpose {MAG}netic system for physical
  modelling in magnetohydrodynamics}, \bibinfo{journal}{Flow Measurement and
  Instrumentation} \bibinfo{volume}{20}~(\bibinfo{number}{6})
  (\bibinfo{year}{2009}) \bibinfo{pages}{241--251}.

\bibitem[{Gorbachev et~al.(1974)Gorbachev, Nikitin, and
  Ustinov}]{ARTICLE_Gorbachev_Nikitin_Ustinov_1974}
\bibinfo{author}{L.~Gorbachev}, \bibinfo{author}{N.~Nikitin},
  \bibinfo{author}{A.~Ustinov}, \bibinfo{title}{Magnetohydrodynamic rotation of
  an electrically conductive liquid in a cylindrical vessel of finite
  dimensions}, \bibinfo{journal}{Magnetohydrodynamics}
  \bibinfo{volume}{10}~(\bibinfo{number}{4}) (\bibinfo{year}{1974})
  \bibinfo{pages}{406--414}.

\bibitem[{Grants and Gerbeth(2001)}]{ARTICLE_Grants_Gerbeth_2001}
\bibinfo{author}{I.~Grants}, \bibinfo{author}{G.~Gerbeth},
  \bibinfo{title}{Stability of axially symmetric flow driven by a rotating
  magnetic field in a cylindrical cavity}, \bibinfo{journal}{Journal of Fluid
  Mechanics} \bibinfo{volume}{431} (\bibinfo{year}{2001})
  \bibinfo{pages}{407--426}.

\bibitem[{Jackson(1975)}]{BOOK_Jackson_Classical_Electrodynamics_1975}
\bibinfo{author}{J.~D. Jackson}, \bibinfo{title}{Classical Electrodynamics},
  \bibinfo{publisher}{Wiley}, \bibinfo{edition}{2} edn., \bibinfo{year}{1975}.

\bibitem[{Beckstein et~al.(2017)Beckstein, Galindo, and
  Gerbeth}]{ARTICLE_Beckstein_Galindo_Gerbeth_2017}
\bibinfo{author}{P.~Beckstein}, \bibinfo{author}{V.~Galindo},
  \bibinfo{author}{G.~Gerbeth}, \bibinfo{title}{Free-surface dynamics in the
  Ribbon Growth on Substrate ({RGS}) process}, \bibinfo{journal}{International
  Journal of Applied Electromagnetics and Mechanics}
  \bibinfo{volume}{53}~(\bibinfo{number}{S1}) (\bibinfo{year}{2017})
  \bibinfo{pages}{43--51}.

\end{thebibliography}


\end{document}